\documentclass[preprint]{aastex}
\shortauthors{JENKINS}
\shorttitle{GAS-PHASE ELEMENT DEPLETIONS IN THE ISM}
\begin{document}
\title{A Unified Representation of Gas-Phase Element Depletions in the 
Interstellar Medium \footnote{Based in large part on published observations 
from (1) the NASA/ESA {\it Hubble Space Telescope\/} obtained at the Space 
Telescope Science Institute, which is operated by the Association of 
Universities for Research in Astronomy, Inc., under NASA contract NAS 5-26555 
(2) the {\it Far Ultraviolet Spectroscopic Explorer (FUSE)\/} mission 
operated by Johns Hopkins University, supported by NASA contract NAS5-32985 
and (3) The {\it Copernicus\/} satellite, supported by NASA grant NAGW-77 to 
Princeton University.}}
\author{Edward B. Jenkins}
\affil{Princeton University Observatory\\
Princeton, NJ 08544-1001}
\email{ebj@astro.princeton.edu}
\begin{abstract}
A study of gas-phase element abundances reported in the literature for 17 
different elements sampled over 243 sight lines in the local part of our 
Galaxy reveals that the depletions into solid form (dust grains) are 
extremely well characterized by trends that employ only three kinds of 
parameters.  One is an index that describes the overall level of depletion 
applicable to the gas in any particular sight line, and the other two 
represent linear coefficients that describe how to derive each element's 
depletion from this sight-line parameter. The information from this study 
reveals the relative proportions of different elements that are incorporated 
into dust at different stages of grain growth. An extremely simple scheme is 
proposed for deriving the dust contents and metallicities of absorption-line 
systems that are seen in the spectra of distant quasars or the optical 
afterglows of gamma-ray bursts. Contrary to presently accepted thinking, the 
elements sulfur and krypton appear to show measurable changes in their 
depletions as the general levels of depletions of other elements increase, 
although more data are needed to ascertain whether or not these findings 
truly compelling.  Nitrogen appears to show no such increase.   The 
incorporation of oxygen into solid form in the densest gas regions far 
exceeds the amounts that can take the form of silicates or metallic oxides; 
this conclusion is based on differential measurements of depletion and thus 
is unaffected by uncertainties in the solar abundance reference scale.
\end{abstract}


\section{Introduction}\label{intro}

\subsection{Brief History}\label{history}

For atomic species in the neutral interstellar medium (ISM), nearly all of 
the transitions out of the ground electronic state of the preferred 
ionization stages for H~I regions occur in the ultraviolet part of the 
electromagnetic spectrum.  While the elements Be and Ti have such transitions 
at visible wavelengths (Dunham 1939; Spitzer \& Field 1955; Habing 1969),
only those from singly-ionized titanium are strong enough to yield detectable
absorption features in the spectra of background stars, leaving this element
as the only one that has been successfully surveyed from ground-based
observatories to derive the atomic abundances in the ISM (Wallerstein \&
Goldsmith 1974; Stokes 1978; Wallerstein \& Gilroy 1992; Welsh et al. 1997;
Prochaska, Tripp, \& Howk 2005; Ellison, Prochaska, \& Lopez 2007).  Attempts
even with powerful telescopes of the present era have failed to show
measurable amounts of Be~II in the ISM
(Boesgaard 1985; H\'ebrard et al. 1997).  Useful kinematic information could
be derived from the strong visible absorption features from atomic species in
stages below the preferred ones, e.g., Na~I and Ca~II 
(Merrill et al. 1937; Adams 1949; M\"unch 1957; M\"unch \& Zirin 1961), but
quantitative abundances were difficult to obtain owing to uncertainties in
our knowledge of the physical conditions and atomic physics parameters 
that govern the ionization balances (Str\"omgren 1948; Herbig 1968).

In large part, for investigations of the abundances of gas-phase atomic 
constituents in the ISM of our Galaxy, it is essential that this research be 
conducted by observatories above the Earth's atmosphere.\footnote{An 
exception was the {\it BUSS\/} balloon-borne payload which could observe over 
limited wavelength intervals in the near-UV (de Boer et al. 1986).} 
The earliest observations of UV stellar spectra at moderate resolution were 
carried out using small photographic spectrographs on sounding rockets, but 
the spectra were only good enough to sense the presence of interstellar 
features (Morton \& Spitzer 1966; Morton, Jenkins, \& Bohlin 1968), 
without permitting derivations any column densities except for that of atomic 
hydrogen based on L$\alpha$ absorption (Jenkins 1970, 1971).  The first
satellite to provide stellar spectra was the second {\it Orbiting
Astronomical Observatory} ({\it OAO-2\/}) that was launched in 1968, but once
again only L$\alpha$ absorption features provided any useful information on
the ISM (Savage \& Jenkins 1972; Jenkins \& Savage 1974).

The era of investigations of UV interstellar absorption features from heavy 
elements began in earnest with a series of observations with the far-UV 
spectrometer (Rogerson et al. 1973a)  aboard the {\it Copernicus\/} 
satellite, a facility that provided stellar spectra of high precision over a 
decade that started 37 years ago (Jenkins et al. 1973; Morton et al. 1973;
Rogerson et al. 1973b; Spitzer et al. 1973); see an early review of
highlights by Spitzer \& Jenkins (1975).  Among the principal findings that
emerged from these studies was that, to varying degrees, the abundances of
heavy elements in atomic form relative to that of hydrogen were below the
solar abundance ratios, which are presumed to approximate the true total
element abundances for the ISM in our part of the Galaxy.  The differences in
these two abundances were taken to represent the loss of atoms into solid
form within dust grains, a picture that was reinforced by the approximate
trend in the strengths of depletions relative to measures of chemical
affinity, such as condensation temperatures in a chemical equilibrium of an
extended stellar atmosphere (Field 1974) or atomic sticking probabilities
that can govern how rapidly different elements are incorporated into the
grains as they grow in the ISM (or, conversely, how easily they are returned
to the gas phase by sputtering).

The {\it International Ultraviolet Explorer\/} ({\it IUE\/}) extended our 
reach to stars that were fainter than those observable with {\it Copernicus} 
and allowed a far greater number of features and stars to be observed [e.g., 
Van Steenberg \& Shull (1988)], but the 
accuracies of the line measurements were not as good as those that were 
obtained previously with {\it Copernicus}.  Later, several new instrumental 
developments brought about considerable progress in the study of atomic 
absorption lines.  First, after the launch of the {\it Hubble Space 
Telescope\/} ({\it HST}) in 1992, we experienced substantial improvements in 
wavelength resolution and the ability to observe faint stars, which in turn 
broadened our knowledge of depletion trends over more elements and sightlines 
 (Savage \& Sembach 1996a).  This facility, in conjunction with the {\it 
Far Ultraviolet Spectroscopic Explorer\/} ({\it FUSE\/}) satellite that came 
later, also increased our coverage to systems outside our galaxy, such as the 
Small and Large Magellanic Clouds (Roth \& Blades 1997; Welty et al. 1997,
2001; Mallouris 2003; Sofia et al. 2006), NGC~1705 (Sahu \& Blades 1997),
I~Zw18 (Aloisi et al. 2003), NGC~625 (Cannon et al. 2005) and  SBS1543+593 
(Bowen et al. 2005).  Second, large aperture telescopes on the ground with
high resolution echelle spectrographs extended our reach to gas systems in
front of bright quasars at high redshift, where the UV transitions could be
viewed at visible wavelengths.  In recent years, new opportunities have arisen
to view gases within or in front of the host galaxies of gamma ray bursts
(GRBs), whose optical afterglows are often bright enough to permit
observations to be performed at high wavelength resolution. With the study of
these new systems came a higher level of complexity, since the results could
be influenced by not only dust depletion but also intrinsic abundance
differences of these objects, many of which are less chemically evolved than
our Galaxy (Pettini et al. 1994, 1997, 1999, 2002; Lu et al. 1996; Prochaska \& Wolfe
2002; Pettini 2003; Khare et al. 2004; Prochaska 2004; Kulkarni et al. 2005;
Wolfe, Gawiser, \& Prochaska 2005; P\'eroux et al. 2006a,b, 2008; Vladilo
et al. 2006; Prochaska et al. 2007; Vladilo, J. 2008; Calura et al. 2009). 
One of the aims of the present paper is to give some guidance on how these
two effects can be separated from each other (\S\ref{QSOALS}).

\subsection{General Findings on Depletions}\label{general_findings}

Aside from the differences in depletions from one element to the next, it has 
been noted that the overall strengths of depletions of many elements 
collectively show large changes over different lines of sight.  Attempts to 
understand these variations have been moderately successful, but not without 
some ambiguity.  For instance, Savage \& Bohlin (1979), Harris, et al.
(1984), Jenkins et al.  (1986) Jenkins (1987) and Crinklaw 
et al. (1994) showed that the depletion strengths 
correlated well with the average density of hydrogen along each sight line.  
This measure is admittedly a crude indication of local conditions, since one 
cannot distinguish a uniform density over a long extent from strongly 
clumped, denser material with a low filling factor.  Nevertheless Spitzer 
 (1985) created a simple model for explaining the 
observations in terms of random mixtures of three different kinds of clouds 
with different densities and depletion strengths.  A slightly different 
tactic was adopted by Savage \& Sembach (1996a), 
who summarized element depletions in terms of averages of warm (presumably 
low density) gas and cool (denser) gas, with some recognition of the 
differences between gas in the disk of the Galaxy and gas in the lower part 
of the halo.  In their study that compared column densities of Si~III and 
Al~III, Howk \& Savage (1999) showed that even 
fully ionized regions exhibited depletions onto dust grains.

Other ways of characterizing depletions for different lines of sight included 
simply comparing them to the hydrogen column density, i.e., not divided by 
the length of the sight line to obtain an average density (Wakker \& Mathis
2000) or, alternatively, comparing them to the fraction of hydrogen atoms
that are in molecular form
(Cardelli 1994; Snow, Rachford, \& Figoski 2002; Jensen,
Rachford, \& Snow 2005; Jensen, \& Snow 2007a,b). The former of the
two methods is one that is useful for lines of sight that extend out far from
the Galactic plane, where the effective length is of order of the scale
height times the cosecant of the Galactic latitude.  The latter represents an
attempt to obtain a more accurate index for sight-line conditions, but it has
the drawback of not factoring in variables other than local density that
govern the molecular fraction, such as the strength of dissociating radiation
field, the self shielding of the H$_2$ transitions, or the finite amount of
time needed to reach an equilibrium state.

All of the above comparisons with external sight-line variables were carried 
out for each element independently, with the recognition that the characters 
of the trends would differ from one case to the next but ultimately could be 
compared with each other.  A shortcoming of this approach is that the 
individual investigations are weakened by two factors that are difficult to 
control: (1) errors in the individual measurements (for both the depletions 
of single elements and the external variables) and (2) a lack of fidelity 
between the external variables and whatever real depletion processes that 
they are supposed to represent.

\subsection{New Approach}\label{new_approach}

The approach of the present study is to concentrate on how the depletions of 
different elements are found to relate to each other, irrespective of any 
external factors, but with the recognition that the severity of the 
depletions generally differ in a systematic way from one location to another 
and from one element to the next.  Our objective is to build a framework that 
describes the depletions in terms of a set of simple parameters that form an 
abstract model that can be used later to investigate some important issues on 
the formation and destruction of dust grains.   The general protocol is 
developed in \S\ref{underlying_strategy} based on an exposition of some 
empirical findings about element depletions; we demonstrate this by using two 
comparisons for three different elements that were chosen to give the most 
instructive examples.  On the basis of these findings, assuming they apply 
generally, one can develop an equation with a few free parameters that offers 
an acceptable fit to the observations (\S\ref{parameters}) and then use a 
comprehensive survey of depletions gathered from measurements that have 
already been published (\S\ref{data}) to derive the best solutions for the 
parameter values (\S\ref{solutions}).  In \S\ref{grain_buildup} we explore 
how these parameters offer insights on the elemental compositions of dust 
grains and how the mix in the buildup of these grains changes as the overall 
severity of depletions increase.  Here, we supply the basic information that 
can be used to build upon the earlier interpretations by Spitzer \& 
Fitzpatrick (1993), Fitzpatrick \& Spitzer 
 (1994), Sofia, Cardelli \& Savage 
 (1994), Sembach \& Savage (1996), Mathis (1996) and Draine (2004) on 
some plausible mixtures of grain compounds that are consistent with the 
amounts of missing gas atoms.

The conclusions presented here on the highly predictable patterns of element 
depletions open the way for investigations of intrinsic abundances in Damped 
Lyman Alpha (DLA) and other gas systems that are seen in absorption in the 
spectra of distant QSOs or the optical afterglows of gamma-ray bursts (GRBs).  
A straightforward method of compensating for the loss of material into dust 
grains in such systems is presented in \S\ref{QSOALS}, based on a simple 
least-squares fit to a linear equation (\S\ref{calc}), which then allows the 
total elemental abundances to be derived from the measured gas-phase column 
densities.  This approach is also useful for the determinations of depletions 
in our Galaxy for which measures of hydrogen are not available, such as 
surveys of matter in front of white dwarf stars that are at distances up to 
about 100~pc from the Sun (\S\ref{WD}).

\section{Concept}\label{concept}

The depletion of an element $X$ in the ISM is defined in terms of (a 
logarithm of) its reduction factor below the expected abundance relative to 
that of hydrogen if all of the atoms were in the gas phase,
\begin{equation}\label{depl_def}
[X_{\rm gas}/{\rm H}]=\log\{ N(X)/N({\rm H})\} -\log(X/{\rm H})_\odot ~,
\end{equation}
which is based on the assumption that solar abundances $(X/{\rm H})_\odot$ 
are good reference values that truly reflect the underlying total abundances.  
In this formula, $N(X)$ is the column density of element $X$ and $N({\rm H})$ 
represents the column density of hydrogen in both atomic and molecular form, 
i.e., $N({\rm H~I})+2N({\rm H}_2)$.  The missing atoms of element $X$ are 
presumed to be locked up in solids within dust grains or large molecules that 
are difficult to identify spectroscopically, with fractional amounts (again 
relative to H) given by
\begin{equation}\label{x_dust}
(X_{\rm dust}/{\rm H})=(X/{\rm H})_\odot (1-10^{[X_{\rm gas}/{\rm H}]})~.
\end{equation}

\subsection{Reference Abundances}\label{reference_abundances}

One issue that has bedeviled investigators of interstellar depletions has 
been the mutually inconsistent results from reasonable sources of information 
on the ``cosmic reference abundances,'' which, as stated above, are taken 
here to be $(X/{\rm H})_\odot$. Viewpoints on these abundance scales have 
changed through the years due either to different opinions on how they should 
be linked to the abundances in different types of stars (or H~II regions) or, 
alternatively, to actual changes in some of the measurement outcomes 
themselves (Mathis 1996; Savage \& Sembach 1996a; Snow \& Witt 1996; Sofia \&
Meyer 2001; Li 2005).  The most dramatic revisions in the recent past (and
some of the most important ones for the ISM) have been for the solar
abundances of the elements C, N and O relative to H, where the values have
decreased by about $-0.2$~dex from previously accepted scales (Grevesse \&
Sauval 1998), but which still exhibit disquieting fluctuations from one
determination to the next
(Holweger 2001; Allende Prieto, Lambert, \& Asplund 2002; Asplund et al.
2004, 2005; Socas-Navarro \& Norton 2007; Allende Prieto 2008; Centeno \&
Socas-Navarro 2008; Mel\'endez \& Asplund 2008; Caffau et al. 2009).  These
newer abundances now seem to be in accord with recent studies of B-star
abundances  (Daflon et al. 2003; Nieva \& Przybilla 2008a,b) but are
inconsistent with the results from helioseismological studies of the sound
speed and depth of the convection zone inside the Sun
(Bahcall et al. 2005a,b; Antia \& Basu 2005, 2006), which favor the
earlier abundances when the opacities are computed (Badnell et al. 2005).

For the study of interstellar depletions presented here, we adopt the 
abundances compiled by Lodders (2003) for the proto-Sun, 
which she estimates to have been higher by 0.07~dex than the current 
abundances in the solar photosphere (due to the fact that some gravitational 
settling may have occurred, thus depleting the photospheric abundances by 
this amount).  These abundances will be adopted to represent our view of the 
expected atomic densities relative to that of hydrogen when there are no dust 
grains or molecules present.  However, they are somewhat higher than the 
abundances of nearby B-type stars determined by Przybilla, Nieva \& Butler 
 (2008), suggesting that the 0.07~dex upward 
corrections may be too large.

One unique aspect of the study here will be the development of a way to learn 
about the depletion of atoms onto dust grains that does not depend on the 
correctness of the reference abundances.  This approach, which looks at 
differential changes in element atomic gas abundances instead of absolute 
depletions, will be presented in \S\ref{grain_buildup}.
\begin{figure}[h!]
\epsscale{0.7}
\plotone{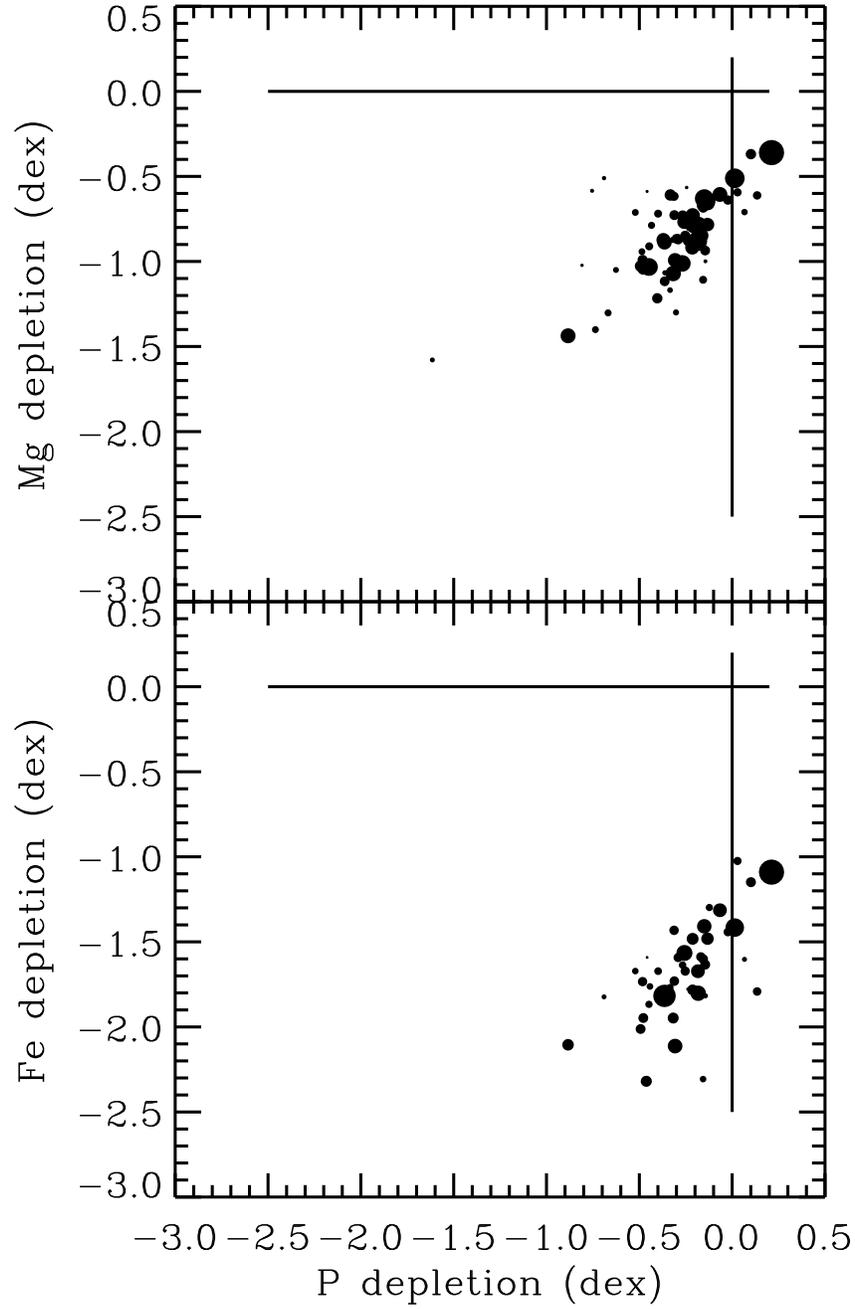}
\caption{Interstellar atomic depletion factors [Mg$_{\rm gas}$/H] ({\it top 
panel\/}) and [Fe$_{\rm gas}$/H] ({\it bottom panel\/}) vs. [P$_{\rm 
gas}$/H].  Measurements with small errors are depicted with disks that have 
large diameters, while those with large errors are more 
point-like.\label{abundf0}}
\end{figure}

\subsection{Underlying Strategy}\label{underlying_strategy}
  
To gain an understanding of how depletions behave under different conditions, 
it is helpful to compare the trends for pairs of elements over different 
lines of sight.  This comparison will lay the groundwork for the analysis 
that will be developed in later sections of this paper.  Figure~\ref{abundf0} 
shows two such comparisons, using depletions extracted from the data 
discussed in \S\ref{data}.  The depletions of Mg and P shown in the upper 
panel of the figure for a collection of individual sight lines exhibit an 
approximately linear relationship with respect to each other, with a slope 
slightly steeper than unity and with an intercept at the axis representing 
zero P depletion being about 0.5~dex below the zero axis for the Mg 
depletion.  This displacement of the zero point indicates that Mg shows a 
small amount of depletion when P is undepleted, but this conclusion is 
dependent on the accuracy of both the respective solar abundances of these 
two elements and the $f$-values of the transitions used to measure their 
interstellar abundances, together with the applicability of the assumption 
that solar abundances are a correct standard to apply to elements in the 
ISM.\footnote{The fact that some measurements of the abundances of P seem to 
be above the solar value demonstrates that depletion measurements are subject 
to these three types of systematic errors.  However, the $f$-value adopted 
for the strongest P line, a line that is most influential in measuring the 
abundance of P for sightlines with low column densities, has recently been 
re-examined by Federman et al. (2007), and their new 
measurements show that the earlier value of the line strength is essentially 
correct.}   While this may be so, the slope of the line has no sensitivity to 
these uncertainties; it depends only on the correctness of the derived column 
densities.  The lower panel of the figure shows that the same behavior is 
seen when the depletions of Fe and P are compared; it is important to note 
that the slope and intercept of a best fit to these points are larger in 
magnitude than for Mg and P.  A diagram similar to these two has been 
presented earlier by Fitzpatrick (1996) for the elements 
Fe and Si vs. S (see his Fig.~2), and the qualitative features are 
fundamentally similar.

\placefigure{abundf0}

By extension from these two examples, one can anticipate that there could be 
a linear relationship between all depletions; that is, in the hyperspace 
whose axes represent the strengths of depletions of all the different 
elements, an acceptable fit to the data for different sight lines will 
approximately conform to a single straight line.   An alternative to this 
representation is to describe the depletion of each element in terms of a 
linear relationship against a generalized depletion parameter that is common 
to all elements through some linear relations, but which can change from one 
line of sight to the next.  This approach will be adopted for this paper. 

\section{Definitions\protect\footnote{Throughout this paper, symbols with 
asterisk subscripts pertain to different sight lines (mnemonic aid: think of 
``*'' representing a target star), and $X$ subscripts denote different 
elements.} and Derivations of Parameters}\label{parameters}

The line-of-sight depletion strength factor, which we denote as $F_*$, 
represents how far the depletion processes have progressed collectively for 
all elements for any given case, i.e., a larger $F_*$ implies a stronger 
depletion for all elements.  Aside from a few exceptions discussed in 
\S\ref{other_considerations}, we are generally not able to sense true local 
depletions for different regions along any sight line, but only the results 
in the form of a single value for each element that represents a composite 
view of such depletions taken from samples seen in projection.  We examine 
the consequences of this kind of blend in a later section 
(\S\ref{outcomes_performance}), where an example with kinematically distinct 
parcels of gas with different depletions is analyzed as a single unit.

The slope of a best-fit line for the observed depletions $[X_{\rm gas}/{\rm 
H}]_{\rm obs}$ for any given element $X$ against $F_*$ may be called $A_X$, 
and this parameter represents the propensity of that element to increase (the 
absolute value of) its particular depletion level as $F_*$ becomes larger.   
For even the smallest observed values of $F_*$, most elements still show some 
depletion, as we learned from Fig.~\ref{abundf0}.  We denote this level of 
depletion as $[X_{\rm gas}/{\rm H}]_0$.

While the nature of $F_*$ has been defined, its numerical scale has not.  
That is, the values of $A_X$ and $[X_{\rm gas}/{\rm H}]_0$ are dependent on 
this scale, which is arbitrary.  We assign a value of 0 for $F_*$ to be that 
which corresponds to a sight line with the lowest collective depletions 
observed in our sample.  At the opposite extreme, we define $F_*=1.0$ to 
conform approximately to the strength of depletions for the low velocity 
component $(v_\odot=-15\,{\rm km~s}^{-1})$ seen toward the star $\zeta$~Oph 
(HD149757), for which the best studied and most detailed information is 
available for the depletions of many different elements (Savage, Cardelli, \&
Sofia 1992). The sight line toward this star has been generally regarded as a
prototype of the strong depletions that are seen in the cold, neutral medium 
(Savage \& Sembach 1996a).

\subsection{Determinations of \boldmath $F_*$}\label{determination_of_F_*}

Now that we have defined two fiducial values of $F_*$, we can state that a 
best fit of the experimental data to differing values of this depletion 
parameter is given by the simple linear relation
\begin{equation}\label{simple_law}
[X_{\rm gas}/{\rm H}]_{\rm fit}=[X_{\rm gas}/{\rm H}]_0+A_XF_*~,
\end{equation}
subject to the normalization conditions for $F_*$ stated earlier.  
Unfortunately, the observations of depletions that are needed to define the 
parameters $[X_{\rm gas}/{\rm H}]_0$, $A_X$ and $F_*$ are sparse.  All 
elements have only a partial coverage of the sight lines, and many sight 
lines have only a few column density determinations.  For this reason, 
coupled with the fact that each quantity is linked to the adopted values for 
the others, solutions for the values of $F_*$ for each of the sight lines had 
to be derived using a series of iterations, described in the next paragraph, 
for converging upon successively more accurate values of this quantity based 
on a set of provisional values $A_X^\prime$ of the fundamental parameters 
$A_X$ (which will be derived more accurately in a later calculation).   The 
values for $A_X^\prime$ are continually adjusted to make them more 
realistically follow the element depletions during the iteration cycles that 
improve upon the $F_*$ solution set.  We must also employ provisional values 
for the zero-point offsets $[X_{\rm gas}/{\rm H}]_0$, which we denote as 
$[X_{\rm gas}/{\rm H}]_0^\prime$.  New values of $[X_{\rm gas}/{\rm 
H}]_0^\prime$ are not derived in successive iteration steps because the 
accuracy of this parameter is not critical in defining $F_*$.  The 
convergence of these iterations was monitored and found to be quite rapid and 
not dependent on the initial values input to the first iteration (except for 
small, uniform increases or decreases in all of the $A_X^\prime$ values from 
one trial case to the next, which are compensated by uniform changes in the 
opposite sense for the $F_*$ set).

The calculations begin with an evaluation of a weighted average of the 
available observed depletions $[X_{\rm gas}/{\rm H}]_{\rm obs}$ for a given 
line of sight, after they have had their respective zero-point depletions 
subtracted and have been normalized to the values of $A_X^\prime$,
\begin{equation}\label{F_*}
F_*={\sum_X\{ W_X([X_{\rm gas}/{\rm H}]_{\rm obs}-[X_{\rm gas}/{\rm 
H}]_0^\prime)/ A_X^\prime\} \over\sum_XW_X}~,
\end{equation}
where the weight factors $W_X$ are set equal to the inverse squares of the 
combined uncertainties of the other terms in the sum, i.e.,
\begin{equation}\label{W_X}
W_X=\Big(\sigma\{ ([X_{\rm gas}/{\rm H}]_{\rm obs}-[X_{\rm gas}/{\rm 
H}]_0^\prime)/A_X^\prime\}\Big)^{-2}
\end{equation}
($W_X=0$ if the observation is unavailable for element $X$.)  For the initial 
evaluation of Eq.~\ref{F_*}, values of $A_X^\prime$ were set to the 
depletions of elements measured toward $\zeta$~Oph ($-15\,{\rm km~s}^{-1}$ 
velocity component) less the respective values of $ [X_{\rm gas}/{\rm 
H}]_0^\prime$, thus insuring that the scale for $F_*$ is approximately 
consistent with the definition stated earlier.  The uncertainties in our 
derived values for $F_*$ are given simply by the relation
\begin{equation}\label{sigma(F_*)}
\sigma(F_*)=\left(\sum_XW_X\right)^{-\onehalf}~.
\end{equation}
The weights $W_X$ in Eq.~\ref{F_*} guarantee that depletion measurements with 
large relative errors have a weak influence in the outcome.   These weights 
make use of error estimates for the quotients of the two terms that appear in 
Eq.~\ref{W_X}, $[X_{\rm gas}/{\rm H}]_{\rm obs}-[X_{\rm gas}/{\rm 
H}]_0^\prime$ and $ A_X^\prime$.  A conventional approach to evaluating these 
errors is to add in quadrature the relative errors of the two terms, yielding 
the relative error of the quotient.  However, this scheme breaks down when 
the error of the denominator $\sigma(A_X^\prime)$ is not very much less than 
the denominator's value $A_X^\prime$.  Appendix~\ref{quotient_errs} shows how 
one may compute errors in the quotients when the errors of the two terms are 
moderately large.

At the point that an initial solution set for $F_*$ is obtained, revised 
values for $A_X^\prime$ can be evaluated from a counterpart to Eq.~\ref{F_*} 
with summations $\sum_*$ over applicable lines of sight,
\begin{equation}\label{A_X^prime} 
A_X^\prime={\sum_*\{ W_*([X_{\rm gas}/{\rm H}]_{\rm obs}-[X_{\rm gas}/{\rm 
H}]_0^\prime)/ F_*\} \over\sum_*W_*}~,
\end{equation}
which, as with Eq.~\ref{sigma(F_*)}, has an uncertainty
\begin{equation}\label{sigma(A_X^prime)}
\sigma(A_X^\prime)=\left(\sum_*W_*\right)^{-\onehalf}
\end{equation}
for
\begin{equation}\label{W_*}
W_*=\Big(\sigma\{ ([X_{\rm gas}/{\rm H}]_{\rm obs}-[X_{\rm gas}/{\rm 
H}]_0^\prime)/F_*\}\Big)^{-2}~.
\end{equation}
Again, the scheme for deriving the errors of quotients is invoked to evaluate 
the error term that appears within the weight factor in Eq.~\ref{W_*}.  Once 
we have defined the improved values for $A_X^\prime$ and their associated 
errors using Eqs.~\ref{A_X^prime}-\ref{W_*}, we start over by repeating the 
evaluations of $F_*$.  We cycle through Eqs.~\ref{F_*} and \ref{A_X^prime} 
until the values of $F_*$ and $A_X^\prime$ stabilize.

\subsection{Determinations of \boldmath $A_X$, $B_X$ and 
$z_X$}\label{determination_of_ABz}

Now that values of $F_*$ and their associated uncertainties for all lines of 
sight have been established, we can improve upon the earlier representations 
of element depletions that made use of the provisional parameters $[X_{\rm 
gas}/{\rm H}]_0^\prime$ and $A_X^\prime$.  We accomplish this by using a more 
direct (noniterative) approach for determining new parameters that describe 
how the gas-phase abundances of element $X$ should be depleted for different 
values of $F_*$.   An improved linear form
\begin{equation}\label{better_equation}
[X_{\rm gas}/{\rm H}]_{\rm fit}=B_X+A_X(F_*-z_X)
\end{equation}
is a modification of Eq.~\ref{simple_law}, where the zero-point reference in 
$F_*$ is displaced to an intermediate value $z_X$, which is unique to element 
$X$ (instead of being at $F_*$ equal to 0), and the depletion at this point 
is called $B_X$.  For each element we can solve for values of the two 
coefficients $A_X$ and $B_X$ through the evaluation of a least-squares fit 
[using the routine {\tt FITEXY} described by Press et al. 
 (2007)] that properly accounts for differing 
errors in both the dependent and independent measurement variables that are 
being fitted, $[X_{\rm gas}/{\rm H}]_{\rm obs}$ and $F_*$.  The reason for 
replacing $[X_{\rm gas}/{\rm H}]_0$ (i.e., the expected depletion for 
$F_*=0$) in the earlier equation with $B_X-A_Xz_X$ is that for a choice
\begin{equation}\label{z_X}
z_X={\sum_* W_*F_*\over \sum_*W_*}~,
\end{equation}
where in this case
\begin{equation}\label{W_*2}
W_*=\{ [\sigma([X_{\rm gas}/{\rm H}]_{\rm 
obs})]^2+[\sigma(F_*)A_X^\prime]^2\} ^{-1}~,
\end{equation}
there is a near zero covariance between the formal fitting errors for the 
solutions of $B_X$ and $A_X$.  This independence for the uncertainties in the 
derived parameters makes them easier to state and comprehend, and it allows 
the errors in quantities that arise from linear combinations of $A_X$ and 
$B_X$ to be calculated in a straightforward fashion (see Eqs. 
\ref{sigma[X/H]_0} and \ref{sigma[X/H]_1} below).  The $W_*$ factors derived 
in Eq.~\ref{W_*2} account for errors in the individual observations which can 
be compounded by the effect of the uncertainties in $F_*$ (which become worse 
if $A_X^\prime$ is large).

In practical circumstances, one must add in quadrature an uncertainty in the 
reference solar abundance $\sigma(X/{\rm H})_\odot$ to the formal error 
$\sigma(B_X)$ that arises from the least-squares fit.  This is done in the 
listing of $B_X$ values for different elements 
(Table~\protect\ref{elem_parameters} that appears in \S\ref{solutions}) 
because the contents of this table are of general practical use, but 
deviations in the fit for each element shown much later in 
Tables~\protect\ref{C} to \protect\ref{Kr} of Appendix~\ref{basic_data} do 
not include the $\sigma(X/{\rm H})_\odot$ term so that one can judge better 
how well individual observations fit the general trend irrespective of any 
overall systematic errors in the solar abundances.  Some of the column 
density measurements were not included in the least-squares fitting; these 
cases are discussed in \S\ref{rejection}.  Results for the fit parameters and 
their errors will be presented in \S\ref{solutions}.

While the formulation given in Eq.~\ref{better_equation} allows us to use 
parameters that have more straightforward errors, it is nevertheless still 
useful to know what the values of $[X_{\rm gas}/{\rm H}]_{\rm fit}$ are for 
two fidicial values of $F_*$, one representing the smallest depletions 
($F_*=0$) and the other representing heavy depletions ($F_*=1$).  These two 
quantities are evaluated from the simple relations,
\begin{equation}\label{[X/H]_0}
[X_{\rm gas}/{\rm H}]_0=B_X-A_Xz_X~,
\end{equation}
with an error
\begin{equation}\label{sigma[X/H]_0}
\sigma([X_{\rm gas}/{\rm H}]_0)=\sqrt{\sigma(B_X)^2+[z_X\sigma(A_X)]^2}~,
\end{equation}
and
\begin{equation}\label{[X/H]_1}
[X_{\rm gas}/{\rm H}]_1=B_X+A_X(1-z_X)~,
\end{equation}
with an error
\begin{equation}\label{sigma[X/H]_1}
\sigma([X_{\rm gas}/{\rm H}]_1)=\sqrt{\sigma(B_X)^2+[(1-z_X)\sigma(A_X)]^2}
\end{equation}

\section{Accumulation and Processing of Data from the Literature}\label{data}

The depletion study can draw upon a substantial and diverse accumulation of 
atomic column density measurements that have been published in the 
astronomical literature over several decades.  One challenge in making use of 
these results is the creation of a good balance between two extremes in 
selecting the data: one being a strong discrimination in favor of the best 
quality results at the expense of obtaining a broad coverage of elements and 
sight lines, as opposed to the alternative of accepting nearly everything, 
good and bad, in an effort to lessen the perturbing effects of natural 
variations and to obtain a fuller representation of conditions in the ISM.   
By necessity, any reasonable compromise between these extremes will still 
entail the use of a very inhomogeneous mixture of data from different 
investigations that had different measurement methodologies and error 
estimation techniques.

To some extent, differences in data quality can be recognized in a proper 
fashion by using the stated uncertainties to govern the weighting of the 
terms used in the parameter estimations, as described earlier in 
\S\S~\ref{determination_of_F_*} and \ref{determination_of_ABz}.  A weakness 
of this approach is that different authors employ different standards for 
estimating errors, which is an effect that may partly undermine the validity 
of the weighting process.   It would be a monumental task to attempt to 
scrutinize each investigator's means of estimating errors and then adjust 
them according to some uniform standard.  Some papers did not state any 
errors in the column densities.  For these situations, conservative estimates 
were made for these works, and they are noted in the notes section of 
Table~\ref{fvals}.  This section also states some special considerations that 
applied to our treatment of the investigation in question.

Elements covered in the present study include carbon, nitrogen, magnesium, 
silicon, phosphorus, sulfur, chlorine, titanium, chromium, manganese, iron, 
nickel, copper, zinc, germanium, and krypton.  Argon was initially included 
in the compilation process, but there were too few reliable measurements 
taken to produce meaningful results, so this element was not considered 
further.  Moreover, the apparent abundance of Ar, as traced by its neutral 
form, is strongly susceptible to being altered in partially ionized regions 
 (Sofia \& Jenkins 1998).

\subsection{Special Treatment of Uncertainties Listed in one 
Survey}\label{JSS86}

The results shown by Jenkins, Savage \& Spitzer (1986) (hereafter JSS86)
were processed in a special manner.    These investigators quoted limits
for the column densities at the $2\sigma$ level of significance, which had
to be replaced by reasonable estimates for the $1\sigma$ results to be
consistent with the limits expressed by other investigators.  For lower
limits based on the strengths of weak lines that had a strong random noise
contribution, JSS86 found it appropriate to quote formal results that were
negative in situations where a chance positive intensity fluctuation
created a negative equivalent width.  This tactic was carried out to allow
anyone to overcome an upward bias in any general average that would arise
from blindly setting these lower limits to zero.  The following rules were
adopted for converting the $2\sigma$ lower limits to approximate
representations of the $1\sigma$ values: if the lower limit was positive, the
logarithm of the adopted new value would be a mean of the stated lower limit
for $\log N$ and the logarithm of the best value, on the presumption that
most of the time the relative likelihoods are symmetric in the logarithms. 
By contrast, for very weak detections these likelihoods are governed by
nearly linear processes, so that if the stated lower limit was negative (and
the best value was stated as a positive number), the mean would be evaluated
in terms of the linear representations of these values and would then be
converted into logarithmic form.  If, after evaluating this average, the
linear form was still negative, then the logarithm of the lower limit was set
to a very low (out of range) number so that subsequent processing would
recognize the best value as really an upper limit, even at the $1\sigma$
level.

For upper limits, the new limit was simply an average of the logarithms of 
the stated best and upper limits, except when the best value was stated as a 
negative (logarithmic) number and a linear average was used instead.  As an 
exception, when JSS86 stated that the upper limit corresponded to a situation 
where the weakest line had a central optical depth greater than 2 (designated 
by an asterisk in their tables), the old upper limit based on the $2\sigma$ 
level of confidence was retained as a hedge against possible errors arising 
from such saturations.  Finally, in the interest of following guideline nr.~2 
stated in \S\ref{rejection} below, any of the best values that had special 
notations (or a negative number) given in the tables of JSS86 were not 
considered for any of the values of generalized element depletions, but 
positive values were retained for expositions in tables and figures. 

\subsection{Criteria for Rejection}\label{rejection}

In recognition of the fact that error estimates in various works have their 
shortcomings, we invoke some quality control measures by implementing a few 
simple rules to bypass observations that may be of questionable validity.  
Moreover, in some cases, we must also reject investigations that include data 
that for good reasons probably misrepresent the trends that are under study 
here.
  
The following censorship rules were adopted for the selection and use of data 
incorporated in the current study:
\begin{enumerate}
\item Data on atomic column densities from observations using the {\it 
International Ultraviolet Explorer\/} ({\it IUE\/}) were not accepted, except 
for measurements of $N({\rm H~I})$ from the L$\alpha$ absorption feature.  
The limitations of wavelength resolution and the maximum achievable S/N of 
this facility decrease the likelihood that, with the usual small velocity 
dispersions found in the ISM\footnote{Exceptions to this are the lines of 
highly ionized species such as Al~III, Si~IV and C~IV (Savage, Meade, \&
Sembach 2001), which have large enough $b$-values to permit reliable
determinations of column densities.  These species are not relevant to the
present study however.}, proper corrections for saturation can be made for
absorption features that have large enough equivalent widths to be measured
accurately. Moreover, Massa et al. (1998) concluded that the {\it 
IUE\/} NEWSIPS reductions had serious photometric inconsistencies.  The {\it 
BUSS\/} balloon-borne payload (de Boer et al. 1986) recorded spectra 
of quality similar to those of {\it IUE\/}, and thus they likewise were not 
included in the current study.
\item There has been a persistent problem with investigators being 
overconfident about the reliability of column density measurements based 
exclusively on the equivalent widths of features on or very near the flat 
portion of the curve of growth.  For this reason, grounds for rejecting 
individual element abundances along any of the sight lines included 
determinations that did not have at least one absorption line that was 
clearly on the linear portion of the curve of growth, or alternatively at 
least 2 lines that were either on or not far above the linear part.  For 
instance, a low resolution observation of a doublet with a 2:1 strength ratio 
was deemed to be acceptable only if the ratio of the two equivalent widths 
was greater than 1.5.\footnote{Even though we are aware of the fact that a 
typical line of sight exhibits many separate velocity components, each with 
different column densities and velocity dispersions, a standard curve of 
growth analysis of the complete features that make up such an ensemble still 
yields an answer for the aggregate column density that is remarkably close to 
the correct answer, provided the features are not badly saturated.  An 
analysis by Jenkins (1986) indicates that our adopted lower 
cutoff for the doublet ratio is a very conservative choice.} However, mildly 
saturated single lines were accepted if the curve of growth was established 
by multiple lines of other species that were expected to behave in a similar 
fashion.\footnote{As shown originally by Routly \& Spitzer (1952), highly
depleted species (Ca~II) have more broadly distributed velocities than less
depleted ones (Na~I).  Hence using one element as a velocity surrogate for
another when multiple components are considered collectively must be done
with caution.  See also Jenkins  (2009).}   Deviations in the standards of
leniency in accepting such results varied from one study to the next, driven
by perceptions on how well the velocity structures were determined from the 
other species.   Cases based on saturated lines that were measured using the 
apparent optical depth method (Savage \& Sembach 1991; Jenkins 1996) for
lines recorded with sufficiently high resolution were deemed to be
acceptable as long as the apparent central optical depths were not too
large.  The application of this rule may have not been fully rigorous in all
cases, since for some investigations it was difficult to determine if
saturation effects created problems.  In some cases, results from saturated
lines were reported as lower limits for the column density; these values are
recorded as such and appear in Tables~\ref{C} to \ref{Kr} and
Figs~\ref{panel_set_1}$-$\ref{panel_set_4}, but they are not included in the
analyses of depletions.  Once in a while, very saturated lines could yield
column densities through the measurement of weak damping wings, and these
cases were accepted. \item If any particular line of sight shows $\log N({\rm
H}) < 19.5$, there is a reasonable possibility that a nonnegligible fraction
of the atoms of any element is in a stage (or stages) of ionization above the
preferred one (i.e., the lowest level with an ionization potential greater
than 13.6~eV), because the gas is not well shielded from ionizing photons
that have energies well above that of the Lyman limit of hydrogen; see a
discussion of this topic by Jenkins (2004).  Aside from this possibility, 
there is also a chance that contributions from fully ionized gas could 
contribute to the column density of an ion in its expected preferred stage 
for an H~I region.  If $N({\rm H~I})$ is small, the relative level of 
contamination from one or more H~II regions (including the one surrounding 
the target star) could be large.  Such regions, of course, do not make any 
contribution to the L$\alpha$ absorption that is used to measure $N({\rm 
H~I})$.  While it is true that in principle one can estimate the probable 
contamination from partially or fully ionized regions by comparing the 
amounts of more highly ionized forms of some elements to their singly-ionized 
stages (Cardelli, Sembach, \& Savage 1995; Sembach \& Savage 1996; Howk \&
Savage 1999; Howk \& Sembach 1999; Prochaska et al. 2002; Howk, Sembach, \&
Savage 2003), a uniform application of any corrections for ionization is not
feasible in a generalized study such as this one, where varying degrees of
coverage of various ions emerge from diverse sources in the literature. 
Aside from ionization corrections, for low column density cases there is also
a possibility that material very close to the star, in the form of either a
shell or disk, could have its own resonance absorption features (Snow,
Peters, \& Mathieu 1979; Oegerle \& Polidan 1984) that could distort the
results for the ISM.  The imposition of a single column density threshold for
all cases may seem like a blunt instrument, given the large variations of
conditions encountered in this survey.  Indeed, we argue later (\S\ref{WD})
that the criterion $\log N({\rm H}) > 19.5$ is probably too conservative for
short sight lines ($d \lesssim 100$~pc).  While this may be so, there is
evidence that over much longer sight lines that could hold multiple absorbing
clouds a higher threshold for $N$(H) may be more appropriate.  For instance,
Howk, Sembach \& Savage (2006) show an example where small systematic
abundance shifts caused by ionization along an extended line of sight can
arise even for $N({\rm H~I})=10^{20}{\rm cm}^{-2}$.  While lines of sight
that had $\log N({\rm H}) < 19.5$ were not used for determining the
parameters $A_X^\prime$, $A_X$, $B_X$ and $z_X$, their values of $F_*$ were
still evaluated so that the results could be shown in the figures and be
tabulated.  The reader is advised to consider these results with some
caution however.  Figure~\ref{abundf3} shows the distribution of total
hydrogen column densities $N$(H) covered in this study.
\placefigure{abundf3}
\begin{figure}
\epsscale{1.}
\plotone{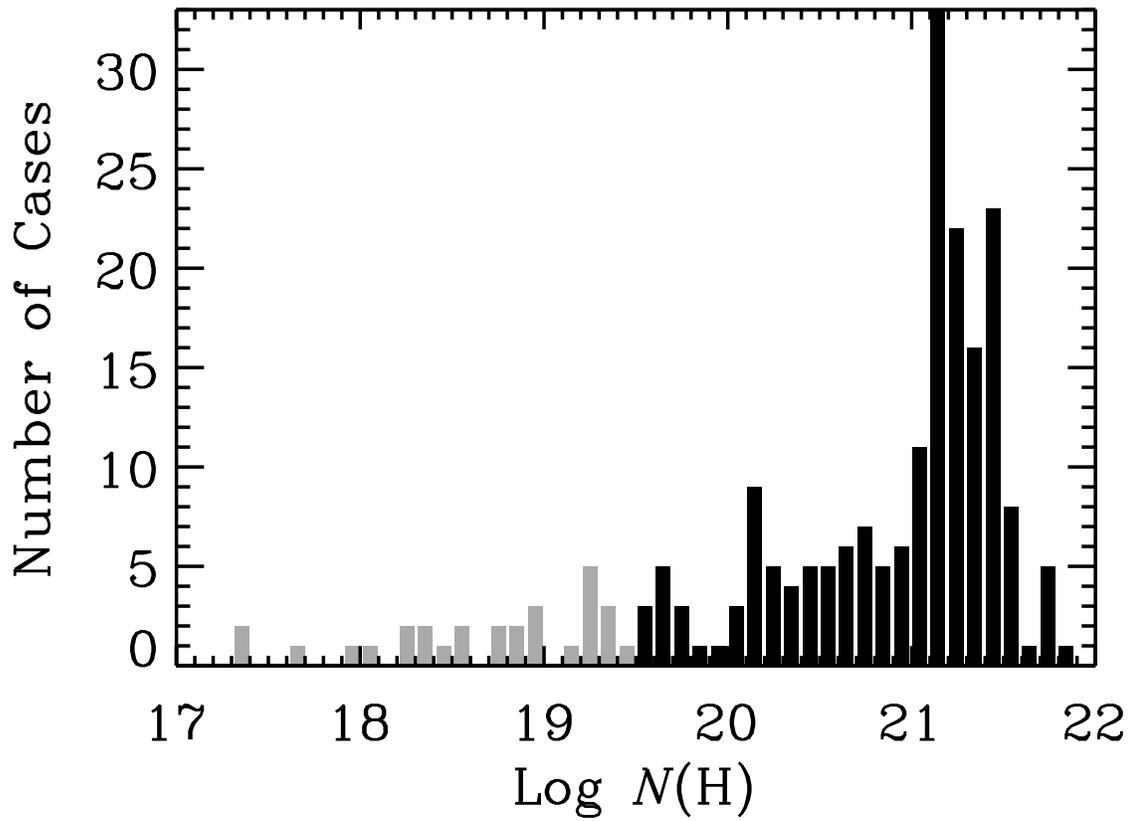}
\caption{The distribution of total hydrogen column densities (atomic and 
molecular) for the sightlines studied in this survey.  The gray bars 
represent cases below the cutoff $N({\rm H})=10^{19.5}{\rm cm}^{-2}$ 
established for defining the best-fit element parameters.\label{abundf3}}
\end{figure}

\item Stars for which only one or two elements were observed were not 
included in the analysis, since they were of little or no value in 
constraining the comparisons of depletions for different elements.  
Nevertheless, as with stars that had $\log N({\rm H}) < 19.5$, their $F_*$ 
values were computed and entries appear in the tables and figures. 
\item Any absorbing systems well outside the disk of our Galaxy are excluded 
(e.g., those in either of the Magellanic Clouds or the Magellenic Stream), 
since their intrinsic element abundances are not known.  Stars in the lower 
halo of the Galaxy are included, but the analysis is restricted to gas at low 
and intermediate velocities.  Infalling gas at high velocities (Wakker 2001)
is not included in this study, since the kinematics of this material point to 
an origin outside the Galactic disk, and the intrinsic abundances of the 
heavy elements are low (Wakker et al. 1999; Murphy et al. 2000; Gibson et al.
2001; Richter et al. 2001; Collins, Shull, \& Giroux 2003; Tripp et al.
2003). \item Gas that is explicitly identified to be in shocks at
extraordinarily high velocity are not included.  (Such clouds generally have
$\log N({\rm H~I}) < 19.5$ anyway, as indicated by their abundances of S~II).
 Grains are usually destroyed in such gas (Jenkins, Silk, \& Wallerstein
1976), and photons generated in the shock front (Shull \& McKee 1979)
can raise the ionization level of the post-shock material (Jenkins et al.
1998). \item A diverse set of observations has revealed that the intrinsic 
abundances of the elements show a gradient with galactocentric distance 
$R_{\rm GC}$ that ranges from about $-0.03$ to $-0.10\,{\rm dex~kpc}^{-1}$
(Shaver et al. 1983; Afflerbach, Churchwell, \& Werner 1997; Gummershach et
al. 1998; Deharveng et al. 2000; Martins \& Viegas 2000; Rolleston et al.
2000; Giveon et al. 2002; Mart\'in-Hern\'andez et al. 2002; Luck, Kovtyukh, \&
Andrievsky 2006).  To limit the influence of these changes on our results, we
do not consider stars outside the range $7<R_{\rm GC}<10\,{\rm kpc}$
(projected onto the plane of the Galaxy) for the determinations of $A_X$,
$B_X$ and $z_X$, however they are plotted in Figs.~\ref{panel_set_1}$-
$\ref{panel_set_4} and appear in Tables~\ref{nh_fstar} and \ref{C}$-
$\ref{Kr}.  (Later, in \S\ref{regional}, we show that an abundance gradient
does not seem to be evident within the entire collection of sight lines
studied here.) \end{enumerate}

\begin{figure}[h!]
\plotone{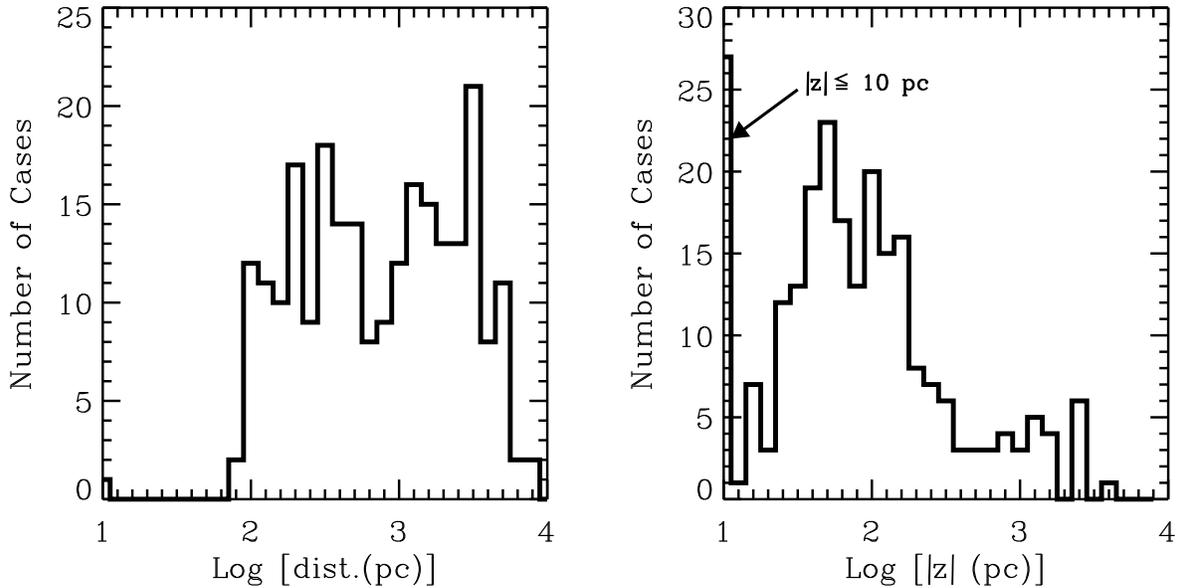}
\caption{The distributions of distances to the target stars considered in 
this survey ({\it left-hand panel}), and their separations away from the 
Galactic plane ({\it right-hand panel}).\label{rz_hist}}
\end{figure}

Figure~\ref{rz_hist} summarizes the distributions of distances and heights 
above or below the Galactic plane for all of the sight lines covered in this 
study.  The absence of stars with distances much below 0.1~kpc arises from 
the fact that the survey was almost completely restricted to data recorded 
for O- and B-type stars.  A good coverage of distances below this limit has 
been provided by surveys of limited selections of elements toward late-type 
stars (Redfield \& Linsky 2004a) and white dwarf stars (Friedman et al.
2002; Kruk et al. 2002; Lehner et al. 2003; Oliveira et al. 2003).  A
special discussion of interstellar depletions toward the white dwarf stars
will appear in \S\ref{WD}.

\placefigure{rz_hist}

\subsection{Other Considerations}\label{other_considerations}

Comparisons of element column densities are made with respect to $N({\rm 
H~I})$ derived from L$\alpha$ absorptions and $N({\rm H}_2)$ derived from 
Lyman band absorptions.  Hence these measurements apply to gas at all 
velocities.  Thus, even for investigations that revealed separate column 
densities of heavy elements over different radial velocity ranges, the 
results had to be combined over all velocities.  However, there are five 
special lines of sight where this consolidation of the column densities over 
all velocities was not applied.  First, Spitzer \& Fitzpatrick (1993) and 
Fitzpatrick \& Spitzer (1997) analyzed exceptionally good observations of 
the stars HD~93521 and HD~215733 taken with the highest resolution mode of 
the GHRS instrument on {\it HST}, and they went to great effort to decompose 
their profiles into separate velocity components.  A principal motivation for 
studying the separate components of these stars for the present study is to 
gain information on the depletions of sulfur, which is a difficult element to 
study because the features are usually strongly saturated.  These two stars 
are at large distances from the Galactic plane ($z=1.56$ and $-1.65\,$pc, 
respectively), and the sulfur profiles were sufficiently spread out in 
velocity that their saturation was not a problem when observed at a good 
velocity resolution.  Also, since there is not much H~I behind these stars, 
we can in principle rely on 21-cm observations to indicate the amount of 
hydrogen at each velocity (while this is true, there is a better method to 
find $N$(H) that will be outlined later in \S\ref{calc}).  Second, for the 
line of sight toward $\gamma^2$~Vel, Fitzpatrick \& Spitzer (1994) showed
that a significant fraction of all of the gas is fully ionized, and they
could identify which velocity components were ionized and which were
neutral.  (This star has a foreground neutral hydrogen column density just
barely above the adopted threshold of $10^{19.5}{\rm cm}^{-2}$).  Thus, for
this star, only column densities associated with H~I regions were included.  
Other stars [e.g.  $\lambda$~Sco  (York 1983) and 23~Ori (Welty et al. 1999)
had components that were identified with H~II regions, but their
contributions were small compared to those from the neutral gas components. 
Finally, the star $\zeta$~Oph had specific velocity components identified
separately because the highly depleted one at $-15\,{\rm km~s}^{-1}$ serves
as a standard for establishing the scale of $F_*$ [$N$(H) for the other
component is small enough to neglect -- see the discussion in
\S\ref{outcomes_performance}], and $\alpha$~CMa had small enough H column
densities that the L$\alpha$ absorption could be decomposed into two
principal components by H\'ebrard et al (1999).

\subsection{Renormalization to Recent \boldmath $f$-values}\label{fvalues}

Many of the earlier determinations of atomic column densities were reported 
using transition $f$-values that have since been revised.  The current 
analysis now includes adjustments to compensate for the differences between 
the old and new values.  In many situations we must acknowledge that this is 
an inexact process since the column densities in the earlier studies were 
determined from several transitions, not all of which may have been revised 
(or had different changes).  

When multiple lines were measured, the task of making the proper correction 
was made easier when the investigator indicated which line was the primary 
source of information for determining a column density.  However, most 
authors did not provide such guidance.   Usually, one can count on the 
weakest lines being the ones that constrain the column density outcome, while 
strong ones only define the velocity dispersion parameter $b$.  However, 
these weakest lines sometimes had to be overlooked if the relative errors in 
the equivalent widths were large.

When two or more lines with different $f$-value corrections were considered 
to be about equally influential in the column density determination, their 
logarithmic corrections were averaged together.  This can be a potentially 
hazardous approach, as illustrated by the following example.  The weaker of 
the two P~II lines analyzed by JSS86 has now had its $f$-value revised 
downwards by 0.127~dex, while the stronger line has remained virtually 
unchanged.  As a result, any new determination of the curve of growth would 
make the new $b$ value lower, the inferred saturation of both lines would 
increase, and $N$(P~II) would need to be revised in an upward direction by 
more than 0.127~dex.  However, this will happen only if indeed the weak line 
is perceived to be saturated, and it is deemed to be well enough measured to 
be useful (i.e., it is not dominated by noise).  If the strong line dominates 
in the determination of $N$(P~II) because its errors are so much smaller than 
those of the weak line and a $\chi^2$ minimization is used (as was the case 
with JSS86), the column density should remain unchanged.  In the particular 
case of P~II in the study by JSS86, the overall distortion of the P~II $b$ 
values is probably small, since there seems to be no systematic offset from 
the $b$ values measured for Fe~II toward the same stars (see their Fig.~1).

All column density adjustments were made on the basis of correcting to the 
$f$-values published in a compilation by Morton (2003).  
While there have been some newer determinations for a few species that have 
appeared in the literature since 2003, we retain Morton's values in the 
interest in having a single standard list that could be consulted in the 
future.  The only new result that was accepted was an $f$-value for the Ni~II 
transition at 1317$\,$\AA\ reported by Jenkins \& Tripp (2006), for which no
line strength was listed by Morton.  A summary of the $f$-value correction
factors for different investigations, presented in logarithmic form, is shown
in Table~\ref{fvals}.  These numbers represent the amounts by which the
originally reported values of $\log N$ were increased for this study.

\begin{deluxetable}{
l  
p{2.0cm}  
c  
c  
c  
c  
c  
c  
c  
c  
c  
c  
c  
c  
c  
c  
c  
c  
c  
l  
}
\tabletypesize{\scriptsize}
\rotate
\tablecolumns{20}
\tablewidth{0pt}
\tablecaption{Column Density Corrections for Revised $f$-Values\label{fvals}}
\tablehead{
\colhead{} & \colhead{} & \multicolumn{17}{c}{Correction Factors (dex)}\\
\cline{3-19}
\colhead{Code\tablenotemark{a}} & \colhead{Reference} & \colhead{C} & \colhead{N} &
\colhead{O} & \colhead{Mg} & \colhead{Si} & \colhead{P} & \colhead{S} &
\colhead{Cl} & \colhead{Ti} & \colhead{Cr} & \colhead{Mn} & \colhead{Fe} &
\colhead{Ni} & \colhead{Cu} & \colhead{Zn} & \colhead{Ge} & \colhead{Kr} &
\colhead{Notes}\\
\colhead{(1)} &
\colhead{(2)} &
\colhead{(3)} &
\colhead{(4)} &
\colhead{(5)} &
\colhead{(6)} &
\colhead{(7)} &
\colhead{(8)} &
\colhead{(9)} &
\colhead{(10)} &
\colhead{(11)} &
\colhead{(12)} &
\colhead{(13)} &
\colhead{(14)} &
\colhead{(15)} &
\colhead{(16)} &
\colhead{(17)} &
\colhead{(18)} &
\colhead{(19)} &
\colhead{(20)}
}
\startdata
  A++03& Andr\'e' et al. 2003& \nodata& \nodata&       0& \nodata& \nodata& \nodata& \nodata& \nodata& \nodata& \nodata& \nodata& \nodata& \nodata& \nodata& \nodata& \nodata&      
  0&        \\   A++93& Albert et al. 1993& \nodata& \nodata& \nodata& \nodata& \nodata& \nodata& \nodata& \nodata&  $-$0.022& \nodata& \nodata& \nodata& \nodata& \nodata&
  \nodata& \nodata& \nodata&        \\   AJS92& Allen, Jenkins, \& Snow 1992& \nodata& \nodata&  $-$0.035&  $-$0.356& \nodata&   0.134& \nodata&   0.025& \nodata& \nodata&  $-$0.257& 
   0.039& \nodata& \nodata& \nodata& \nodata& \nodata&        \\   B++83& Bohlin et al. 1983& \nodata& \nodata& \nodata& \nodata& \nodata& \nodata& \nodata& \nodata& \nodata&
  \nodata& \nodata& \nodata& \nodata& \nodata& \nodata& \nodata& \nodata&       1\\   BFM07& Burgh, France, \& McCandliss 2007& \nodata& \nodata& \nodata& \nodata& \nodata&
  \nodata& \nodata& \nodata& \nodata& \nodata& \nodata& \nodata& \nodata& \nodata& \nodata& \nodata& \nodata&       2\\
 BLWY84& Barker et al. 1984&        & \nodata& \nodata& \nodata&   0.251& \nodata& \nodata& \nodata& \nodata& \nodata& \nodata& \nodata& \nodata& \nodata& \nodata& \nodata&
 \nodata&        \\
  BSD78& Bohlin, Savage, \& Drake 1978& \nodata& \nodata& \nodata& \nodata& \nodata& \nodata& \nodata& \nodata& \nodata& \nodata& \nodata& \nodata& \nodata& \nodata& \nodata&
  \nodata& \nodata&       1\\   C++91& Cardelli et al. 1991&  $-$0.028& \nodata&   0.033&  $-$0.399& \nodata&   0.266& \nodata& \nodata& \nodata&   0.143&   0.011&   0.161&   0.231&
   $-$0.342&   0.011& \nodata& \nodata&       3\\
\enddata
\end{deluxetable}

\placetable{fvals}

\placetable{Stellar_data}

\placetable{MK_refs}
\begin{deluxetable}{
r  
c  
c  
c  
c  
c  
c  
c  
c  
c  
c  
c  
c  
c  
c  
c  
c  
c  
c  
}
\tabletypesize{\tiny}
\rotate
\tablecolumns{19}
\tablewidth{0pt}
\tablecaption{Stellar Data and Line of Sight Information\label{stellar_data}}
\tablehead{
\colhead{} & \colhead{} & \multicolumn{2}{c}{Gal. Coord.} &
\colhead{} & \multicolumn{2}{c}{Spectral Type} & \colhead{} & \colhead{dist.} &
\colhead{$z$} & \multicolumn{4}{c}{$\log N$(H I)} & \colhead{} &
\multicolumn{4}{c}{$\log N({\rm H}_2)$}\\
\cline{3-4} \cline{6-7} \cline{11-14} \cline{16-19}
\colhead{HD} & \colhead{Name} & \colhead{$\ell$} & \colhead{$b$} &
\colhead{$V$} & \colhead{Value} & \colhead{ref.\tablenotemark{a}} & \colhead{$E(B-V)$} &
\colhead{(kpc)} & \colhead{(kpc)} &
\colhead{l.l.} & \colhead{best} & \colhead{u.l.} & \colhead{Ref.\tablenotemark{b}} &
\colhead{~} &
\colhead{l.l.} & \colhead{best} & \colhead{u.l.} & \colhead{Ref.\tablenotemark{b}}\\
\colhead{(1)} &
\colhead{(2)} &
\colhead{(3)} &
\colhead{(4)} &
\colhead{(5)} &
\colhead{(6)} &
\colhead{(7)} &
\colhead{(8)} &
\colhead{(9)} &
\colhead{(10)} &
\colhead{(11)} &
\colhead{(12)} &
\colhead{(13)} &
\colhead{(14)} &
\colhead{} &
\colhead{(15)} &
\colhead{(16)} &
\colhead{(17)} &
\colhead{(18)}
}
\startdata
\objectname[        HD1383]{          1383}&        HD 1383&119.02&    $ -0.89$&\phm{-} 7.63&             B1 II&  JM53&\phm{-} 0.37&     2.94&    $ -0.05$&  21.33&  21.42&  21.51& CLMS04&&  20.38&  20.45&  20.52& CLMS04\\
\objectname[        HD2905]{          2905}&   $\kappa$ Cas&120.84&\phm{-} 0.14&\phm{-} 4.19&           B0.7 Ia&   W72&\phm{-} 0.29&     1.18&\phm{-} 0.00&  21.05&  21.20&  21.32&  BSD78&&  20.09&  20.27&  20.45&  BSD78\\
\objectname[        HD5394]{          5394}&   $\gamma$ Cas&123.58&    $ -2.15$&\phm{-} 2.39&            B0 IVe& B++08&\phm{-} 0.21&     0.17&    $ -0.01$&  20.06&  20.16&  20.24&  BSD78&&  16.48&  16.53&  16.58&    S98\\
\objectname[       HD12323]{         12323}&       HD 12323&132.91&    $ -5.87$&\phm{-} 8.92&              O9 V& B++08&\phm{-} 0.21& 4.4\phn &    $ -0.45$&  21.09&  21.18&  21.27& CLMS04&&  20.24&  20.32&  20.40& CLMS04\\
\objectname[       HD13268]{         13268}&       HD 13268&133.96&    $ -4.99$&\phm{-} 8.18&            O8 Vnn& B++08&\phm{-} 0.23& 2.1\phn &    $ -0.18$&  21.23&  21.32&  21.41& CLMS04&&  20.35&  20.42&  20.49& CLMS04\\
\objectname[       HD13745]{         13745}&       HD 13745&134.58&    $ -4.96$&\phm{-} 7.90&          O9.7 IIn& B++08&\phm{-} 0.34& 3.2\phn &    $ -0.28$&  21.15&  21.25&  21.35&   DS94&&  20.42&  20.47&  20.52& S++08a\\
\objectname[       HD14434]{         14434}&       HD 14434&135.08&    $ -3.82$&\phm{-} 8.59&         O5.5 Vnfp& B++08&\phm{-} 0.38& 3.5\phn &    $ -0.23$&  21.28&  21.37&  21.46& CLMS04&&  20.40&  20.47&  20.54& CLMS04\\
\objectname[       HD15137]{         15137}&       HD 15137&137.46&    $ -7.58$&\phm{-} 7.86&      O9.5 II-IIIn& B++08&\phm{-} 0.24& 3.5\phn &    $ -0.46$&  20.95&  21.11&  21.27&   DS94&&  20.24&  20.27&  20.30& S++08a\\
\enddata
\end{deluxetable}
\begin{deluxetable}{
l  
l  
}
\tablewidth{0pt}
\tablecaption{Refrence Codes for Spectral Types Listed in Table~\protect\ref{stellar_data}\label{MK_refs}}
\tablehead{
\colhead{Code} & \colhead{Reference}
}
\startdata
   A81& Abt 1981\\
ACNP85& Andersen et al. 1985\\
 B++08& Bowen et al. 2008\\
   B54& Bidelman 1954\\
   B58& Bertiau 1958\\
 BFM07& Burgh, France, \& McCandliss 2007\\
CLMS04& Cartledge et al. 2004\\
 D++06& Dufton et al. 2006\\
   F58& Feast 1958\\
 FTW55& Feast,
 Thackeray, \&
 Wesselink
 1955\\
  GG94& Garrison
   \& Gray
  1994\\  
  GK76& Garrison
   \& Kormendy
  1976\\  
  GMB77&Golay, Mandwewala, \& Bartholdi 1977\\
  H56a& Hoffleit
   1956\\  
  H56b& Hiltner
  1956\\
   H70& Hill
   1970\\
 HGS69& Hiltner,
  Garrison, \&
 Schild 1969\\
   J81& Jensen,
   1981\\
  JJ92& Jaschek
  \& Jaschek
  1992\\  
  JM53& Johnson
  \& Morgan
  1953\\  
  KM89& Kilkenny
   \& Muller
  1989\\
  KW99& Kohoutek \& Wehmeyer 1999\\
   L68& Lesh
   1968\\   
   L72& Levato
   1972\\
  LA76& Levato
  \& Abt 1976\\
   M61& Morris
   1961\\
 MCW55& Morgan,
 Code, \& Whitford 1955\\ 
 MHG71& Morgan,
 Hiltner, \&
 Garrison 1971\\
  MR50& Morgan
  \& Roman
  1950\\
   P43& Popper
   1943\\   
   S68& Simonson
    1968\\   
   S82& Slettebak 1982\\
  SF90& Strassmeier \& Fekel
  1990\\
 SMM85& Savage,
 Massa, \& Meade
 1985\\  SMS01&
 Savage, Meade,
 \& Sembach
 2001\\
   T71& Thackeray 1971\\   
   W71& Walborn
   1971\\   
   W72& Walborn
   1972\\   
   W73& Walborn
   1973\\
\enddata
\end{deluxetable}

\placefigure{fdist}
\clearpage
\subsection{Characteristics of the Sight Lines and Target 
Stars}\label{characteristics}

Table~\ref{stellar_data} lists some fundamental information on the stars 
included in this investigation, along with determinations of the amount of 
foreground hydrogen in both atomic and molecular form.   The stars are 
identified by both their HD numbers and their alternate representations 
[columns~(1) and 2], followed by their Galactic coordinates [columns~(3) and 
(4)].  Column~(5) lists visual magnitudes of the stars taken from the Simbad 
database.  Except for a few stars, these V magnitudes were not used for 
calculating the reddenings or spectroscopic parallaxes discussed in 
\S\ref{dist}; the tabulated values are therefore of uncertain origin and 
meant only as an approximate guide.  Column~(6) lists the spectral types of 
the stars, followed by codes in column~(7) that designate their sources in 
the literature according to the matches to references given in 
Table~\ref{MK_refs}.  Column~(8) lists the $B-V$ color excess toward each 
star, which is one indication of the total amount of dust along the sight 
line.  Column~(9) shows the distances toward the stars, computed according to 
the principles given below in \S\ref{dist}, followed by the corresponding 
distances $z$ from the Galactic plane [column~(10)].  Columns~(11) through 
(18) show the lower limits, best values, and upper limits for atomic and 
molecular hydrogen, in each case followed by codes that give the sources for 
these values.  Note that many atomic hydrogen column densities had 
adjustments applied to account for stellar contamination of the L$\alpha$ 
profile, as will be explained later in \S\ref{stellar_lalpha}.

\begin{figure}
\plotone{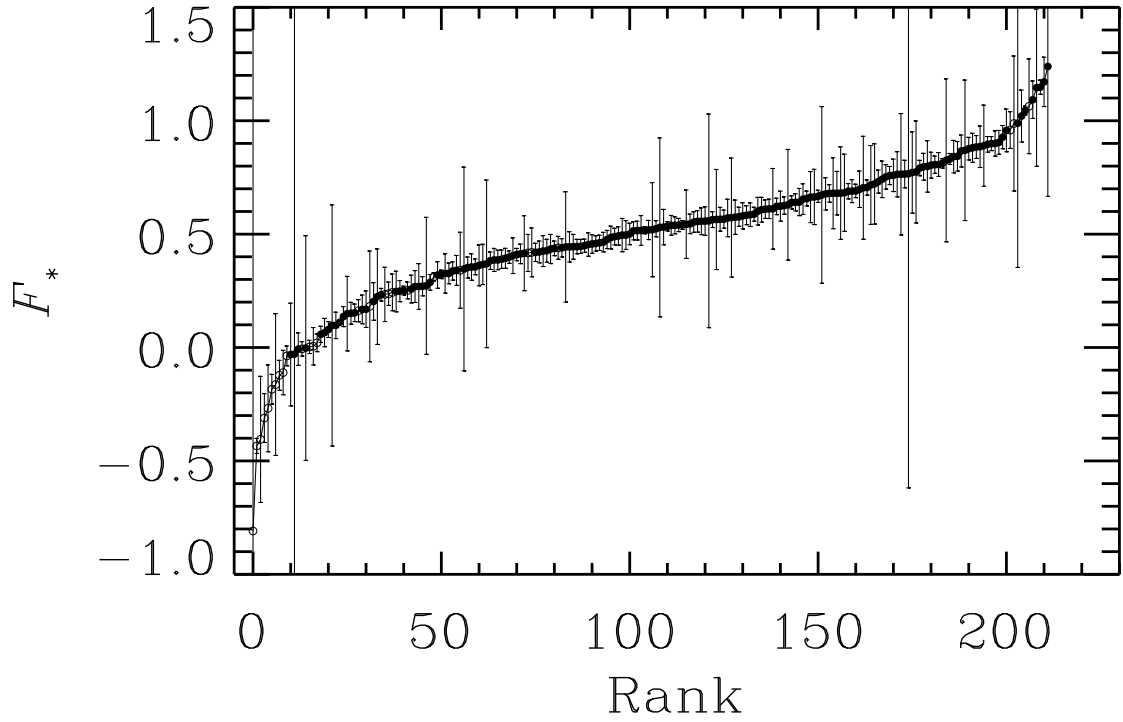}
\caption{The outcomes for $F_*$ and their respective $1\sigma$ errors in this 
study, arranged in a sequence according to their rank in value.  Solid points 
apply to cases where $N({\rm H})>10^{19.5}{\rm cm}^{-2}$, and hollow ones 
indicate measurements below this range.\label{fdist}}
\end{figure}

Figure~\ref{fdist} shows the distribution of $F_*$ values for all of the 
stars considered, including those which did not meet the eligibility 
requirements for helping to determine the element parameters $A_X$, $B_X$, 
and $z_X$ as outlined under point nrs. 3, 4, and 7 in \S\ref{rejection}.  
Note that there are values of $F_*$ below zero, but the sightlines for these 
cases have $\log N({\rm H})<19.5$, which violates one of the conditions 
needed for participation in the estimation of the element parameters.  Actual 
numerical values of $F_*$ (and their uncertainties) toward the specific 
target stars appear later in Table~\ref{nh_fstar}.
\subsubsection{Calculations of Distances and Reddenings}\label{dist}

For determining the distances to early-type stars, one must make use of 
different sources of information, depending on the circumstances.  At the 
most basic level, we note that trigonometric parallaxes provide the most 
accurate measures of distance for nearby stars, but when the errors in these 
parallaxes are not considerably smaller than the measured values, it is 
better to rely on spectroscopic parallaxes.  For trigonometric parallaxes, we 
rely on the second reduction of the Hipparcos data (nr. I/311 in the 
Strasbourg CDS VizieR on-line catalog)\footnote{The faulty version of this 
catalog that appeared during July to Sept 2008 was not used.} (van Leeuwen
2007), but accept the values only if $\pi/\sigma(\pi)>10$.  For stars that do
not meet this standard, we revert to distance derivations based on
spectroscopic parallaxes, as described below.

As a part of an investigation of O~VI absorptions in the Galactic disk, Bowen 
et al. (2008) (hereafter B08) carried out a rigorous 
process for determining the distances to stars based on spectroscopic 
parallaxes with many additional refinements, as described in Appendix~B of 
their article.  We adopt here many of the principles that they did, but with 
some simplifications since the standard of accuracy here is not as demanding 
as that for one of the objectives of the O~VI survey.  Some of the stars in 
the current survey were listed by B08.   For those cases, 
Table~\ref{stellar_data} simply duplicates the values of distance and 
reddening that they listed, except for cases where they adopted Hipparcos 
parallaxes.  The reason for rejecting their distances based on Hipparcos data 
is that (1) they used results from the earlier, less accurate solutions that 
were incorporated into the first catalog (Schrijver 1997), and (2) 
their acceptance threshold was set to $\pi/\sigma(\pi)>5$ instead of 10.

Stars for which neither the Hipparcos parallaxes nor the determinations by 
B08 were applicable had their spectroscopic parallaxes determined with the 
standard formula for the distance $d$ (in kpc),
\begin{equation}\label{dist_formula}
\log d=(m_V-A_V-M_V-10)/5
\end{equation}
where  $m_V$ is the apparent $V$ magnitude of the star, $A_V$ is the 
extinction by dust in the visible (assumed to be $3.1E(B-V)$), and $M_V$ is 
the absolute $V$ magnitude of the star.  We follow the recommendation of B08 
in adopting, when available, the two-color photometric measurements from the 
{\it Tycho} Starmapper catalog (Schrijver 1997) to obtain values for 
the $B$ and $V$ magnitudes, after a suitable transformation from the {\it 
Tycho\/} magnitude system $B_T$ and $V_T$ (see Appendix B1 of B08).  The 
premise here is that these magnitudes represent a uniform set of measurements 
where both colors were determined at a single epoch (which can be important 
for variable stars).  If {\it Tycho\/} magnitudes were not available, we use 
values of $B$ and $V$ listed in the catalog of Johnson et al. (1966). 
Finally, if neither of the above two sources had listings for the stars, we
used the magnitudes given by the Simbad web site.

Column~(6) of Table~\ref{stellar_data} lists the adopted spectral 
classifications of the target stars, along with the respective sources of the 
assignments in the column that follows.  (Many of these classifications were 
taken from either B08 or Savage, Meade \& Sembach (2001), who made judgments
on the most reliable sources and listed them in their tables.)  For these
chosen spectral types, we link them to values of $M_V$ in the consolidation
from many sources made by B08 and adopt the intrinsic colors from Wegner
(1994).  A fraction of the stars are recognized as spectroscopic binaries and
thus must have their distances adjusted outward to compensate for the fact
that their brightnesses are greater than that of the primary star alone.  B08
outline a procedure for implementing this correction (see their
Appendix~B1.2).

\subsection{Determinations of \boldmath $N$(H)}\label{NH}

For most stars, we can draw upon measurements of both H~I and H$_2$ and 
evaluate the quantity $N({\rm H~I})+2N({\rm H}_2)$ to find the total number 
of hydrogen atoms $N$(H) along a given sight line.  Since the errors in 
$N$(H~I) and $N({\rm H}_2)$ are uncorrelated, we can simply add them in 
quadrature to find the net uncertainty in $N$(H).  Unfortunately, for some of 
the stars of potential value for this survey, there were no measurements of 
the foreground H~I, H$_2$ (or both), or there were just upper limits thereof.  
For other stars, corrections had to be made for the upward shift in the 
apparent $N$(H~I) caused by L$\alpha$ absorption in the stellar atmospheres.  
The handling of these special circumstances is discussed in the following 
subsections.

\subsubsection{Missing Information}\label{missing_info}

At the most fundamental level, sight lines without information on both H~I 
and H$_2$ were not suitable for measuring depletions directly.  However, for 
about half of these stars $N$(H~I) is known, and values of $N({\rm H})$ could 
be salvaged on the premise that empirically the fraction of hydrogen in 
molecular form is very small ($f({\rm H}_2)\equiv 2N({\rm H}_2)/[N({\rm 
H~I})+2N({\rm H}_2)]<0.1$) when the star's $B-V$ color excess is less than 
0.05 (Savage et al. 1977); see also Rachford et al. (2009).  An 
equivalent cutoff can be established on the basis of just the measurement of 
$N({\rm H~I})$: In a survey of stars using the {\it Copernicus\/} satellite, 
Bohlin, Savage \& Drake (1978) found that
\begin{equation}\label{Htot/E(B-V)}
\langle \{ N({\rm H~I})+2N({\rm H}_2)\} /E(B-V)\rangle=5.8\times 
10^{21}\,{\rm atoms~cm}^{-2}{\rm mag}^{-1}~,
\end{equation}
which is approximately consistent with
\begin{equation}\label{HI/E(B-V)}
\langle N({\rm H~I})/E(B-V)\rangle=5.2\times 10^{21}\,{\rm atoms~cm}^{-2}{\rm 
mag}^{-1}~,
\end{equation}
derived for an {\it IUE\/} survey conducted by Shull \& Van~Steenberg 
 (1985) for a larger number of stars.   For such stars with no explicit 
H$_2$ data, but which had $\log N({\rm H~I}) < 20.4$ the value for $N({\rm 
H})$ was set equal to $N({\rm H~I})$, but with an increase in the upper error 
bar by 0.04~dex to allow for the fact that an unseen additional 10\% of the 
hydrogen atoms could be in molecular form.

Many stars had no reported values for $N$(H~I) because either the L$\alpha$ 
absorption was not observed or the star was of a late enough spectral type 
that there was a good chance that the L$\alpha$ feature was strongly 
dominated by a stellar contribution.  It is important to emphasize that for 
such cases we did not use $E(B-V)$ to determine $N$(H) on the basis of the 
empirical relationship given in Eq.~\ref{Htot/E(B-V)}.  This choice makes the 
survey immune to possible misleading effects caused by real deviations from 
the general connection between selective extinction and the amount of gas 
present. This can be important for future studies that might attempt to 
relate depletions to various observable properties of the dust (i.e., $A_V$, 
$R_V$, wavelength of maximum polarization, etc.).  It also avoids our being 
deceived by photometric errors arising from the occasional presence of 
emission lines in the spectra of stars.

We avoided the practice of estimating $N$(H) simply by using the column 
density of a supposedly undepleted element and assuming a solar abundance 
ratio, since the lack of any depletion for that element could be subject to 
question.  For example, we did not accept values of $N$(H) listed by 
Cartledge et al. (2004, 2006) based on the column densities of Kr.  (Later,
it will be shown that Kr may exhibit some very mild depletion.)

Ultimately, stars for which $N$(H) could not be recovered from data in the 
literature were not useless.  A means for calculating an indirect, synthetic 
value for this quantity is discussed in \S\ref{calc}, and it will be 
demonstrated that this outcome offers a reasonably accurate replacement for 
an observed value of $N$(H).

\subsubsection{Corrections for Stellar L$\alpha$ 
Absorption}\label{stellar_lalpha}

A significant proportion of the sight lines covered in this study (102 cases) 
made use of stars that had spectral types B1 or cooler.  In such instances, 
there is a danger that the equivalent width of the observed L$\alpha$ 
absorption feature is enhanced by an underlying stellar contribution 
 (Savage \& Panek 1974). It is important to account for this effect, 
since there will be a systematic shift in the measurement of $N$(H~I), 
sometimes quite small, to some amount that is above the true value that 
belongs to the ISM.  Many investigators who reported or used values of 
$N$(H~I) recognized this problem and provided cautions that some of their 
results were probably contaminated by a stellar contribution, but they did 
not attempt to apply compensations 
(Bohlin, Savage, \& Drake 1978; Shull \& Van Steenberg 1985; Andr\'e et al.
2003; Cartledge et al. 2004, 2008).  However Bohlin, et al. (1983) appear to
have overlooked this problem when they derived additional values of $N$(H~I).

In a study of L$\alpha$ absorption toward a large number of stars, Diplas \& 
Savage (1994) devised a means for estimating the 
stellar L$\alpha$ absorption and corrected many of their measurements to 
compensate for it.  Their method was based on the findings of Savage \& Panek 
 (1974) with some additional guidance from NLTE 
stellar atmosphere calculations, and it used as a yardstick the 
reddening-corrected measure of the Balmer discontinuity based on a 
combination of narrow-band Str\"{o}mgren photometric indices 
$[c_1]=c_1-0.2(b-y)$, where the uncorrected Balmer jump index 
$c_1=(u-v)-(v-b)$.

In making corrections for stellar L$\alpha$ absorption in the present study, 
we assume that the stellar profile is well approximated by a Lorentzian 
shape, as did Diplas \& Savage, so that we may simply subtract from the 
observed $N$(H~I) the equivalent column density for the stellar line to 
obtain the interstellar value.  In estimating the strength of the stellar 
line, we follow exactly the recipe given by Diplas \& Savage.  Our only 
departure from their practice was that we did not exclude from consideration 
cases where $\log N({\rm H~I})_{\rm obs.} - \log N({\rm H~I})_{\rm stellar} 
\leq 0.5~{\rm dex}$.  We justify this action on the grounds that the larger 
uncertainties are well accounted for in the error estimation technique 
described below, which then influences the weight factors in the parameter 
estimations described in \S\S\ref{determination_of_F_*} and 
\ref{determination_of_ABz} without totally discarding the results at an 
arbitrary level.

After subtraction of $N({\rm H~I})_{\rm stellar}$ from $N({\rm H~I})_{\rm 
obs.}$ to obtain $N({\rm H~I})_{\rm ISM}$, we define the error in the result 
$\sigma[N({\rm H~I})]_{\rm ISM}$ in terms of a combination of errors in both 
the observed column density and the estimate for the stellar contribution, 
given by the relation
\begin{equation}\label{err_N_ISM}
\sigma[N({\rm H~I})]_{\rm ISM}=\sqrt{\sigma_{\pm}[N({\rm H~I})]_{\rm 
obs.}^2+[N({\rm H~I})_{\rm stellar}(1-10^{\mp\Delta})]^2}~,
\end{equation}
where $\sigma_{\pm}[N({\rm H~I})]_{\rm obs.}$ represents the differences 
between the best values of the observed column densities and their respective 
upper and lower bounds, and $\Delta$ is the logarithm of the relative 
uncertainty in $N({\rm H~I})_{\rm stellar}$.  We adopted a value 
$\Delta=0.20~{\rm dex}$ for all of the correction calculations.\footnote{The 
best value of $\Delta$, an error parameter that must include both random and 
systematic errors, is difficult to quantify with much precision.  Our choice 
of $\Delta=0.20~{\rm dex}$ is a conservative one based on 3 considerations: 
(1) An estimate by Savage \& Panek (1974) that 
their rms errors in equivalent widths of the stellar L$\alpha$ feature are 
about 20\%, which translates into (+0.16, -0.19)~dex errors in $N({\rm 
H~I})_{\rm stellar}$, (2) An rms deviation of approximately 0.08~dex in 
$N$(H~I) at $[c_1]\approx 0.3$ on either side of the theoretical line shown 
in Fig.~2 of Diplas \& Savage (1994) (where the 
stellar line probably dominates over the interstellar contribution), but with 
4 outliers elsewhere that were more than three times this value in the 
negative direction, and (3) the size of the transition between the two 
discrete choices for factoring in the stellar surface gravity, one at $\log 
g=3$ and the other at $\log g=4$, in the recipe of Diplas \& Savage.}

For most of the stars that needed an evaluation of $N({\rm H~I})_{\rm 
stellar}$, values of the critical parameter $[c_1]$ could be retrieved from 
the catalog of Hauck \& Mermilliod (1998).\footnote{The photometric data are
available in the Strasbourg CDS VizieR on-line catalog nr. II/215.}  In a few
cases, other sources were needed, as indicated in the endnotes of
Table~\ref{stellar_data}.  The photometry for stars whose spectral
classifications indicated the presence of emission lines (i.e., with an ``e''
appended) are probably untrustworthy.  For these stars, as well as others for
which no measurements of $[c_1]$ could be found, the estimates for $N({\rm
H~I})_{\rm stellar}$had to be based on the stars' spectral types, using mean
values of $[c_1]$ found for other stars with similar classifications.  From
the dispersion of individual results about these means, we judge that the
uncertainty of any outcome using only the spectral classification is about
0.38~dex; hence we used this value for $\Delta$ in Eq.~\ref{err_N_ISM} for
the small number of cases where a spectral type had to be used instead of
$[c_1]$.  These stars with reduced accuracy are also identified explicitly in
the table.

In a number of instances, we found that $\sigma[N({\rm H~I})]_{\rm ISM} > 
N({\rm H~I})_{\rm ISM}$, but $N({\rm H~I})_{\rm ISM}>0$.  When this happened, 
lower limits were not stated in Table~\ref{stellar_data} and should thus be 
considered to be zero (the preferred values and upper limits were retained 
however).  In other cases, $N({\rm H~I})_{\rm ISM}<0$ but $N({\rm H~I})_{\rm 
ISM}+\sigma[N({\rm H~I})]_{\rm ISM}>0$; under these circumstances only upper 
limits set equal to $N({\rm H~I})_{\rm ISM}+\sigma[N({\rm H~I})]_{\rm ISM}$ 
were stated.  Finally, there were 6 instances where $N({\rm H~I})_{\rm 
ISM}+\sigma[N({\rm H~I})]_{\rm ISM}<0$; when this happened, only upper limits 
were stated and they were simply set equal to the upper limits for $N({\rm 
H~I})_{\rm obs.}$.  The fact that his occurred for only 6 out of the 95 cases 
considered for the correction offers a rough indication that the estimates 
for the $1\sigma$ uncertainties in $N({\rm H~I})_{\rm ISM}$ are probably not 
unrealistically small.

There are a few determinations of $N$(H~I) for the cooler stars that could be 
accepted at their stated values because either (1) their interstellar 
features could be seen as distinct absorptions at the bottoms of the 
photospheric features ($\alpha$~CMa, $\alpha$~Vir,
$\beta$~Cen and $\lambda$~Sco) (York \& Rogerson 1976; York 1983; H\'ebrard et
al. 1999) or (2) the H~I column density was determined by the observed shape
of the star's energy distribution in the EUV after accounting for hydrogen
absorption in the star's photosphere ($\beta$ and $\epsilon$~CMa) (Cassinelli
et al. 1995, 1996).

\section{Solutions for the Element Coefficients}\label{solutions}

Compilations of the atomic column density measurements (corrected for 
$f$-value changes) and their sources in the literature appear in 
Appendix~\ref{basic_data} with a series of tables organized according to the 
different elements studied in this survey.  These same tables also show the 
outcomes for $F_*$, together with information on how well the individual 
measurements conform to the best-fit solutions within our generalized 
framework.

For the depletion parameters that pertain to the just the elements, 
Table~\ref{elem_parameters} presents the outcomes of the weighted least 
squares fits described in \S\ref{determination_of_ABz}. (Sulfur is an element 
that presents special challenges and will be handled separately in 
\S\ref{sulfur}.) Column (2) of this table lists the assumed reference 
abundances taken from Lodders (2003) for the proto-Sun (see 
\S\ref{reference_abundances}).  The fundamental parameters of the linear fits 
are $A_X$, $B_X$ and $z_X$ listed in columns (3) to (5), but the secondary 
quantities $[X_{\rm gas}/{\rm H}]_0$ and $[X_{\rm gas}/{\rm H}]_1$ in columns 
(6) and (7) allow us to understand how these parameters translate into the 
expected depletions near the two extremes of $F_*$, $F_*=0$ and 1 (values of 
$F_*$ greater than 1 do show up for a few stars however -- see 
Figure~\ref{fdist}).  The last three columns of the table present information 
on how well the observations fit their respective best-fit trends.  For each 
element, we can use the values of $\chi^2$ [column (8)] with their 
appropriate degrees of freedom $\nu$ (number of observations minus 2) listed 
in the next column to compute the probability shown in column (10) that the 
fit could have been worse than what we obtained.  These probabilities are 
based on the assumptions that (1) the basic model for depletions expressed in 
Eq.~\ref{better_equation} is correct and that (2) the errors in the observed 
depletions were estimated correctly.  For Mn, this probability value seems 
rather low, which may indicate that either there are complicating factors 
that render the model as inappropriate for this element or that the errors in 
measuring column densities were underestimated (or both).  Conversely, 
unreasonably high values for these probabilities (e.g., Mg, Fe, Cu, and Ge) 
indicate that the measurement errors have probably been overestimated.

\placetable{elem_parameters}
\begin{deluxetable}{
c  
c  
c  
c  
c  
c  
c  
r  
r  
c  
}
\rotate
\tablewidth{0pt}
\tablecaption{Element Depletion Parameters\tablenotemark{a}\label{elem_parameters}}
\tablehead{
\colhead{} & \colhead{} & \colhead{} & \colhead{} & \colhead{} & \colhead{} & \colhead{} &
\colhead{} & \colhead{} & \colhead{Prob.}\\
\colhead{Elem.} & \colhead{Adopted} & \colhead{} & \colhead{} & \colhead{} & \colhead{} & \colhead{} & \colhead{} &
\colhead{} & \colhead{worse}\\
\colhead{$X$} & \colhead{$(X/{\rm H})_\odot$\tablenotemark{b}} & \colhead{$A_X$} &
 \colhead{$B_X$\tablenotemark{c}} & \colhead{$z_X$} & \colhead{$[X_{\rm gas}/{\rm H}]_0$\tablenotemark{c}} &
\colhead{$[X_{\rm gas}/{\rm H}]_1$\tablenotemark{c}} & \colhead{$\chi^2$} & \colhead{$\nu$} & \colhead{fit}\\
\colhead{(1)} &
\colhead{(2)} &
\colhead{(3)} &
\colhead{(4)} &
\colhead{(5)} &
\colhead{(6)} &
\colhead{(7)} &
\colhead{(8)} &
\colhead{(9)} &
\colhead{(10)}
}
\startdata
  C&$8.46\pm 0.04$&$        -0.101\pm  0.229$&$        -0.193\pm  0.060$&  0.803&$        -0.112\pm 0.194$&$        -0.213\pm 0.075$&  3.7&  8& 0.881\\
  N&$7.90\pm 0.11$&$        -0.000\pm  0.079$&$        -0.109\pm  0.111$&  0.550&$        -0.109\pm 0.119$&$        -0.109\pm 0.117$& 28.8& 32& 0.628\\
  O&$8.76\pm 0.05$&$        -0.225\pm  0.053$&$        -0.145\pm  0.051$&  0.598&$        -0.010\pm 0.060$&$        -0.236\pm 0.055$& 75.0& 64& 0.164\\
 Mg&$7.62\pm 0.02$&$        -0.997\pm  0.039$&$        -0.800\pm  0.022$&  0.531&$        -0.270\pm 0.030$&$        -1.267\pm 0.029$& 79.0&103& 0.962\\
 Si&$7.61\pm 0.02$&$        -1.136\pm  0.062$&$        -0.570\pm  0.029$&  0.305&$        -0.223\pm 0.035$&$        -1.359\pm 0.052$& 19.4& 16& 0.247\\
  P&$5.54\pm 0.04$&$        -0.945\pm  0.051$&$        -0.166\pm  0.042$&  0.488&$\phm{-}  0.296\pm 0.049$&$        -0.649\pm 0.050$& 69.5& 65& 0.330\\
 Cl&$5.33\pm 0.06$&$        -1.242\pm  0.129$&$        -0.314\pm  0.065$&  0.609&$\phm{-}  0.442\pm 0.102$&$        -0.800\pm 0.082$& 38.9& 44& 0.688\\
 Ti&$5.00\pm 0.03$&$        -2.048\pm  0.062$&$        -1.957\pm  0.033$&  0.430&$        -1.077\pm 0.043$&$        -3.125\pm 0.049$& 50.7& 43& 0.195\\
 Cr&$5.72\pm 0.05$&$        -1.447\pm  0.064$&$        -1.508\pm  0.055$&  0.470&$        -0.827\pm 0.062$&$        -2.274\pm 0.064$& 24.1& 20& 0.239\\
 Mn&$5.58\pm 0.03$&$        -0.857\pm  0.041$&$        -1.354\pm  0.032$&  0.520&$        -0.909\pm 0.038$&$        -1.765\pm 0.038$&106.3& 83& 0.043\\
 Fe&$7.54\pm 0.03$&$        -1.285\pm  0.044$&$        -1.513\pm  0.033$&  0.437&$        -0.951\pm 0.038$&$        -2.236\pm 0.041$& 48.5& 66& 0.948\\
 Ni&$6.29\pm 0.03$&$        -1.490\pm  0.062$&$        -1.829\pm  0.035$&  0.599&$        -0.937\pm 0.051$&$        -2.427\pm 0.043$& 30.7& 34& 0.630\\
 Cu&$4.34\pm 0.06$&$        -0.710\pm  0.088$&$        -1.102\pm  0.063$&  0.711&$        -0.597\pm 0.089$&$        -1.307\pm 0.068$& 15.3& 32& 0.995\\
 Zn&$4.70\pm 0.04$&$        -0.610\pm  0.066$&$        -0.279\pm  0.045$&  0.555&$\phm{-}  0.059\pm 0.058$&$        -0.551\pm 0.054$& 25.6& 19& 0.142\\
 Ge&$3.70\pm 0.05$&$        -0.615\pm  0.083$&$        -0.725\pm  0.054$&  0.690&$        -0.301\pm 0.078$&$        -0.916\pm 0.059$& 12.4& 24& 0.975\\
 Kr&$3.36\pm 0.08$&$        -0.166\pm  0.103$&$        -0.332\pm  0.083$&  0.684&$        -0.218\pm 0.109$&$        -0.384\pm 0.089$& 18.9& 26& 0.839\\
\enddata
\tablenotetext{a}{As defined in Eqs.~\protect\ref{better_equation},
\protect\ref{z_X}, \protect\ref{[X/H]_0} \& \protect\ref{[X/H]_1}.
Coefficients for S do not appear in this table because a nonstandard
approach was required.  The coefficients are given in the text of \S\protect\ref{sulfur}.}
\tablenotetext{b}{On a logarithmic scale with H = 12. Values and their errors taken
from the recommended solar abundances of Lodders (2003).}
\tablenotetext{c}{Unlike the convention for listing errors in the fit
outcomes in Tables~\protect\ref{C} to \protect\ref{Kr},
the uncertainties with the terms listed here include both
the formal errors of the fit coefficients and the
error in the adopted value of $(X/{\rm H})_\odot$, added together
in quadrature.}
\end{deluxetable}

\section{The Buildup of Dust Grains}\label{grain_buildup}

The strengths of chemical bonds for compounds that are most likely to form in 
dust grains vary over a large range.  As a consequence, the propensity of 
different elements to condense into solid form, or the likelihood that they 
can subsequently be liberated back into the gas phase, are strongly dependent 
on physical conditions and time scales for creating or destroying the 
compounds.  One popular paradigm is that the most refractory compounds are 
formed early in the nucleation process (possibly in the mass-loss outflows of 
stars or in the ejecta of supernovae), forming a core of the dust grain, and 
this is followed by the accumulation in dense molecular clouds of more 
loosely bound compounds that form a mantle around this core (Greenberg 1989;
Jones, Duley, \& Williams 1990; Mathis 1990; Dwek 1998; Tielens 1998; Draine
2003a).  However, in approaching the issue of relative depletions in
different regions of space, we can bypass the question of how the grains are
structured, i.e., whether they have a core-mantle assembly or a more
amorphous configuration, and simply focus on the empirical relationships
between different element abundances when the overall severity of the
depletions change.

Since we define lines of sight with $F_*=0$ to represent the circumstances 
that exhibit the minimum general level of depletion, we can regard values of 
$[X_{\rm gas}/{\rm H}]_0$ to represent a ``base depletion'' that, by virtue 
of it being found everywhere, probably indicates the composition of the most 
durable constituents of grains (or in the parlance of the core-mantle 
picture, the disappearance of elements in the gas phase to make up the 
``core'' of a grain).  As the composition of the grains evolve from being 
dominated by refractory compounds to more volatile ones, different elements 
increase the absolute values of their depletions at different rates.  One way 
to characterize the makeup of the more developed grains that have 
incorporated these volatile compounds might be to consider the depletions at 
some much larger level of depletion, say at $F_*=1$.

An important drawback of any declaration of an absolute level of depletion is 
that it depends on the assumed abundance of an element in the ISM when no 
grains exist at all, for which there have been some inconsistent quantitative 
conclusions, as discussed earlier in \S\ref{reference_abundances}.  As a 
result, there have been conflicting views on the makeup of the grains, which 
in turn have created some challenges in constructing representations of the 
number, sizes and compositions of dust grains that had to be reconciled with 
the observed absorption, scattering and polarization at visible, UV and X-ray 
wavelengths (Mathis 1996; Smith \& Dwek 1998; Draine 2003b,c).

While our expressions of the base depletions $[X_{\rm gas}/{\rm H}]_0$ must 
depend on the adopted values of the reference abundances, we can dispense 
with this relationship for more strongly developed depletions by not 
attempting to characterize the total composition of grains in the more 
advanced stages of growth, but instead simply by measuring the additional 
consumptions of different elements as they are incorporated into the newly 
formed grain materials.  That is, by determining how rapidly the abundances 
of different elements decrease as $F_*$ advances, we become insensitive to 
ambiguities that arise from uncertainties in the reference abundances.

If we substitute the right-hand side of Eq.~\ref{better_equation} for 
$[X_{\rm gas}/{\rm H}]$ into Eq.~\ref{x_dust} and differentiate it with 
respect to $F_*$, we find that
\begin{eqnarray}\label{differential_grain_comp}
d(X_{\rm dust}/{\rm H})/dF_*&=&-(\ln 10)(X/{\rm H})_\odot 
A_X10^{B_X+A_X(F_*-z_X)}\nonumber\\
&=&-(\ln 10)A_X(X_{\rm gas}/{\rm H})_{F_*}
\end{eqnarray}
The first equality gives the result in terms of variables defined earlier in 
this paper, while the second shows that this outcome is independent of the 
adopted solar abundances –-- only the actual expectation of $(X_{\rm 
gas}/{\rm H})$ and its slope ($A_X$) with $F_*$ matter.  (Note that the term 
$(X_{\rm gas}/{\rm H})$ refers to the actual gas-phase abundance of an 
element $X$ relative to H, whereas the notation used in earlier equations, 
$[X_{\rm gas}/{\rm H}]$, refers to the logarithm of the element's depletion 
factor.)

Figures~\ref{panel_set_1} through \ref{panel_set_4} show two fundamental 
results for all of the elements except sulfur.  (Again, sulfur is a difficult 
case that will be treated separately in \S\ref{sulfur}.)  For each element, 
the upper panel depicts the observed depletions as a function of $F_*$.  
Individual observed depletions are plotted as points with diameters that 
indicate their respective levels of accuracy.  Dashed lines follow the linear 
trends with $F_*$ represented by the best fits defined by the parameters 
$A_X$, $B_X$, and $z_X$ listed in Table~\ref{elem_parameters}.  The 
quantities $[X_{\rm gas}/{\rm H}]_0$ and $[X_{\rm gas}/{\rm H}]_1$ in columns 
(6) and (7) in that table are equal to the intercepts of these lines at 
$F_*=0$ and $F_*=1$, respectively; see Eqs.~\ref{[X/H]_0} through 
\ref{sigma[X/H]_1}.  The lower panels show the differential grain 
compositions, expressed in terms of the number of atoms per H atom that 
condense onto the grains per unit change in $F_*$.  The cross-hatched regions 
show the allowed combinations of this differential composition for 1 and 
$2\sigma$ deviations in the errors for $A_X$ and $B_X$. (Note that the 
portion of $\sigma(B_X)$ that is attributable to $\sigma(X/{\rm H})_\odot$ 
drops out of Eq.~\ref{differential_grain_comp}, thus leaving only the formal 
uncertainty in the least-squares solution for the intercept of the fit at 
$z_X$.  These values of $\sigma(B_X)$ can be recovered by subtracting in 
quadrature the error values listed in column (2) of 
Table~\ref{elem_parameters} from those listed in column (4) of the same 
table.)

Clearly, the slopes of the logarithms of the consumption rates of free atoms 
exhibit large variations from one element to the next, indicating that as the 
gas becomes more depleted the composition of the grains must change (or, put 
differently, that the material in the outer portions of the grain mantles 
differs from that in or near the cores). Our outlook on plausible mixtures of 
compounds within the grains must be constrained by not only the consumption 
information presented here, but also the chemical properties of the compounds 
themselves (Mathis 1996; Draine 2003a, 2004).

\placefigure{panel_set_1}
\placefigure{panel_set_2}
\placefigure{panel_set_3}
\placefigure{panel_set_4}
\clearpage
\begin{figure}
\caption{(Shown on next page) {\it Top row of panels:\/} Measured depletions 
(points) and the linear trends defined by the parameters $A_X$, $B_X$ and 
$z_X$ in  Eq.~\protect\ref{better_equation} (dashed lines), as listed in 
Table~\protect\ref{elem_parameters}, shown as a function of the generalized 
depletion parameter $F_*$ for the elements C, N, O, Mg and Si.  Solid points 
have $N({\rm H})>10^{19.5}{\rm cm}^{-2}$, while open ones have $N$(H) values 
below this range. Gray points represent sightlines that had only 3 elements 
to define their $F_*$ parameters, while black ones represent those that had 4 
or more elements.  Upper and lower limit measurements are depicted with 
arrows (and were not included in any analysis).  Sight lines that crossed the 
galactocentric limits $R_{\rm GC}<7\,$kpc or $R_{\rm GC}>10\,$kpc are 
overlaid with crosses (+) or x's ($\times$), respectively, to indicate that 
they were not used to define the linear trends for the elements.  {\it Bottom 
row of panels:\/}  Differential consumptions of elements by number (relative 
to hydrogen) by dust grains for small changes in $F_*$, again plotted as a 
function of $F_*$.  The trend lines that follow 
Eq.~\protect\ref{differential_grain_comp} with the best values of $A_X$ and 
$B_X$ are shown with dark lines, while the allowable changes that can arise 
from the uncertainties in $A_X$ and $B_X$ are shown by the shaded regions.  
Uncertainties at the $1\sigma$ level are shown by the cross-hatched areas, 
while the envelopes for $2\sigma$ deviations have simple line shading in only 
one direction.\label{panel_set_1}}
\end{figure}
\clearpage
\addtocounter{figure}{-1}
\begin{figure}
\plotfiddle{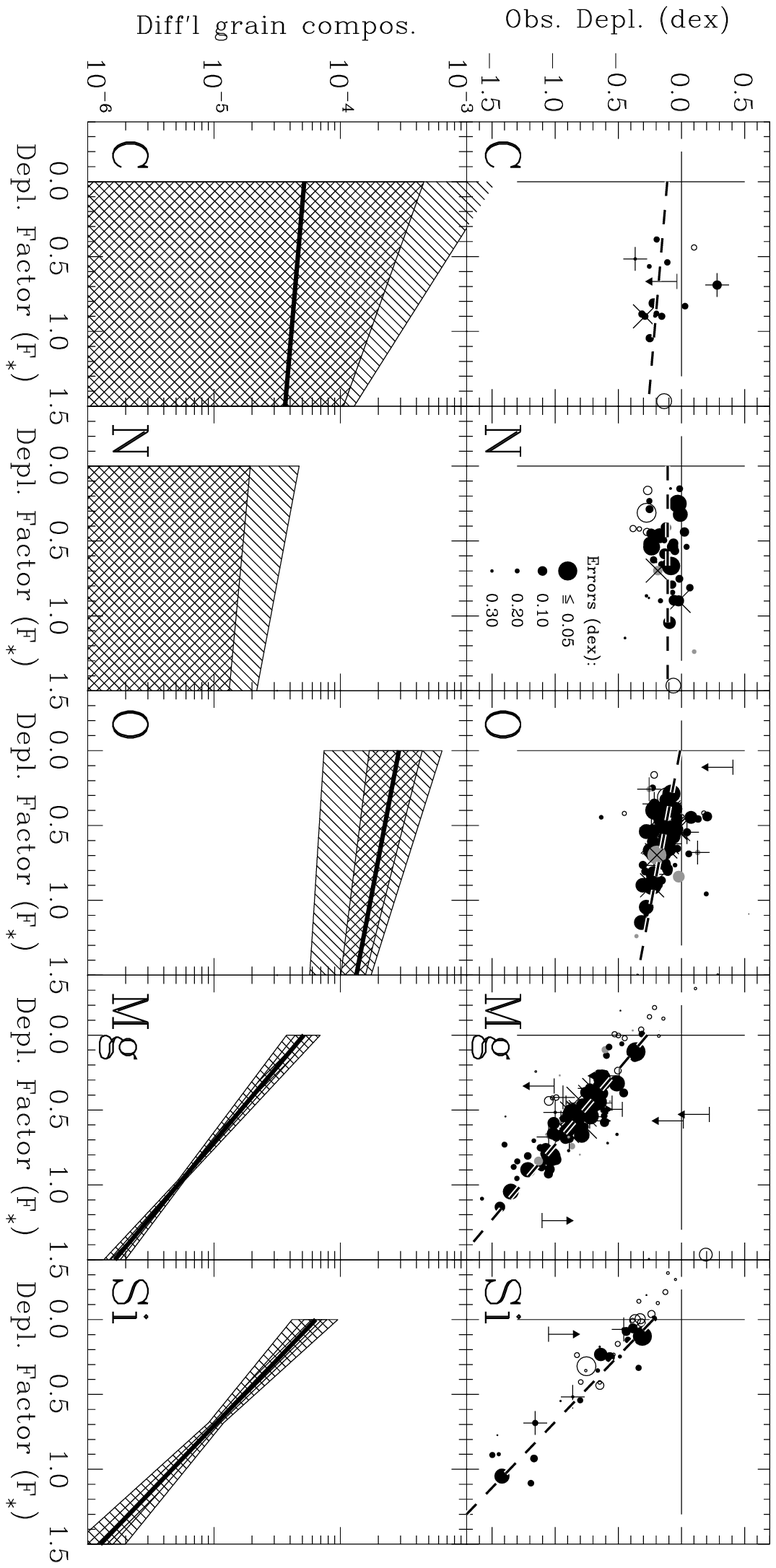}{-1in}{180}{600}{630}{-100}{300}
\caption{(Caption on previous page.)}
\end{figure}
\begin{figure}
\includegraphics[scale=1.,keepaspectratio=true,angle=90]{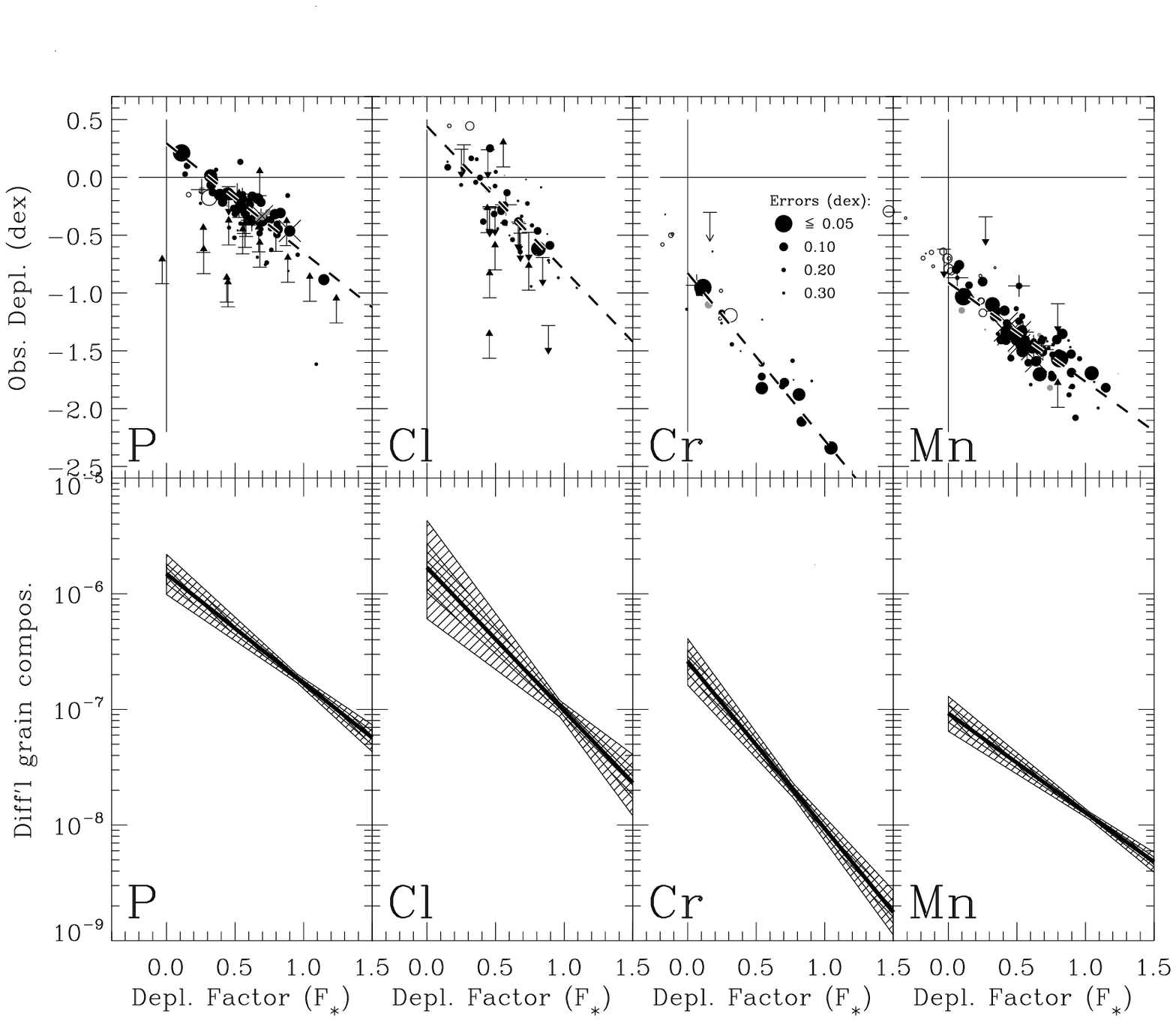}
\caption{Same as for Fig.~\protect\ref{panel_set_1} for the elements P, Cl, 
Cr, and Mn.\label{panel_set_2}}
\end{figure}
\begin{figure}
\includegraphics[scale=1.,keepaspectratio=true,angle=90]{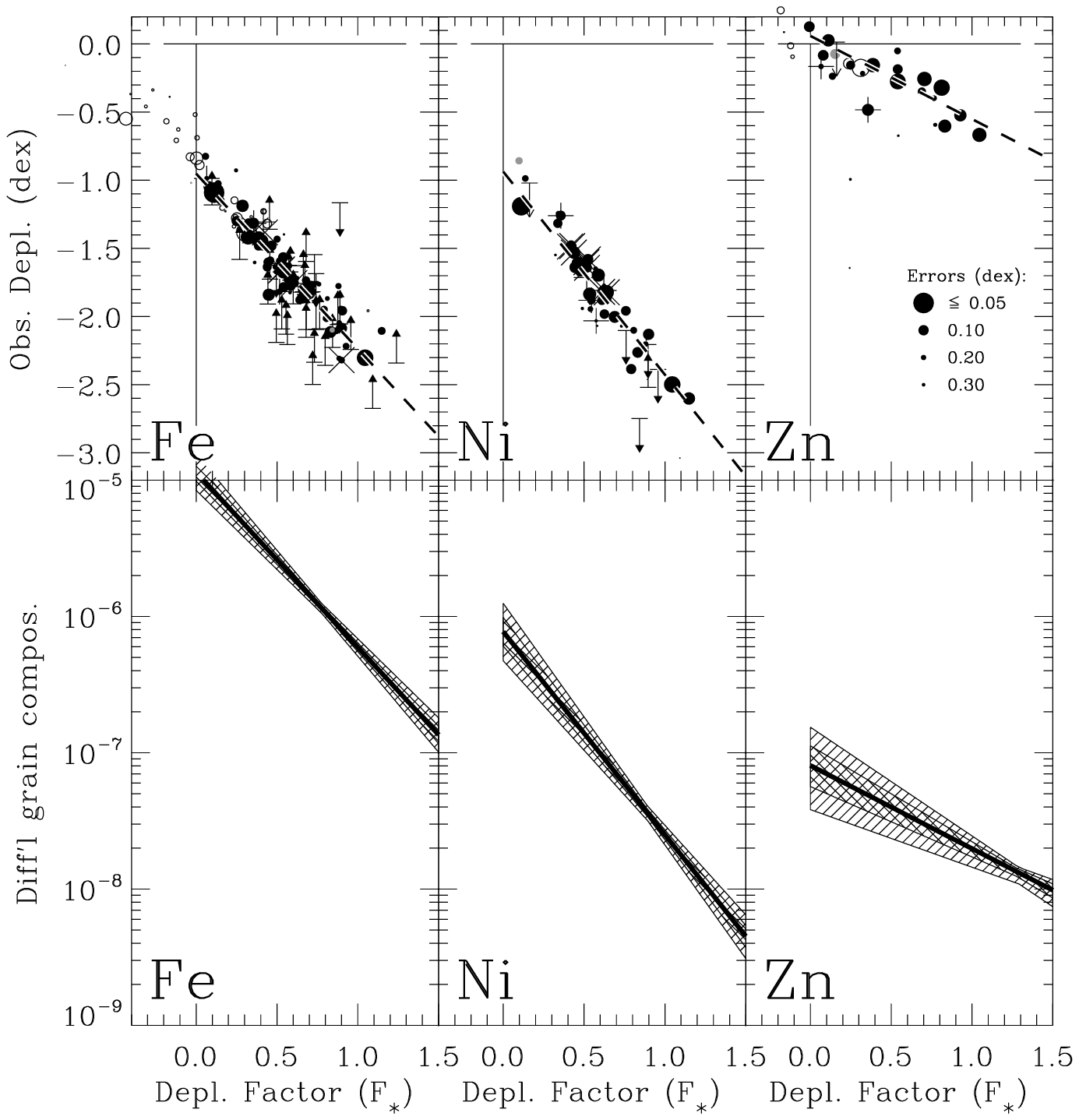}
\caption{Same as for Fig.~\protect\ref{panel_set_1} for the elements Fe, Ni, 
and Zn.\label{panel_set_3}}
\end{figure}
\begin{figure}
\includegraphics[scale=1.,keepaspectratio=true,angle=90]{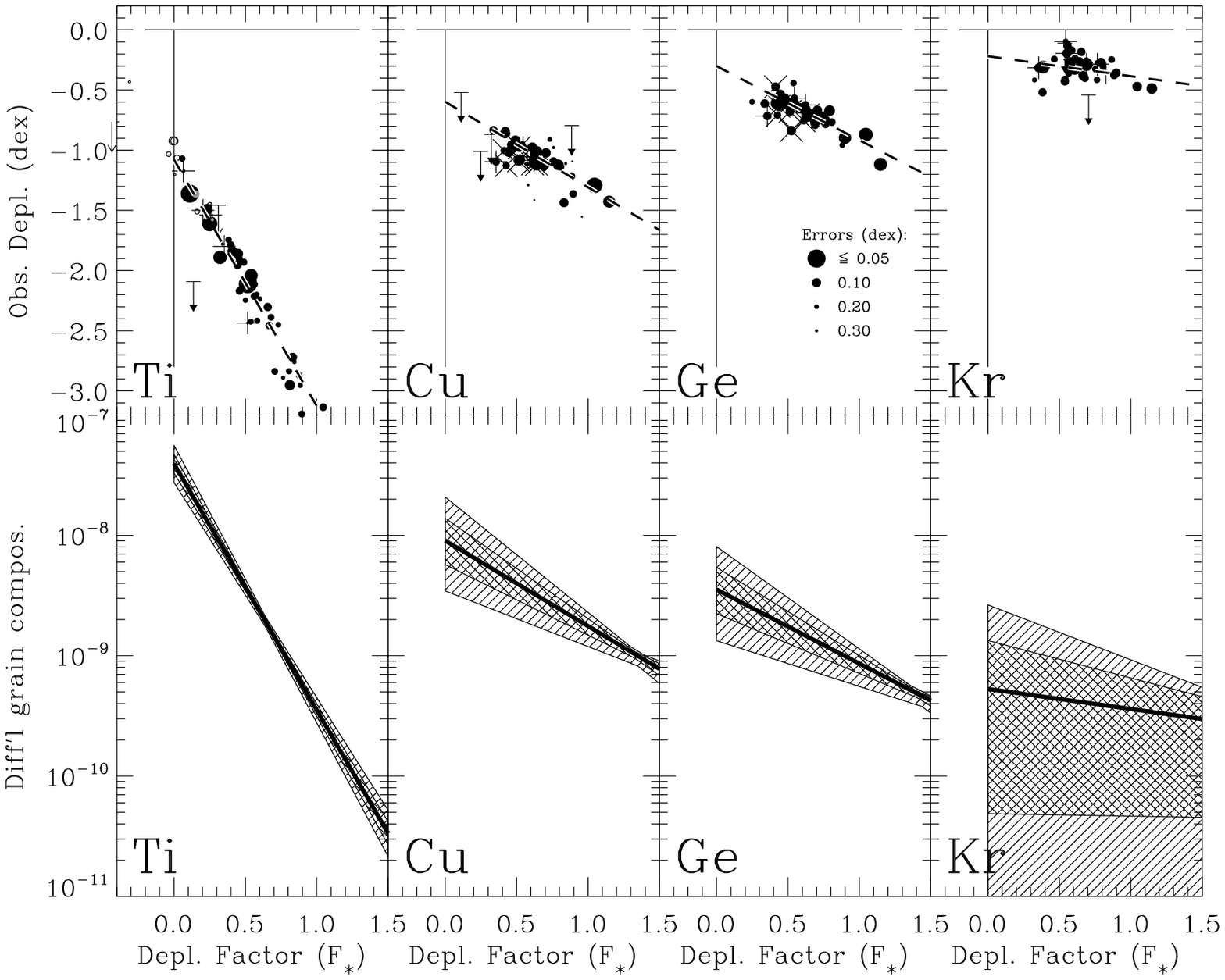}
\caption{Same as for Fig.~\protect\ref{panel_set_1} for the elements Ti, Cu, 
Ge, and Kr.\label{panel_set_4}}
\end{figure}
\clearpage

\section{Derivation of \boldmath $N$(H) and $F_*$ when $N$(H) is not 
Observed}\label{calc}

\subsection{Method}

There are a number of applications where we can use the information on the 
depletion trends either to make up for the fact that $N$(H) is not known, or, 
if it is known, to make an independent determination of the overall total 
abundances of heavier elements in the gas (usually referred to as the 
metallicity of the gas).  In this section, we describe a means for processing 
information on the relative gas phase abundances to recover both $N$(H), 
assuming a metallicity equal to the solar value, and the depletion strength 
$F_*$. Later, we will touch upon the relevance of this analysis for specific 
issues ranging from determinations of depletions of gas not far from the Sun 
(\S\ref{WD}), to interpretations of the metallicity of gas out to several kpc 
from the Sun (\S\ref{regional}), to the general behavior of sulfur depletions 
(\S\ref{sulfur}), and, finally to the metallicity in absorption systems at 
high redshifts (\S\ref{QSOALS}).  Even within the current survey, a 
reconstruction of $N$(H) and $F_*$ is of some utility: out of all of the 
sightlines (or specific velocity components) studied here, there are 97 cases 
where information is missing on the observed column densities of H~I, H$_2$, 
or both.  For 39 of them, the fact that $N({\rm H}_2)$ was not observed did 
not present a problem, since H$_2$ was unlikely to contribute much to $N$(H), 
as discussed in \S\ref{missing_info}.  For the remaining 58, the lack of 
information on the observed $N$(H) prevented the determination of $[X_{\rm 
gas}/{\rm H}]_{\rm obs}$ for any of the elements, values of which are 
essential for calculating $F_*$ through the use of Eq.~\ref{F_*}.

If it were true that all of the elements depleted logarithmically in unison 
as the general level of depletion became more severe (i.e., their values of 
$A_X$ were virtually identical), without information on $N$(H) it would be 
impossible to distinguish between a line of sight with a modest level of 
depletion and a certain estimated $N$(H), as opposed to a situation where the 
depletions were very strong and $N$(H) was much higher.  Fortunately, this is 
usually not the case.  Given that many sight lines have measured abundances 
of elements with different $A_X$ values, it is possible to estimate with 
reasonable accuracy the quantities $N$(H) and $F_*$, even when $N$(H) is not 
known from the observations.

If we substitute $[X_{\rm gas}/{\rm H}]_{\rm obs}$ for $[X_{\rm gas}/{\rm 
H}]_{\rm fit}$ in Eq.~\ref{better_equation}, we obtain
\begin{equation}\label{modified_better_equation}
[X_{\rm gas}/{\rm H}]_{\rm obs}=B_X+A_X(F_*-z_X)~.
\end{equation}
Noting that
\begin{equation}
[X_{\rm gas}/{\rm H}]_{\rm obs}=\log N(X)-\log N({\rm H})-\log (X/{\rm 
H})_\odot
\end{equation}
(this is simply a restatement of Eq.~\ref{depl_def}), we can rearrange the 
terms in this equation to obtain a simple linear expression
\begin{mathletters}
\begin{equation}\label{xy}
y=a+bx~,
\end{equation}
where
\begin{equation}
y=\log N(X)-\log (X/{\rm H})_\odot-B_X+A_Xz_X~,
\end{equation}
\begin{equation}
x=A_X~,
\end{equation}
and the coefficients of the equation have the meaning
\begin{equation}
a=\log N({\rm H})~,
\end{equation}
and
\begin{equation}
b=F_*~.
\end{equation}
\end{mathletters}
This set of equations allows us to derive the most likely values of $\log 
N({\rm H})$ and $F_*$ through the method of finding a (weighted) 
least-squares fit, once again using the routine {\tt FITEXY} 
 (Press et al. 2007) that recognizes the existence of errors 
in the measurements of both $x$ and $y$ when minimizing the $\chi^2$ of the 
fit.  Note that the uncertainty in $\log (X/{\rm H})_\odot$ does not 
contribute to the errors in $y$, since excursions in this term are exactly 
canceled by opposing changes in the derived values of $B_X$.  Any combination 
of elements that does not give large differences for the $A_X$ values will 
yield a value for $b$ (i.e., $F_*$) that is very uncertain.  For this reason, 
lines of sight where the $A_X$ values span a range of less than 0.5 were not 
evaluated.  Also, at least 3 elements were required for the analysis to 
proceed. (In principle, only 2 elements are needed to obtain a solution, but 
then we have no information on the goodness of fit.)  The element sulfur was 
not considered. Henceforth, we will refer to the outcomes of the above set of 
equations as synthetic versions of $N$(H) and $F_*$.  

\subsection{Outcomes and Performance}\label{outcomes_performance}

Table~\ref{nh_fstar} allows us to compare the observed values of $\log N({\rm 
H})$ and values of $F_*$ computed according to Eq.~\ref{F_*} in 
\S\ref{determination_of_F_*} [columns~(3)$-$(6)] to the synthetic ones 
obtained from the best-fit calculations described above [columns~(7) and 
(8)].  This table also lists synthetic values for these two quantities for 
the stars that had incomplete or no information on the observed $N$(H).  
Stars for which $F_*$ could not be computed either from Eqs.~\ref{F_*} or 
\ref{xy} do not appear in the table.

\placetable{nh_fstar}

As stated in \S\ref{parameters}, the highly depleted velocity component at 
$-15\,{\rm km~s}^{-1}$ in the direction of $\zeta$~Oph was adopted as an 
approximate fiducial point for defining the scale of $F_*$.  In the initial 
analysis, we had assumed that this component is responsible for most of the 
hydrogen along the sight line.  We are now in a position to test this 
assumption.  The least-squares fit outcome for Eq.~\ref{xy} evaluated for the 
other component at $-27\,{\rm km~s}^{-1}$ yields $\log N({\rm H})_{\rm 
syn.}=19.53\pm 0.06$, which is well below $\log N({\rm H})_{\rm obs.}=21.15$ 
determined from the damped L$\alpha$ profile and the Lyman series lines of 
H$_2$ in the spectrum of this star.  This value is even lower than the 
estimate of $\log N({\rm H})=19.74$ that was adopted by Savage, Cardelli \& 
Sofia (1992).

For most of the determinations the trends of $y$ vs. $x$ appear to be well 
defined, and the scatter of $y$ values on either side of the best-fit line 
are consistent with the measurement errors.  However, on some occasions the 
minimum $\chi^2$ values indicated that the fit was poor, as shown by small 
values for the probabilities of a worse fit given in column (9) of the table.  
When this happens, we must be cautious about the reliability of the outcomes 
for the synthetic $N$(H) and $F_*$.  Two examples illustrate some common 
reasons for this sort of outcome.

\begin{deluxetable}{
r  
c  
c  
c  
c  
c  
c  
c  
c  
c  
l  
}
\tabletypesize{\small}
\rotate
\tablecolumns{11}
\tablewidth{0pt}
\tablecaption{Observed and Synthetic $\log N$(H) and $F_*$\label{nh_fstar}}
\tablehead{
\colhead{} & \colhead{} & \multicolumn{4}{c}{Observed} &
\colhead{} & \multicolumn{2}{c}{Synthetic} & \colhead{Prob.}\\
\cline{3-6} \cline{8-9}
\colhead{} & \colhead{} & \multicolumn{3}{c}{$\log N$(H)} &
\colhead{} & \colhead{} & \colhead{} & \colhead{} & \colhead{worse}\\
\cline{3-5}
\colhead{HD} & \colhead{Name} & \colhead{l.l.} &
\colhead{best} & \colhead{u.l.} & \colhead{$F_*$} &
\colhead{~} & \colhead{$\log N$(H)} & \colhead{$F_*$} &
\colhead{fit} & \colhead{Elements Considered}\\
\colhead{(1)} &
\colhead{(2)} &
\colhead{(3)} &
\colhead{(4)} &
\colhead{(5)} &
\colhead{(6)} &
\colhead{} &
\colhead{(7)} &
\colhead{(8)} &
\colhead{(9)} &
\colhead{(10)}
}
\startdata
          1383&        HD 1383&  21.43&  21.50&  21.58&   $\phm{-}  0.61\pm 0.04$&&$21.49\pm 0.08$&   $\phm{-}  0.59\pm 0.08$&  0.510&  O Mg Mn Ni Cu Ge\\
          2905&   $\kappa$ Cas&  21.17&  21.29&  21.40&   $\phm{-}  0.58\pm 0.06$&&$21.00\pm 0.26$&   $\phm{-}  0.41\pm 0.14$&  0.989& Mg Cl Ti Mn Fe Ni Cu\\
          5394&   $\gamma$ Cas&  20.06&  20.16&  20.24&   $\phm{-}  0.52\pm 0.04$&&$20.15\pm 0.05$&   $\phm{-}  0.51\pm 0.05$&  0.306&  N  O Mg  P Cl Ti Mn Fe\\
         12323&       HD 12323&  21.21&  21.29&  21.36&   $\phm{-}  0.52\pm 0.04$&&$21.24\pm 0.08$&   $\phm{-}  0.48\pm 0.07$&  0.011&  O Mg Mn Fe Ni Cu Ge\\
         13268&       HD 13268&  21.35&  21.42&  21.49&   $\phm{-}  0.51\pm 0.04$&&$21.38\pm 0.07$&   $\phm{-}  0.48\pm 0.08$&  0.147&  O Mg Mn Ni Cu Ge\\
         13745&       HD 13745&  21.30&  21.37&  21.45&   $\phm{-}  0.43\pm 0.07$&&        \nodata&                   \nodata&\nodata& Mg Fe\\
         14434&       HD 14434&  21.40&  21.47&  21.54&   $\phm{-}  0.52\pm 0.04$&&$21.45\pm 0.09$&   $\phm{-}  0.51\pm 0.09$&  0.301&  O Mg Mn Ni Cu\\
         15137&       HD 15137&  21.10&  21.22&  21.35&   $\phm{-}  0.37\pm 0.09$&&        \nodata&                   \nodata&\nodata& Mg Fe\\
\enddata
\end{deluxetable}
\clearpage

\begin{figure}
\plotone{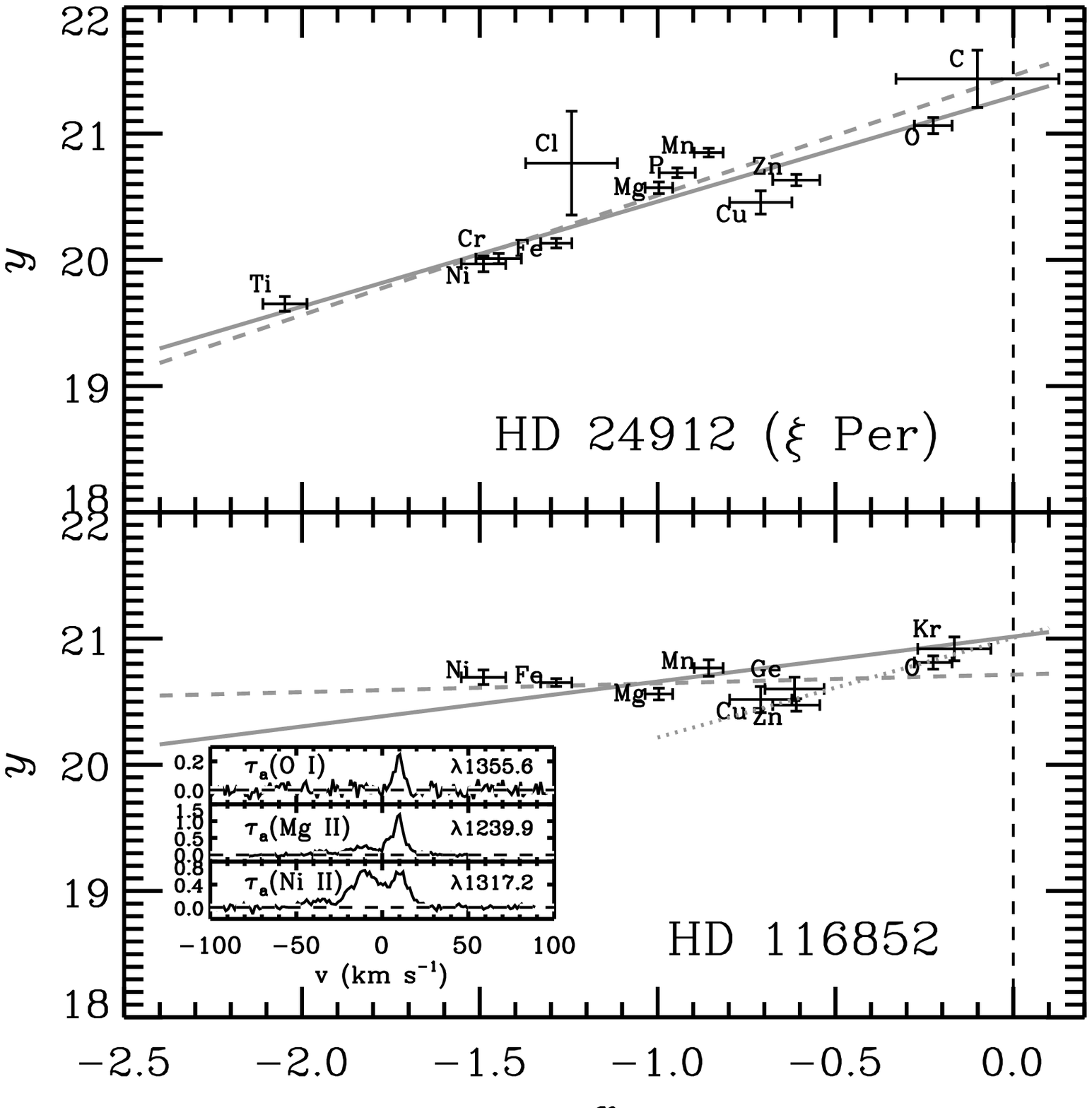}
\caption{Illustrations of the fits obtained from the use of 
Eq.~\protect\ref{xy}, which can be used to estimate $N$(H) and $F_*$ when 
$N$(H) is not observed.  The dashed gray line in each case shows the best 
fit: its slope yields $F_*$ and the $y$-intercept at $x=0$ yields $N$(H).  
The solid gray line shows the slope and intercept obtained through from the 
observed value of $N$(H) and the value of $F_*$ derived using the formulae 
given in \S\protect\ref{determination_of_F_*}. For HD~116852 (lower panel), a 
fit (dotted gray line) is also shown for just the elements O, Cu, Zn, Ge and 
Kr. The inset in this panel shows the shapes of the apparent optical depths 
vs. heliocentric radial velocity for the elements O~I, Mg~II, and Ni~II.  The 
two sight lines depicted in this figure highlight some special considerations 
that resulted in the poor fits discussed in the 
text.\label{nh_fstar_examples}}
\end{figure}

First, as the top panel of Fig.~\ref{nh_fstar_examples} shows for the star 
$\xi$~Per, the poor fit may simply be due to the fact that many of the 
observational errors for the column densities are remarkably small 
 (Cardelli et al. 1991), which results in a large value for $\chi^2$ 
for the fit even when the overall trend seems to be reasonably well defined.  
Here, there is a moderate but tolerable disagreement between the intercepts 
and slopes of two lines, one representing the trend that conforms to $\log 
N({\rm H})_{\rm obs}=21.29\pm 0.08$ and $F_*=0.83\pm 0.02$ derived using 
Eq.~\ref{F_*} (solid gray line) and the other arising from the best-fit 
solution to Eq.~\ref{xy} (dashed gray line), yielding $\log N({\rm H})_{\rm 
syn.}=21.46\pm 0.06$ and $F_{*\,{\rm syn.}}=0.95\pm 0.05$. (It is important 
to note that the quoted uncertainties in the results are based on only the 
measurement errors, with no reference to how much the points scatter about 
the fit line.  If one repeats the analysis using just the measurements with 
no errors and assumes that the model is perfectly correct, the anticipated 
uncertainties in $\log N({\rm H})_{\rm syn.}$ and $F_{*\,{\rm syn.}}$ 
increase to 0.14 and 0.12, respectively.)

\placefigure{nh_fstar_examples}

A second reason for a poor outcome is that there may be fundamental problems 
arising from the fact that different regions with markedly different 
depletion levels are being grouped together.  For example, the line of sight 
toward the star HD~116852 in the lower halo of the Galaxy traverses regions 
at different radial velocities that have markedly different relative 
abundances (Sembach \& Savage 1996).  When such a mixture is 
considered as a whole, the basic premise that the relative abundances should 
obey the simple relation given by Eq.~\ref{better_equation} starts to break 
down.  It is clear from the lower panel of Fig.~\ref{nh_fstar_examples} that 
the disagreements between the two variable pairs is far worse: $\log N({\rm 
H})_{\rm obs}=21.02\pm 0.08$ vs. $\log N({\rm H})_{\rm syn.}=20.71\pm 0.05$ 
(i.e., a factor of 2 in column density, which is well outside the quoted 
errors) and $F_*=0.36\pm 0.04$ from Eq.~\ref{F_*} vs. $F_{*\,{\rm 
syn.}}=0.07\pm 0.05$ from Eq.~\ref{xy}.  The inset in the figure showing the 
apparent optical depths $\tau_{\rm a}$ as a function of velocity for three 
species shows that the small changes in depletion for the lightly depleted 
elements O and Mg contrast sharply with strong changes in the usually heavily 
depleted Ni.  If we could actually see a velocity profile for H, it would 
probably not look much different than the one for O.  The broad peaks seen in 
the Ni profile centered at $v=-10$ and $-35\,{\rm km~s}^{-1}$ probably have a 
low depletion and an almost negligible amount of hydrogen associated with 
them, but they are conspicuous because they are seen alongside the component 
centered on $v=+10\,{\rm km~s}^{-1}$ where the Ni is highly depleted.  If we 
repeat the analysis using just the information derived from the elements 
O,Cu, Zn, Ge and Kr, the $y$-intercept corresponds to $\log N({\rm H})_{\rm 
syn.}=21.01\pm 0.09$ (dotted line), which almost exactly (fortuitously) 
equals the measured value.  With the analysis restricted to this subset of 
elements with only light to moderate depletions, which gives greater emphasis 
to the velocity component that has the most hydrogen, the probability of a 
worse fit comes out at a satisfactory value of 0.674.

\begin{figure}
\plotone{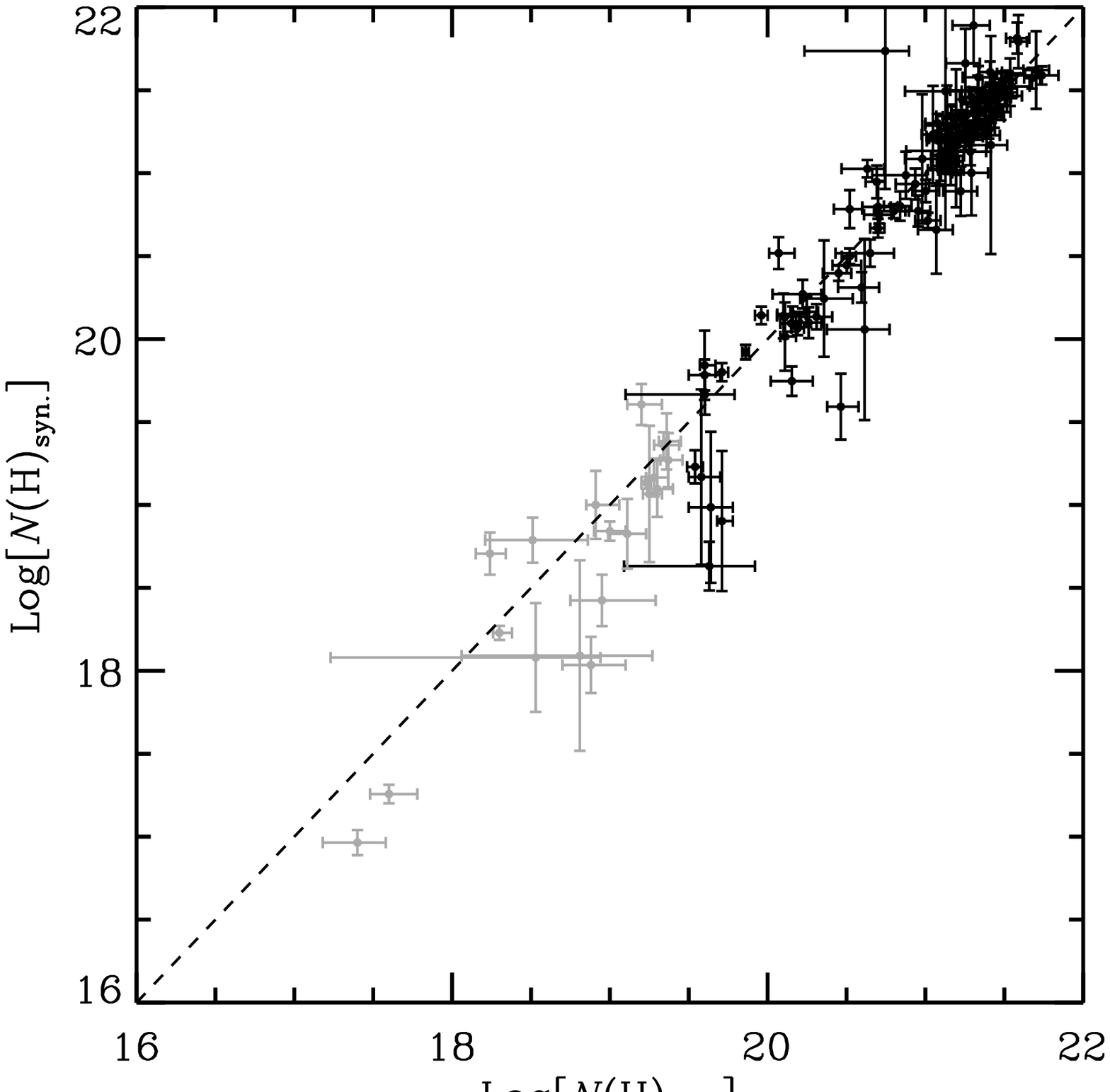}
\caption{A comparison of the synthetic values of $N$(H) calculated using 
Eq.~\ref{xy} ($y$ axis) against their actual observed values ($x$ axis), when 
known.  See Columns~(3$-$5) and Column~(7) of Table~\protect\ref{nh_fstar}. 
Gray points and error bars signify cases where $\log N({\rm H})_{\rm 
obs.}<19.5$ to emphasize the fact that their abundance measurements may not 
reliably indicate the true gas phase abundances in H~I 
regions.\label{compare_nh}}
\end{figure}
\begin{figure}
\plotone{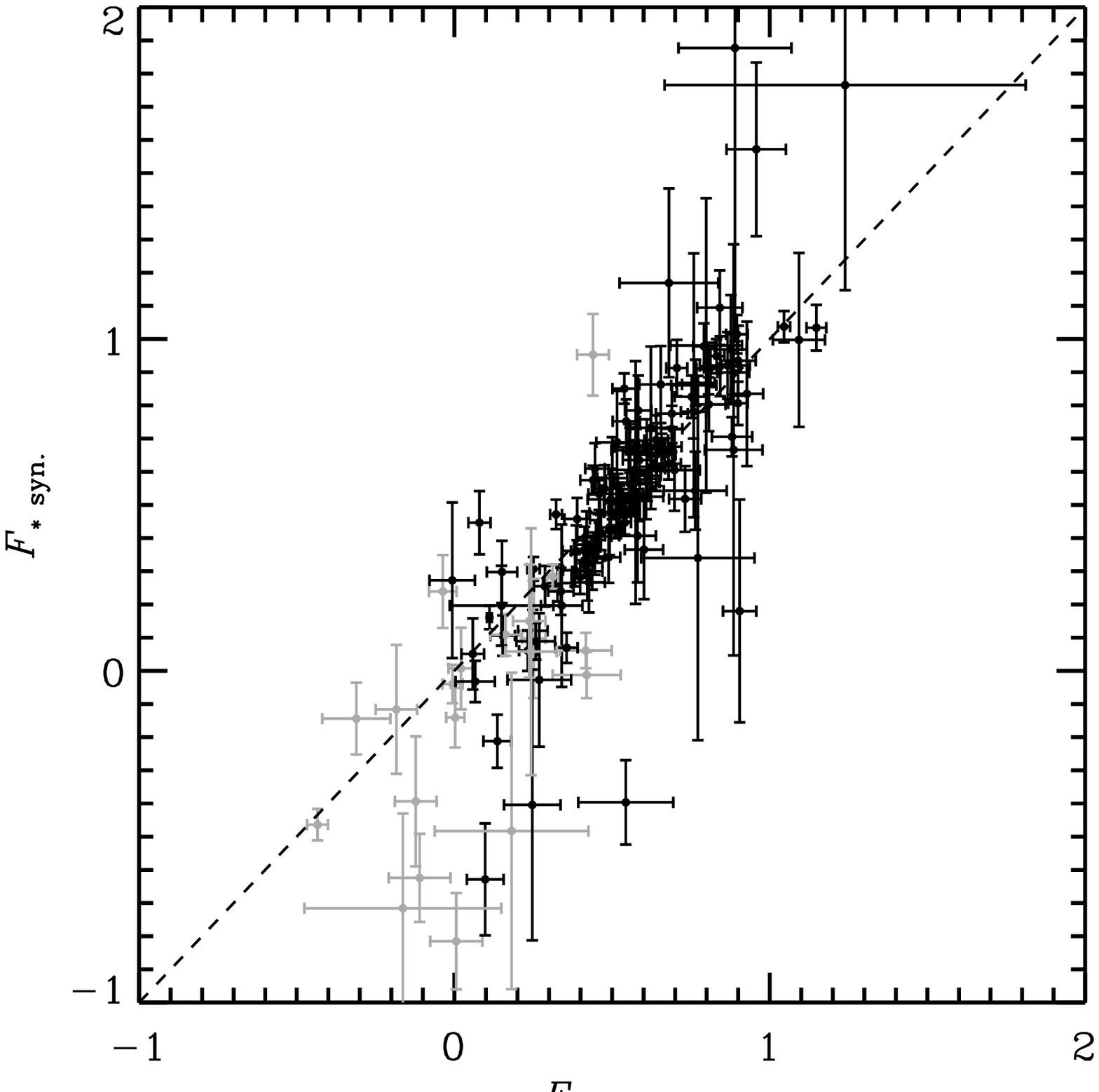}
\caption{A comparison of the synthetic values of $F_*$ computed using 
Eq.~\protect\ref{xy} ($y$ axis), which would be used if $N$(H) were unknown, 
against computations of $F_*$ obtained from Eq.~\protect\ref{F_*} for all 
cases where $N$(H) has actually been observed.  As in 
Fig.~\protect\ref{compare_nh}, gray symbols show cases where $\log N({\rm 
H})_{\rm obs.}<19.5$.\label{compare_fstar}}
\end{figure}
\begin{figure}
\epsscale{.95}
\plotone{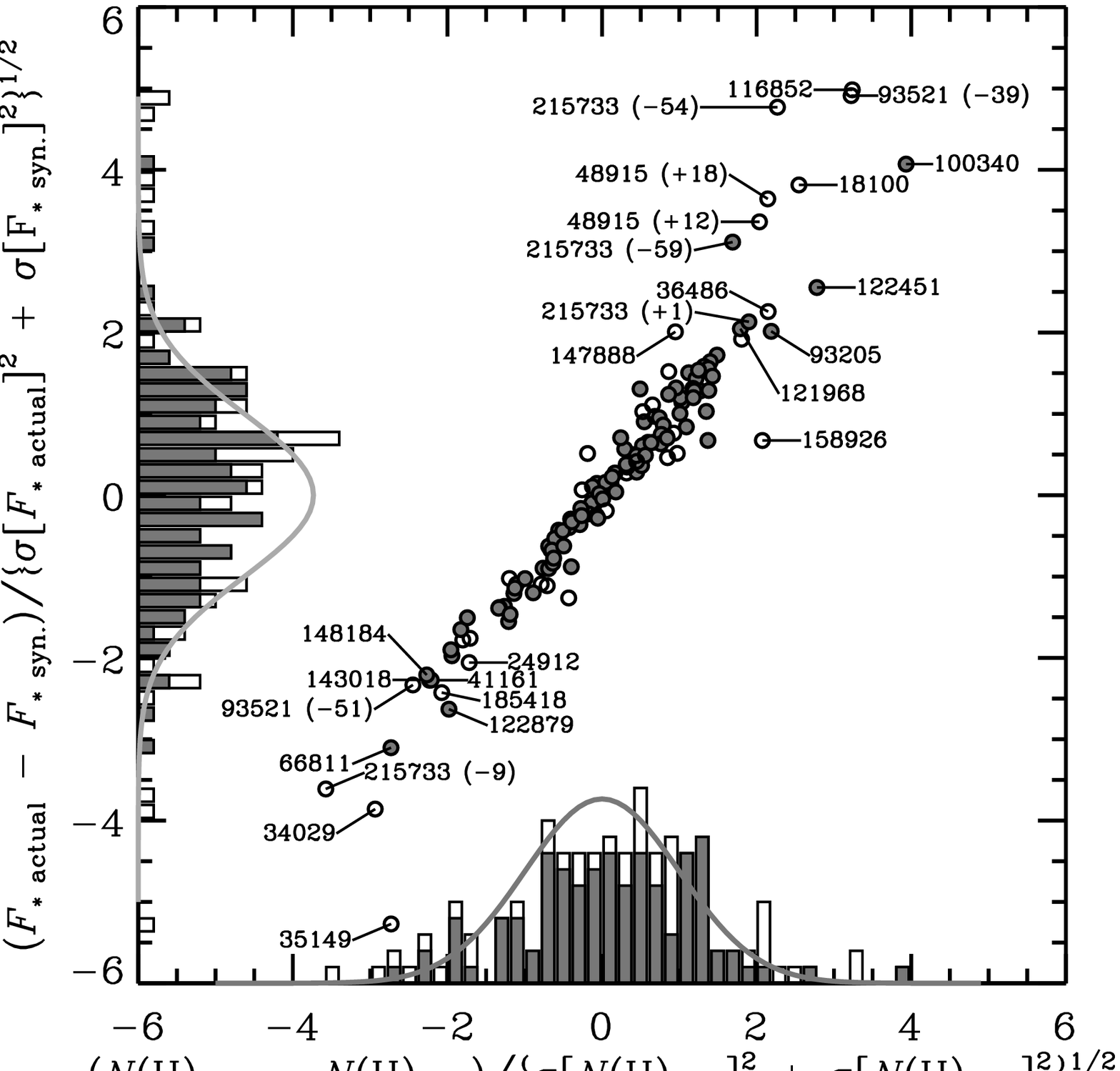}
\caption{Deviations of $N$(H) and $F_*$ (shown in Columns~(3$-$6) of 
Table~\protect\ref{nh_fstar}) from their synthetic counterparts derived from 
Eq.~\protect\ref{xy} (Columns~7 and 8 of the same table), divided by the 
respective uncertainties for the differences of the two quantities as defined 
by denominators shown in the $x$ and $y$ axis labels, for all cases where 
$N$(H) is known from observations.  Points that are filled in have 
probabilities of worse fit (see Column~(9) of Table~\protect\ref{nh_fstar}) 
greater than 0.05, while open points are below this threshold.  The HD 
numbers of stars (and the velocity components in parentheses, if appropriate) 
are labeled for outlier points that have deviations for either case that are 
more than $2\sigma$ away from zero.  Histograms showing the distributions of 
all deviations are shown next to the $x$ and $y$ axes, with an overlay of a 
Gaussian distribution with a zero mean and standard deviation of 1 for 
comparison purposes. Solid and open portions of the bars are coded in the 
same manner as the points. \label{nh_fstar_deviations}}
\end{figure}

The two examples highlighted in the previous two paragraphs were carefully 
chosen to demonstrate two principal reasons for finding probability of worse 
fit outcomes at extremely low values (in these cases, lower than 0.001, as 
can be seen for the entries in column (9) of Table~\ref{nh_fstar} for the 
respective stars).  Other reasons for poor fits may include errors in the 
column densities that are well outside the quoted uncertainty limits (i.e., 
mistakes or poor judgment on the part of the observer) or perhaps some 
unusual effects that cause deviations in the element abundances so that they 
no longer conform to the simplified, general depletion trends among the 
elements being described here.

In making judgments on how well the applications of Eq.~\ref{xy} are working, 
we can examine all cases where $N({\rm H})_{\rm obs}$ is known and for which 
the right combination of elements were measured, and then simply compare the 
observed and synthetic values of $N$(H).  We can also do the same for the two 
methods for determining $F_*$.  Figures \ref{compare_nh} and 
\ref{compare_fstar} show these comparisons.  For both variables, a vast 
majority of stars show a good agreement between the two methods.

\placefigure{compare_nh}
\placefigure{compare_fstar}

We can also examine the overall behavior of differences in the two ways of 
deriving both variables divided by the expected errors of such differences.  
This is shown in Figure~\ref{nh_fstar_deviations}.  It is clear that the 
errors are highly correlated, which comes as no surprise if one imagines how 
errors in the slopes of the lines can have a large leverage in creating 
deviations for the locations of the $y$-intercepts at $x=0$ (toward the far 
right-hand sides in the plots shown in Fig.~\ref{nh_fstar_examples}), 
especially if there is no representation by elements with small depletions.  
(If one examines the errors in the synthetic $N$(H) values listed in 
Table~\ref{nh_fstar}, it is clear that they are generally larger than average 
if C, N, O or Kr are not represented in the list of elements shown in the 
last column.)  Histograms on the sides of the plot box in the figure show how 
the deviations are distributed, with a Gaussian curve having a unit standard 
deviation and zero mean overlaid for comparison.  While the distribution of 
the bars of these histograms seem generally consistent with this Gaussian 
curve, it is clear that the number of outliers beyond $2\sigma$ is larger 
than expected.  Those cases are identified by their HD numbers (and 
velocities, if distinct).  For both the histogram bars and the circle points 
in the diagram, the filled-in cases have probabilities of a worse fit that 
are larger than 0.05, while those that are open have lower probabilities.

\placefigure{nh_fstar_deviations}

\section{Depletions toward White Dwarf Stars in the Local Bubble}\label{WD}

It is clear from Fig.~\ref{rz_hist} that our coverage of distances exhibits 
an abrupt lower limit at about 100~pc.  The volume of space out to about 
100~pc from the Sun has an uncharacteristically low density; it is a region 
known as the Local Bubble (Cox \& Reynolds 1987; Breitschwerdt et al. 1996;
Vallerga 1996; Ferlet 1999; Lallement et al. 2003; Frisch 2007), and it
contains a collection of isolated, warm clouds (Redfield \& Linsky 2004a, b)
embedded within and confined by a hot plasma (Breitschwerdt \& de Avillez
2006; Savage \& Lehner 2006).  It was probably created by a series of
supernova explosions that arose from an association of early-type stars that
passed through our vicinity about 14 Myr ago
(Ma\'iz-Appel\'aniz 2001; Bergh\"ofer \& Breitschwerdt 2002; Fuchs et
al. 2006).  Many white dwarf stars embedded in the Local Bubble were observed
by the {\it Far Ultraviolet Spectroscopic Explorer\/} ({\it FUSE\/})
satellite.  However, for all but a few of these targets, $N$(H~I) was not
observed.  Thus, if we wish to obtain an understanding about the strength of
the depletions toward these objects, we must rely on the analysis of
synthetic $F_*$. 

A large fraction of the sight lines studied in the Local Bubble probably have 
$\log N({\rm H~I})<19.5$, and thus they violate the restriction that we 
imposed to lessen the chances that partial ionization effects could influence 
the outcomes (see rule nr.~3 in \S\ref{rejection}).  Indeed, along the sight 
lines to white dwarf stars in the Local Bubble, it is not unusual to find 
that $N({\rm N~II})>N({\rm N~I})$ (Kruk et al. 2002; Lehner et al. 2003). 
However, for the elements Si and Fe, we can see from Figs~\ref{panel_set_1}
and \ref{panel_set_3} that most measurements with $\log N({\rm H~I})<19.5$
(shown as open circles in the figures) seem to lie reasonably close to the
trend established by the higher column density cases (or an extrapolation of
it for $F_*<0$), suggesting that, for these particular elements, our initial
hesitation to include these low column density cases was more cautious than
necessary.  Moreover, ionization models of gas within the Local Bubble shown
in Figure~6 of Lehner et al.  (2003) indicate that for $\log N({\rm
H~I})\gtrsim 18$, the influence of photoionization within the Local Bubble
on the abundances of O~I, P~II, Si~II and Fe~II relative to H~I is small.  
Comparisons of O~I to H~I are rather secure under almost all conditions 
because the ionizations of these two elements are coupled to each other by a 
strong charge exchange reaction (Field \& Steigman 1971; Chambaud et al.
1980; Stancil et al. 1999).

In light of the above statements, we bypass here the column density 
restriction and proceed with derivations of synthetic values of $N$(H~I) and 
$F_*$ for sight lines toward white dwarf stars in the Local Bubble by using 
the prescription outlined in \S\ref{calc}.  In order to obtain a satisfactory 
spread in $A_X$, we require that information is available for O~I and one or 
more of the species P~II, Si~II, and Fe~II.  Table~\ref{lism_tbl} lists 
results for the stars that satisfy these conditions, based on observations 
reported by Oliveira et al (2003) and Lehner et al 
 (2003).  If errors in $y$ of Eq.~\ref{xy} were set to 
their formal values, unreasonably large values of $\chi^2$ were obtained.  
This probably reflects the fact that the accuracy of the observations of 
$N(X)$, along with the formal errors in $A_X$ and $B_X$, are far better than 
the precision of the assumption that Eq.~\ref{xy} truly applies for gas 
parcels in the Local Bubble.  For this reason, errors in $y$ were 
artificially increased by 0.2~dex (added in quadrature to the original 
errors) to make the worse fit probabilities, as reflected by the values of 
$\chi^2$, evenly distributed over the interval 0 to 1.

\begin{deluxetable}{
r  
c  
c  
c  
c  
c  
c  
c  
c  
l  
}
\tabletypesize{\footnotesize}
\tablecolumns{10}
\tablewidth{0pt}
\tablecaption{Synthetic $\log N$(H) and $F_*$ for Stars in the Local Bubble\label{lism_tbl}}
\tablehead{
\colhead{} & \colhead{} & \multicolumn{2}{c}{Gal. Coord.} & \colhead{} & \colhead{dist.\tablenotemark{a}}\\
\cline{3-4}
\colhead{WD nr.} & \colhead{Alt. Name} & \colhead{$\ell$} & \colhead{$b$} &
\colhead{$V$} & \colhead{(pc)} & \colhead{$\log N({\rm H})_{\rm syn.}$} &
\colhead{$F_{\rm *~syn.}$} & \colhead{$\chi^2$} & \colhead{Elem. Considered\tablenotemark{b}}\\
\colhead{(1)} &
\colhead{(2)} &
\colhead{(3)} &
\colhead{(4)} &
\colhead{(5)} &
\colhead{(6)} &
\colhead{(7)} &
\colhead{(8)} &
\colhead{(9)} &
\colhead{(10)}
}
\startdata
0004+330&           GD 2&112.48&      $-28.69$& 13.8&  97&           $19.69\pm 0.30$&   $\phm{-}  0.58\pm 0.33$&   0.61&  O  P Fe\\
0050-332&         GD 659&299.14&      $-84.12$& 13.4&  58&           $18.53\pm 0.26$&   $\phm{-}  0.03\pm 0.11$&\nodata&  O Fe\\
0131-163&          GD984&167.26&      $-75.15$& 14.0&  96&           $19.21\pm 0.26$&   $\phm{-}  0.32\pm 0.27$&   2.49&  O Si  P Fe\\
0455-282&  MCT 0455-2812&229.29&      $-36.17$& 14.0& 102&           $18.09\pm 0.65$&          $ -0.18\pm 0.45$&   0.48&  O Si Fe\\
0501+527&       G191-B2B&155.95&$\phm{-}$ 7.10& 11.8&  69&           $18.05\pm 0.26$&          $ -0.22\pm 0.26$&   1.32&  O Si Fe\\
0549+158&          GD 71&192.03&      $ -5.34$& 13.1&  49&           $17.52\pm 0.28$&          $ -0.06\pm 0.20$&\nodata&  O Si\\
0621-376&               &245.41&      $-21.43$& 12.1&  78&           $18.56\pm 0.26$&   $\phm{-}  0.15\pm 0.34$&\nodata&  O Fe\\
0715-703&               &281.62&      $-23.49$& 14.2&  94&           $19.16\pm 0.27$&   $\phm{-}  0.28\pm 0.27$&   2.77&  O Si  P Fe\\
1017-138&               &256.48&$\phm{-}$34.74& 14.6&  90&           $18.93\pm 0.46$&   $\phm{-}  0.11\pm 0.39$&   4.72&  O Si  P Fe\\
1202+608&       Feige 55&133.12&$\phm{-}$55.66& 13.6& 200&           $18.86\pm 0.25$&          $ -0.11\pm 0.23$&   1.54&  O Si Fe\\
1211+332&          HZ 21&175.04&$\phm{-}$80.02& 14.7& 115&           $18.97\pm 0.25$&   $\phm{-}  0.07\pm 0.21$&   0.84&  O Si  P Fe\\
1234+481&               &129.81&$\phm{-}$69.01& 14.4& 129&           $18.99\pm 0.27$&   $\phm{-}  0.40\pm 0.31$&   0.07&  O Si  P Fe\\
1314+293&         HZ 43A& 54.11&$\phm{-}$84.16& 12.7&  68&           $17.82\pm 0.25$&   $\phm{-}  0.22\pm 0.28$&   0.37&  O Si Fe\\
1528+487&               & 78.87&$\phm{-}$52.72& 14.5& 140&           $19.12\pm 0.27$&   $\phm{-}  0.39\pm 0.28$&   1.38&  O Si  P Fe\\
1615-154&       EGGR 118&358.79&$\phm{-}$24.18& 12.4&  55&           $19.01\pm 0.28$&   $\phm{-}  0.17\pm 0.28$&   1.64&  O  P Fe\\
1631+781&  IES 1631+78.1&111.30&$\phm{-}$33.58& 13.0&  67&           $19.29\pm 0.27$&   $\phm{-}  0.48\pm 0.29$&   0.05&  O Si  P Fe\\
1634-573&     HD 149499B&329.88&      $ -7.02$&  9.8&  37&           $18.79\pm 0.25$&   $\phm{-}  0.35\pm 0.27$&   1.77&  O Si  P Fe\\
1636+351&               & 56.98&$\phm{-}$41.40& 14.9& 109&           $18.77\pm 0.33$&          $ -0.31\pm 0.32$&   3.00&  O Si  P Fe\\
1800+685&               & 98.73&$\phm{-}$29.78& 14.6& 159&           $19.35\pm 0.29$&   $\phm{-}  0.10\pm 0.25$&   1.34&  O Si  P Fe\\
1844-223&               & 12.50&      $ -9.25$& 14.0&  62&           $19.23\pm 0.31$&   $\phm{-}  0.28\pm 0.31$&   1.14&  O Si  P Fe\\
2004-605&               &336.58&      $-32.86$& 13.4&  58&           $18.93\pm 0.26$&   $\phm{-}  0.26\pm 0.29$&   0.75&  O Si  P Fe\\
2011+395&EUVE J2013+40.0& 77.00&$\phm{-}$ 3.18& 14.6& 141&           $19.32\pm 0.26$&   $\phm{-}  0.31\pm 0.26$&   2.90&  O Si  P Fe\\
2111+498&               & 91.37&$\phm{-}$ 1.13& 13.1&  50&           $18.26\pm 0.26$&          $ -0.07\pm 0.25$&   5.86&  O Si Fe\\
2124-224&               & 27.36&      $-43.76$& 14.6& 224&           $19.05\pm 0.25$&          $ -0.20\pm 0.25$&   6.04&  O Si  P Fe\\
2152-548&               &339.73&      $-48.06$& 14.4& 128&           $18.75\pm 0.27$&   $\phm{-}  0.03\pm 0.24$&   1.18&  O Si Fe\\
2211-495&               &345.79&      $-52.62$& 11.7&  53&           $18.54\pm 0.25$&          $ -0.01\pm 0.24$&   2.29&  O Si  P Fe\\
2247+583&         Lan 23&107.64&      $ -0.64$& 14.3& 122&           $20.11\pm 0.33$&   $\phm{-}  0.56\pm 0.35$&\nodata&  O Fe\\
2309+105&         GD 246& 87.26&      $-45.12$& 13.0&  79&           $18.94\pm 0.25$&   $\phm{-}  0.29\pm 0.27$&   1.44&  O Si  P Fe\\
2331-475&  MCT 2331-4731&334.84&      $-64.81$& 13.5&  82&           $18.71\pm 0.26$&   $\phm{-}  0.12\pm 0.24$&   1.75&  O Si  P Fe\\
\enddata
\tablenotetext{a}{Taken from the papers that described the FUSE observations.}
\tablenotetext{b}{All column density data from Lehner et al. (2003), except for
HZ~21 (Oliveira et al. 2003).}
\end{deluxetable}
\begin{figure}
\plotone{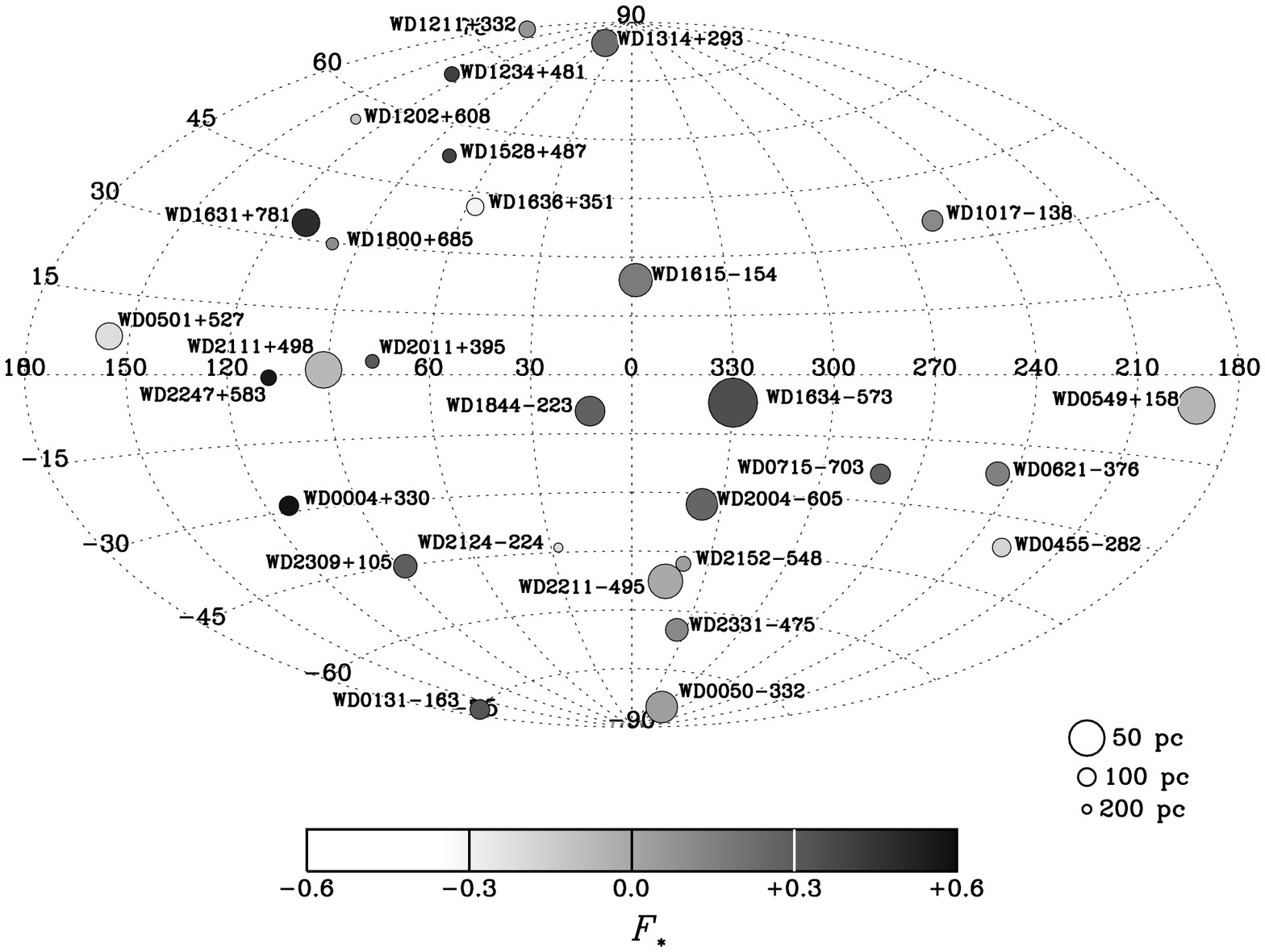}
\caption{A depiction of $F_{*\,{\rm syn.}}$ for the sight lines toward white 
dwarf stars observed with {\it FUSE}.  The positions of the circles indicate 
the Galactic coordinates ($\ell=0$, $b=0$ at the center) and their sizes 
indicate the distances to the stars, according to the legend shown in the 
lower right-hand portion of the figure.  The darknesses of the filled in 
portions of the circles indicate the derived values of $F_{*\,{\rm syn.}}$ 
according to the gray scale at the bottom.\label{lism}}
\end{figure}

Values of $\log N({\rm H~I})_{\rm syn.}$ listed in Table~\ref{lism_tbl} range 
from 17.5 to 20.1.  For three cases, observations of $N$(H~I) are available 
for spot checks on the accuracy of the synthetic values: HZ~43A $\log N({\rm 
H~I})_{\rm obs.}=17.93\pm 0.03$ (Kruk et al. 2002) vs. $\log N({\rm 
H~I})_{\rm syn.}= 17.82\pm 0.25$; Lan~23 $\log N({\rm H~I})_{\rm 
obs.}=19.89^{+0.25}_{-0.04}$ (Wolff, Koester, \& Lallement 1999) vs. $\log
N({\rm H~I})_{\rm syn.}=20.11\pm 0.33$; GD~246 $\log N({\rm H~I})_{\rm 
obs.}=19.11\pm 0.03$ (Oliveira et al. 2003) vs. $\log N({\rm 
H~I})_{\rm syn.}=18.94\pm 0.25$.  In all cases, the agreements are well 
within the estimated errors.

\placetable{lism_tbl}

Figure~\ref{lism} shows the distribution in the sky of the {\it FUSE\/} white 
dwarf observations along with the distances and derived values of $F_{*\,{\rm 
syn.}}$.  The construction of this figure is very similar to Fig.~1 of Lehner 
et al. (2003) so that one can easily compare their 
values of $\log N({\rm O~I})$ (which are not very different from $N({\rm 
H~I})_{\rm syn.}$ aside from a constant factor) with our values of 
$F_{*\,{\rm syn.}}$.  It is also similar to Figs 7$-$12 of Redfield \& Linsky 
 (2004a), which show column densities of various 
species toward mostly late-type stars in the Local Bubble (these targets are 
generally much closer to us than the white dwarf stars).  Our Fig.~\ref{lism} 
indicates that there is a mild degree of coherence in the depletions: stars 
in the general direction of the south Galactic pole have light depletions, 
those toward the Galactic center have moderate depletions, and distant stars 
over a broad range of Galactic latitudes and at longitudes of around 
90\arcdeg have the largest values of $F_{*\,{\rm syn.}}$.  Six of the stars 
have 
$F_{*\,{\rm syn.}}\geq 0.30$ and yet their values of $N({\rm H})_{\rm syn.}$ 
are below our fiducial lower limit of $10^{19.5}{\rm cm}^{-2}$ for the 
general study, indicating that, provided we are not being misled by 
ionization effects, moderately strong depletions can be found in the Local 
Bubble.  This conclusion is consistent with the findings of Kimura et al. 
 (2003) for very local, warm gas that surrounds our 
heliosphere.

\placefigure{lism}

\section{Sulfur: A Troublesome Element}\label{sulfur}

Our consideration of sulfur has been deferred until now because this element 
presents difficulties that warrant special treatment.  A study of the 
depletion behavior of sulfur is especially important, since many studies of 
gas in both our Galaxy and very distant systems have relied on this element 
as a standard for what should be virtually zero depletion.  We need to 
re-examine whether or not this is true.  For instance, Calura et al. 
 (2009) have summarized some recent findings reported 
in the literature that may revise our understanding of the status of sulfur 
depletions or possible lack thereof.

The singly-ionized form of sulfur is detected only through a triplet with 
transitions at 1250.6, 1253.8 and 1259.5$\,$\AA, with $f$-values of 0.00543, 
0.0109, and 0.0166, respectively (Morton 2003).  Unfortunately, for 
many lines of sight that contain moderately large amounts of gas, even the 
weakest line is partly or strongly saturated.  Observations taken at low 
resolution (e.g., with the G160M configuration of the GHRS instrument on {\it 
HST\/}) often exhibited ratios for the equivalent widths of the two weaker 
lines of less than 1.5, and thus they were deemed to be saturated enough to 
violate one of the censorship guidelines for this study (rule nr.~2 in 
\S\ref{rejection}).  However, a small number of observations showed that the 
saturations were not so severe, or that the S~II column densities could be 
extracted from moderately saturated features recorded at high resolution and 
analyzed through their apparent optical depths.

Prominent in the limited selection of S~II measurements were the individual 
velocity components of the stars HD~93521 and HD~215733 analyzed by Spitzer 
\& Fitzpatrick (1993) and Fitzpatrick \& 
Spitzer (1997).  In a few instances, the 
velocity separations of adjacent components were small, which could lead to 
errors in the assignments of column densities between them.  An additional 
drawback of using the individual velocity components toward these stars is 
the reliance on 21-cm emission line measurements of H~I instead of L$\alpha$ 
absorption.

Table~\ref{S} in Appendix~\ref{basic_data} shows the stars that had 
measurements of S~II that were accumulated in the current study (but not all 
of which were suitable for determining depletion coefficients).  Relative to 
the stars that could be used for other elements, they are few in number (12) 
and a substantial majority of them have hydrogen column densities that are 
either unknown or below the threshold $N({\rm H})=10^{19.5}{\rm cm}^{-2}$ 
that qualifies them for consideration.  As with the measurements of 
depletions of other elements, a reference abundance was adopted from Lodders 
 (2003), $\log ({\rm S/H})_\odot + 12 = 7.26\pm 0.04$.
\begin{figure}
\epsscale{1.2}
\plottwo{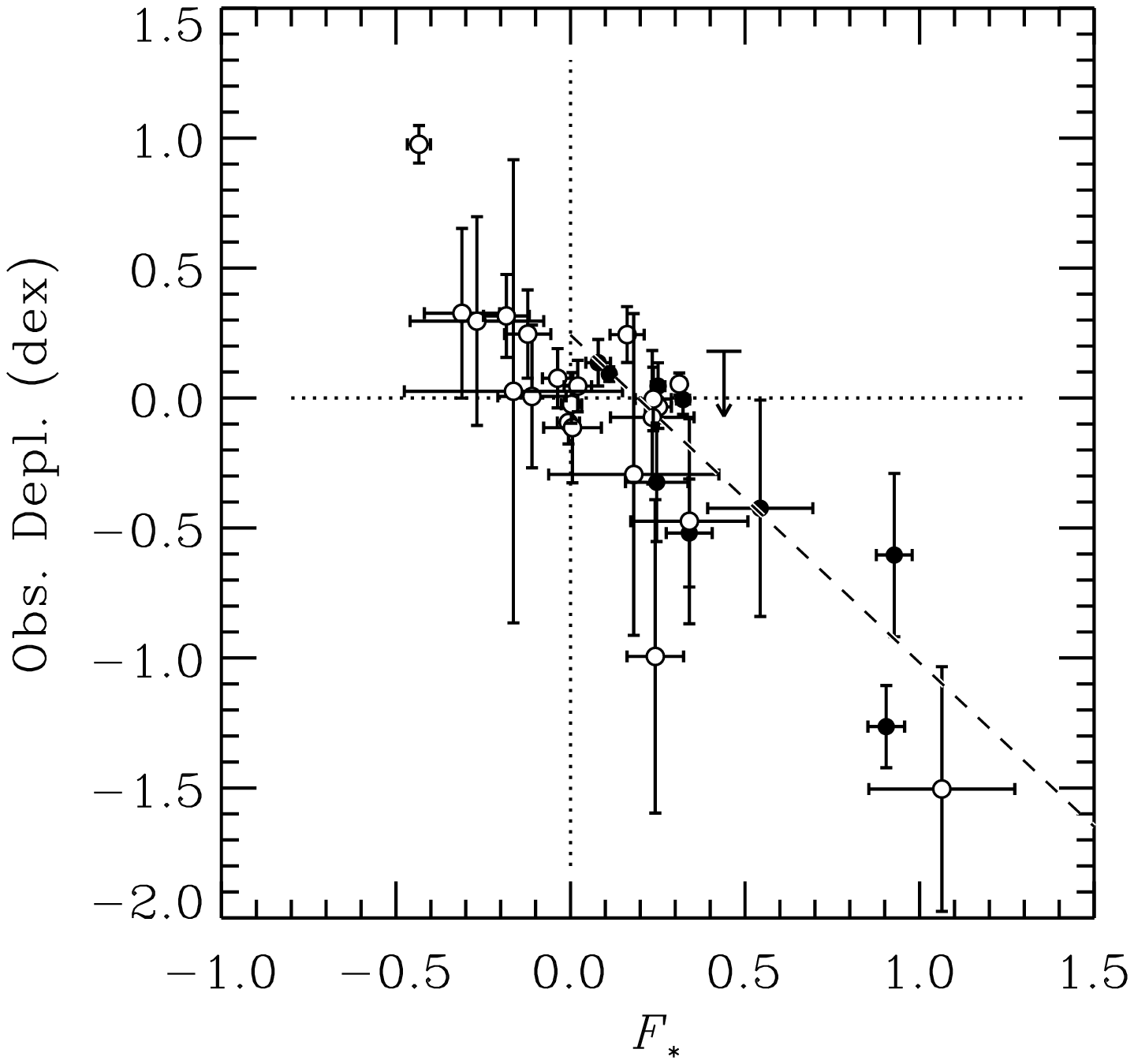}{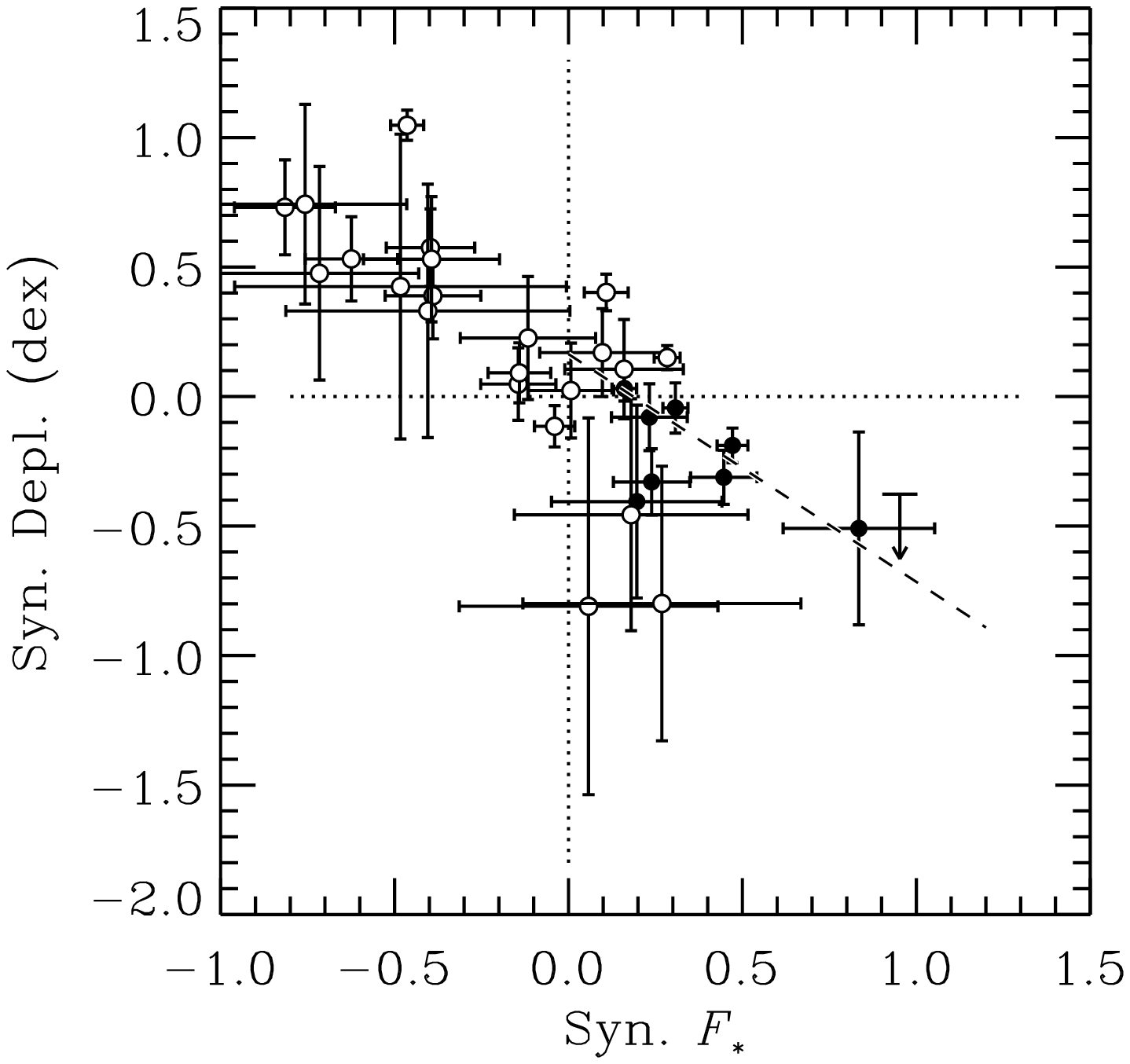}
\caption{{\it Left-hand panel:\/} Observed depletions of S as a function of 
$F_*$.  Cases with $N({\rm H})<10^{19.5}{\rm cm}^{-2}$ are shown with open 
symbols, while cases with larger values of $N$(H) have solid symbols.  The 
dashed line shows the best least-squares fit to the solid points.  {\it 
Right-hand panel:} Same as the left-hand panel, except that $N({\rm H})_{\rm 
syn.}$ was used to derive the depletions and $F_*$ derived from 
Eq.~\protect\ref{F_*} was replaced by $F_{*\,{\rm syn.}}$ calculated from the 
best fits to Eq.~\protect\ref{xy}.\label{sulfur_panels}}
\end{figure}

The left-hand panel of Fig.~\ref{sulfur_panels} shows the observed depletions 
as a function of $F_*$.  A least-squares best fit to the observations with 
$\log N({\rm H})>19.5$ (the 8 solid points) yields $A_{\rm S}=-1.261\pm 
0.165$, $B_{\rm S}=0.028\pm 0.047$, and $z_{\rm S}=0.170$, whose linear 
equation is represented by the dashed line.  However, the probability of a 
worse fit is exceptionally low: $p=0.005$.

\placefigure{sulfur_panels}

There is a strong possibility that the 21-cm measurements that were used to 
set the H~I column densities for the velocity components toward HD~93521 and 
HD~215733, which represent 26 of the points shown in the figure, are giving 
misleading outcomes.  For instance, the beam width of the telescope was large 
(21\arcmin), and some H could be positioned inside the coverage of the beam 
but not in front of a star.  Alternatively, some gas in front of the star 
that subtends a small solid angle in the sky could suffer severe beam 
dilution.  Finally, some of the gas registered at 21~cm could be behind the 
star.  (It was for these reasons that Fitzpatrick \& Spitzer 
 (1997) chose to disregard the hydrogen 
measurements and instead determined the depletions of elements other than S 
by making comparisons to the respective determinations of $N$(S~II), on the 
presumption that S was always undepleted.)

To overcome the uncertainties in $N$(H~I), we can resort to the tactic of 
deriving the synthetic versions of this quantity through the method outlined 
in \S\ref{calc} that made use of the relative abundances of other species.  
Likewise, $F_*$ determinations that used $N({\rm H})_{\rm obs.}$ could be 
declared as suspect and instead we can use the synthetic values of $F_*$ 
derived in conjunction with the calculations of $N$(H).  The right-hand panel 
of Fig.~\ref{sulfur_panels} shows the outcomes with these different synthetic 
values of $N$(H) and $F_*$ that were obtained from the fit to Eq.~\ref{xy}.  
Here, we find that $A_{\rm S}=-0.879\pm 0.284$, $B_{\rm S}=-0.091\pm 0.042$, 
and $z_{\rm S}=0.290$.  For the two fiducial points of $F_*$, we obtain from 
Eqs.~\ref{[X/H]_0} to \ref{sigma[X/H]_1} depletions $[{\rm S}_{\rm gas}/{\rm 
H}]_0=0.163\pm 0.092$ and $[{\rm S}_{\rm gas}/{\rm H}]_1=-0.715\pm 0.206$.  
The chance of obtaining a worse fit for the 8 valid points in this case is 
0.482 (from $\chi^2=5.5$ with 6 degrees of freedom), a considerable 
improvement over the earlier value that used the conventional determination 
of $F_*$ and the observed values of $N$(H).  Based on misgivings about the 
$N$(H) values derived from the 21-cm data and the improved fit using the 
synthetic values of $N$(H) and $F_*$, the numbers given immediately above are 
probably more reliable than those that were listed earlier, which were based 
on $N({\rm H})_{\rm obs}$.

The ionization potential of ${\rm S}^+$ is 23.4~eV; which, except for carbon, 
is the highest of the singly ionized species considered in this survey.  For 
this reason, sulfur may be especially prone to the problem that its apparent 
abundance could be enhanced by its singly ionized form appearing in regions 
where the hydrogen is fully ionized by starlight photons (especially within 
H~II regions created by stars with only moderately high effective 
temperatures, so that low energy photons that can ionize H are plentiful, 
while more energetic ones that are needed to ionize S$^+$ are not).  One can 
therefore imagine that the appearance of readings of $[{\rm S}_{\rm gas}/{\rm 
H}] > 0$ shown in Fig.~\ref{sulfur_panels} for $\log N({\rm H})<19.5$ (open 
circles) is caused by the invisibility of the accompanying hydrogen.

We could certainly benefit from some future, far more comprehensive survey of 
sulfur abundances.  Not only do we find that the number of trustworthy 
determinations is small, but there may be a formidable selection bias that 
favors cases where $[{\rm S}_{\rm gas}/{\rm H}]$ is lower than normal because 
we had to avoid cases where the sulfur absorption lines were not saturated 
and yet still satisfy our requirement that $\log N({\rm H})>19.5$.

\section{Discussion}\label{discussion}
\subsection{The Composition of Interstellar Dust}\label{composition_dust}
\subsubsection{General Remarks}

The current study of depletions departs from most of the traditional ones 
that have emphasized the behavior of one or a few elements and how their 
abundance ratios change with such external factors as $\langle n(H)\rangle$, 
$f({\rm H}_2)$, and location.  Our new perspective ignores these factors 
(except for retrospective studies reported later in 
\S\ref{relationship_other}) and recognizes that depletions change markedly 
from one sight line to the next but that among different elements they are 
well correlated.  The objective now is to define how strongly each element 
participates in this collective behavior.  In doing so, we can concentrate on 
only two properties of element depletions: (1) An initial depletion value 
$[X_{\rm gas}/{\rm H}]_0$, which serves as a minimum strength for all sight 
lines and (2) an index $A_X$ that informs us about how rapidly individual 
elements deplete beyond this initial depletion compared to the others.

The existence of nonzero initial depletions $[X_{\rm gas}/{\rm H}]_0$ is an 
effect that has been known for some time (Jenkins, Savage, \& Spitzer 1986;
Joseph 1988; Fitzpatrick 1996), and it has commonly been identified with the
gas phase abundances in a warm, low density medium  (Spitzer 1985; Savage \&
Sembach 1996a).  We can surmise that these minimum depletions indicate the
composition of grains (through Eq.~\ref{x_dust}) that either have not had a
chance to grow in dense interstellar media or have been partially stripped of
certain elements by the recent passage of a shock at some point after such a
growth phase.  

The correctness of the values for $[X_{\rm gas}/{\rm H}]_0$ depend not only 
on the measured interstellar abundances, but also on the adopted values for 
the intrinsic abundance of the gas and dust put together, which in turn 
depend on the appropriateness and accuracy of the pre-solar abundance scale 
used in the current study.  Also, it is important to remember that the 
definition of the zero point for $F_*$ is probably strongly driven by our 
having adopted $N({\rm H})>10^{19.5}{\rm cm}^{-2}$ as an artificial boundary 
condition in our assessment of the least severe levels of depletion.  This 
requirement was imposed to minimize distortions in the abundance levels 
caused by the effects of ionization.  As we look back to the open circles 
shown in the upper panels in Figs.~\ref{panel_set_1} to \ref{panel_set_4}, it 
appears that for the most part this rule may have been too conservative.  
Many of these open circles seem to be close to the trend lines or 
extrapolations thereof to negative values of $F_*$.

In contrast to the stated values of $[X_{\rm gas}/{\rm H}]_0$, the 
derivations of the progressive amounts of depletion represented by the 
various slopes $A_X$ of elements $X$ multiplied by their respective actual 
interstellar abundances (see Eq.~\ref{differential_grain_comp}) have no 
sensitivity to the choices for the reference abundances; they depend only on 
the magnitude of the adopted scale factor for the index $F_*$, which is an 
index that we have created to define the collective depletion levels for the 
individual lines of sight.  These differential rates for both the growth and 
destruction of grains allow us to determine the composition of the material 
that has been added to the cores responsible for the initial depletions.  The 
elemental composition of this outer material (by number, not by mass) per 
unit change in the depletion index $F_*$, relative to the amount of hydrogen 
gas present, is represented by the lines and their error envelopes shown in 
the lower panels of Figs.~\ref{panel_set_1} to \ref{panel_set_4}.

Values of $A_X$ listed in column (3) of Table~\ref{elem_parameters} range 
between the extremes of zero (for $A_{\rm N}$) to a large value for $A_{\rm 
Ti}$, which is equal to 2.4 times the median value for all elements of 
$-0.85$.  The strong variability of $A_X$ among the different elements allows 
us to rule out two explanations for the probable root cause of depletion 
differences in different sight lines: (1) errors in $N$(H), which would cause 
all depletions in any given sight line to rise or fall in unison and (2) the 
dilution of elements in the gas within the Galactic disk caused by the infall 
and mixing of low metallicity material from the halo (Meyer et al. 1994;
Meyer, Jura, \& Cardelli 1998), which likewise would cause the logarithms of
all element abundances to decrease in lock step with each other.

In \S\ref{stellar_lalpha}, we discussed the corrections to $N({\rm H~I})_{\rm 
obs.}$ to account for increases in the strengths of the L$\alpha$ absorption 
profiles caused by the stellar absorption features in stars of spectral type 
B1 and cooler.  The median rank in $F_*$ for these stars was 137, while for 
the hotter stars that did not need such a correction the median rank was 103.  
As a consequence, for elements that exhibited very mild progressive 
depletions (i.e., small absolute values of $A_X$), these corrections operated 
in a fashion to increase very slightly the numerical values of $A_X$ and thus 
decrease the apparent contributions to grain growth.  These systematic shifts 
were always smaller than the random errors associated with the fitting 
processes to derive $A_X$, as stated in Table~\ref{elem_parameters}, and, 
except for Kr, were not of much importance (see the footnote in 
\S\ref{krypton}).

In the following subsections, we discuss some noteworthy points about the 
depletions of a few specific elements.

\subsubsection{Carbon}\label{carbon}

Carbon is a major constituent of grains, yet our knowledge of the actual 
differential consumption of C by grains still remains rudimentary, as is 
indicated by the large shaded region in Fig.~\ref{panel_set_1}.  The paucity 
of data for C in our survey toward distant stars can be traced to the need to 
observe a weak intersystem transition at 2325$\,$\AA\ (Sofia et al. 2004) at
a high S/N, since the allowed transitions at 1036 and 1335$\,$\AA\ are always
strongly saturated.  The strong transitions reveal reliable C~II column
densities for only a modest fraction of stars that are much closer to the
Sun, i.e., those that have very low values of $N$(H) (Wood \& Linsky 1997;
Lehner et al. 2003; Redfield \& Linsky 2004a).  There has been some concern
expressed in the literature that not enough carbon is depleted to explain the
optical properties of dust (Kim \& Martin 1996; Mathis 1996; Dwek 1997).  If
we disregard the present large uncertainty in $A_{\rm C}$ and focus our
attention on $B_{\rm C}$ to obtain a nominal carbon depletion (at $F_*\approx
0.8$), we infer that the amount of C that is available for interstellar dust
(and small molecules) is only $10^{-4}$ times the amount of H by number. 
About twice this amount is needed to explain the dust extinction (Draine
2003a).  However, a recent, preliminary study by Sofia \& Parvathi (2009)
suggests that the strength of the intersystem line may be about twice as
strong as that considered previously, which would lower the interstellar
abundances by the same factor (and raise the dust carbon content by about
90\%).
 
\subsubsection{Nitrogen}\label{nitrogen}

The abundance of nitrogen is $-0.109\pm 0.111$~dex below its reference solar 
system abundance, regardless of the value of $F_*$, i.e., $A_N=0.00\pm 0.08$.  
Knauth, et al. (2003) claimed to have detected 
progressively stronger depletions of nitrogen as $N$(H) increased, but this 
effect may be indirectly related to an apparent enhancement of ${\rm N}_{\rm 
gas}/{\rm O}_{\rm gas}$ within 500~pc of the Sun discussed by Knauth et al. 
 (2006). The latter result highlights the possible 
influence of regional differences in relative proportions of outputs from 
different nucleosynthesis sources (Type~II SNe vs. AGB stellar winds) coupled 
with incomplete mixing in the ISM.  However, if an effect such as this one 
were influential for our findings for N, it would probably have resulted in a 
poor outcome for the $\chi^2$ value of the fit, which appears not to be the 
case (see Table~\ref{elem_parameters}).  The fact that $A_N\approx 0$ 
suggests that the negative value of $B_N$ might arise from the adopted value 
for $({\rm N/H})_\odot$ being too high, although the apparent deviation of 
$B_N$ from zero is only at the $1\sigma$ level of significance.

Gail \& Sedlmayr (1986) have pointed out that the 
condensation of N into any sort of solid compound could be inhibited by the 
production of N$_2$, which is very stable.  The saturated bond of this 
molecule results in a high activation energy barrier for gas phase reactions 
to form other molecules.  While this may be an important theoretical 
consideration, the fact remains that the abundance of N$_2$ in the diffuse 
ISM is small (Knauth et al. 2004, 2006).

\subsubsection{Oxygen}\label{oxygen}

Cartledge et al. (2004) found that the abundance of 
oxygen exhibited a weak, but convincing downward trend of its abundance with 
respect to $\langle n({\rm H})\rangle$.  Thus, the fact that our value of 
$A_{\rm O}=-0.225\pm  0.053$ differs significantly from zero is not 
unexpected.  However, what comes as a surprise is the finding that the 
extraction of oxygen from the gas phase seems, for the larger values of 
$F_*$, far out of proportion to the consumption of other, less abundant 
elements that can be thought to form solid compounds with oxygen.  For 
instance, from Eq.~\ref{differential_grain_comp} we find that when $F_*=0$, 
$d({\rm O}_{\rm dust}/{\rm H})/dF_*$ is 1.6 times the sum of the solid phase 
accumulation rates (measured the same way) of Mg, Si and Fe, i.e., $d({\rm 
Mg+Si+Fe}_{\rm dust}/{\rm H})/dF_*$, and a factor of 16 greater when $F_*$ 
reaches 1.0.  Yet the conventional view is that oxygen is mostly incorporated 
into such refractory compounds as metallic oxides and amorphous silicates.  
However, even the most oxygen-rich of these compounds, magnesium silicate 
(enstatite) Mg$\,$Si$\,$O$_3$, has only 3/2 times as much O as the other 
elements.  Considering the uncertainties in the O consumption at $F_*=0$, the 
3/2 ratio, or even a somewhat lesser amount, could be satisfied.  However, 
the divergence between the O consumption and those of the other elements 
makes this equality rapidly vanish when $F_*$ becomes somewhat larger than 0.  
At even the lower edge of $-2\sigma$ error zone for the O consumption at 
$F_*=1$ shown in Fig.~\ref{panel_set_1}, we find that O atoms are taken out 
of the gas phase at a rate that is 6 times that of Mg + Si + Fe.

If we now switch our attention to absolute depletions instead of differential 
ones (this now relies on the premise that the pre-solar abundances are 
correct), our calculation of the value of $({\rm O}_{\rm dust}/{\rm H})$ at 
$F_*=1$ is $2.41^{+0.74}_{-0.66}\times 10^{-4}$; the nominal value here is 
larger than a limit of $1.8\times 10^{-4}$ that is established by the {\it 
total\/} availability of other elements that can be incorporated into either 
the silicates, metallic oxides, or some combination of the two 
 (Cardelli et al. 1996).  However, the negative ($1\sigma$) error 
limit is consistent with this number.  The uncertainty calculated for $({\rm 
O}_{\rm dust}/{\rm H})$ includes the presumption that $({\rm O/H})_\odot$ has 
a possible error as large as 0.05~dex.

We are drawn to the conclusion that O must be locked up in either some carbon 
or hydrogen compound (or as O$_2$), given that these are the only reactive 
elements with a sufficiently large cosmic abundance.  (While it is abundant, 
the consumption of N during grain growth is nowhere near as much as O.)  An 
initially attractive prospect was that O is incorporated in the form of 
amorphous H$_2$O ice on the grain surfaces (Ioppolo et al. 2008).  
Slightly more than 5\% of the available oxygen atoms are found in the form of 
water in the material (with large extinction values) toward young stellar 
objects (van Dishoeck 1998), but various surveys indicate that the 
strength of the $3.05\mu\,{\rm m}$ ice band shows a linear trend that 
extrapolates to zero when $A_V$ decreases to values of around 2.6 to 5 
(Whittet et al. 1988; Eiroa \& Hodapp 1989; Smith, Sellgren, \&
Brooke 1993).  The star Cygnus OB2 No. 12\footnote{Sometimes called VI Cyg
No.~12.} has an extinction $A_V=10.2\pm 0.3$ but shows no detectable ice band
in its spectrum ($\tau<0.02$) (Gillett et al. 1975; Whittet et al. 1997).  It
may be possible that long term irradiation by cosmic ray particles and UV
radiation modifies this ice layer in a way that inhibits the appearance of
the infrared absorption feature (Greenberg 1982; Palumbo 2006).  Other simple
oxygen-bearing molecules in solid form, such as CO, CO$_2$ and O$_2$
generally have smaller abundances than that of H$_2$O, but they are not
entirely negligible (van Dishoeck 2004).

An entirely separate method of determining what fraction of the oxygen is 
locked up in compounds is to examine the structure of absorptions in the 
vicinity of the K absorption edge at 23$\,$\AA\ in the spectra of x-ray 
binaries recorded by the grating spectrometers aboard {\it Chandra\/} and 
{\it XMM-Newton}.  Molecular bonds shift the energies of bound-bound and 
bound-free transitions and create such complex structures, but unfortunately 
ionization of the atoms can play a similar role, which makes the analysis 
ambiguous (Costantini, Freyberg, \& Predehl 2005).  As a result, different 
investigators have arrived at differing interpretations of the observations.  
For instance, Paerels et al. (2001), Schulz et al 
 (2002), and Takei et al (2002) 
have viewed their results on the O-edge structures in terms of specific 
compounds in the ISM, but these conclusions were later criticized by Juett, 
Schulz \& Chakrabarti (2004), who interpret the 
discrete features seen in a number of x-ray sources as arising simply from 
O~I, O~II, and O~III.  If their identification of absorption by compounds is 
correct, Takei et al. (2002) found that the amount of O 
in free atoms toward Cyg-X2 is $(8.6\pm 2.8)\times 10^{17}{\rm cm}^{-2}$, 
while O in bound form corresponds to $(5.2\pm 2.2)\times 10^{17}{\rm 
cm}^{-2}$, which is consistent with a depletion of atomic O of $-0.205\,$dex.  
This amount of depletion is what we would expect for $F_*=0.86$.  However, 
Cunningham, McCray \& Snow (2004) found that the 
total x-ray absorption by oxygen in all forms toward X~Per (a sight line 
included here with $F_*=0.90\pm 0.06$; see Table~\ref{O}) is consistent with 
just the gas-phase measurement, leaving no room for an appreciable amount of 
additional O in solid form.

A different approach is to use the x-ray absorption results to compare the 
total abundance of O with those of other elements, on the premise that 
perhaps large amounts of oxygen are locked within dust grains that have 
diameters of order or greater than $1\,\mu$m. Grains this large have been 
detected in the local ISM by the dust sensors aboard the {\it Galileo\/} and 
{\it Ulysses\/} spacecrafts
(Frisch et al. 1999; Landgraf et al. 2000; Kr\"uger et al. 2006), 
and these measurements indicate that the large grains contribute a 
substantial portion of the total mass budget of material in solid form.  The 
largest grains become optically thick to x-rays at energies near the K 
absorption edge.  Takei et al (2002) found that 
$\log({\rm Ne/O_{gas~+~dust}})=-0.84\pm 0.18$ is consistent with the solar 
value of $-0.81\pm 0.11\,$dex (Lodders 2003), indicating that all of 
the O absorption was evident in the observed edge absorption.  In contrast, 
higher values that appear to indicate that some of the O is hidden have been 
derived in other investigations: $\log({\rm Ne/O})=-0.52\pm 0.21$ (Yao \&
Wang 2006), $-0.66\pm 0.08$ (Yao et al. 2009), and an outcome even as high as
0.1 (Paerels et al. 2001). Moreover, Ueda et al.  (2005) found that Si/O and
Mg/O were about 0.6$\,$dex higher than their respective solar values (as
adopted here, not as expressed in their article).  The fact that the latter
result is at variance with the UV absorption line data is consistent with the
idea that while Si and Mg both reside mostly within small grains that
individually have small optical depths for x-rays, significant amounts of O
could be incorporated into grains or complexes thereof that are large enough
to be fully opaque in x-rays.

Large dust grains having thick mantles of water ice should be extremely hard 
to detect by most astronomical methods.  They contribute little to the 
extinction at visible and IR wavelengths, and their effectiveness in creating 
a distinctive 3.07$\,\mu$m ice band absorption feature would be limited if 
the grain diameters exceeded $1\,\mu$m (B.~T.~Draine, private communication).  
About the only way to detect such large particles, if they are present, might 
be through very small angle scattering of x-rays from point sources
(Smith \& Dwek 1998; Witt, Smith, \& Dwek 2001) or possibly even very 
faint scattering at larger angles at visible wavelengths for particles that 
are more than a few $\mu$m in diameter (Socrates \& Draine 2008).  
However, various interpretations of the x-ray observations so far seem to 
indicate that the small-angle x-ray scattering data are consistent with dust 
grain populations composed of refractory compounds over a distribution of 
sizes smaller than 0.25$\,\mu$m in diameter
(Draine \& Tan 2003; Dwek et al. 2004; Costantini, Freyberg, \& Predehl
2005; Xiang, Zhang, \& Yao 2005; Smith et al. 2006; Ling et al. 2009;
Smith 2008). 

\subsubsection{Phosphorous}\label{phosphorus}

Phosphorous depletes more rapidly than oxygen, as is clear from the 
significant difference between the $A_{\rm O}$ and $A_{\rm P}$ values and 
their errors listed in Table~\ref{elem_parameters}.  This finding is in 
conflict with the findings of Lebouteiller, Kuassivi \& Ferlet 
 (2005), who claim that $N$(P~II)/$N$(O~I) always 
appears to be consistent with the solar abundance ratio over a large range of 
O~I column densities.

\subsubsection{Chlorine}\label{chlorine}

The abundance trend of Cl is not as regular as for the other elements, as is 
evident from the scatter of points shown in Fig.~\ref{panel_set_2} and the 
larger than usual errors for the parameters of Cl in 
Table~\ref{elem_parameters}.  It should also be clear from the small dot 
sizes for Cl in Fig.~\ref{panel_set_2} that the measurement errors are larger 
than usual.  Some of the scatter in the Cl measurements may be due to the 
large fractions of Cl being in neutral form, compared to the singly ionized 
form.  Jura (1974) has shown that chlorine ions can react with 
H$_2$ to produce HCl$^+$~+~H and eventually, through a chain of reactions 
that follow, revert to an amount of neutral chlorine that can even surpass 
the remaining concentration of chlorine ions.  Indeed, there are a few lines 
of sight in the survey of JSS86 that indicate that $N({\rm Cl~I})>N({\rm 
Cl~II})$.  If these cases are predominantly at the largest values of $F_*$ 
(which seems likely, since there will be a higher relative concentration of 
H$_2$), the slope of the best-fit line may be too steep (i.e., $A_X$ is at 
too large a negative value), and this may be the dominant cause for our 
having found that $[{\rm Cl}_{\rm gas}/{\rm H}]_0$ is positive.

\subsubsection{Krypton}\label{krypton}

One apparently remarkable outcome is that Kr seems to show not only an offset 
below its solar system value, as indicated by the fact that $B_{\rm 
Kr}=-0.332\pm 0.083$, but that also there is an indication, not an entirely 
conclusive one, that there is some progressive depletion as $F_*$ increases 
($A_{\rm Kr}=-0.166\pm 0.103$).  However it is clear that the nonzero value 
of this parameter is significant only at the 1.6$\sigma$ level of 
significance.\footnote{In an earlier phase of this investigation, corrections 
for stellar L$\alpha$ absorption had not yet been implemented.  The 
significance of the negative value for $A_{\rm Kr}$ was higher at that time.  
After the correction was added, a number of cases that supported stronger 
depletions of Kr at large $F_*$ vanished because the calculated values of 
$N({\rm H~I})_{\rm stellar}$ were about equal to the uncorrected $N({\rm 
H~I})_{\rm obs.}$.}   A Pearson correlation coefficient of the 33 values 
$[{\rm Kr}/{\rm H}]_{\rm obs.}$ vs. their respective $F_*$, which does not 
take into account the measurement errors, is only $-0.225$.  This value 
differs from zero correlation only at the 89.5\% confidence level (we use a 
one-tail test here, since we reject the possibility that $A_{\rm Kr}>0$), 
which again supports the notion that this is a weak result.  (Note that the 
determinations of $F_*$ included measurements of Kr, but the weight factors 
$W_X$ for Kr given in Eq.~\ref{W_X} are extremely small compared those of 
other elements because the depletions have relatively large errors.  Thus, 
there is a negligible influence of the Kr measurements on $F_*$, which in 
principle could have further weakened the conclusion on the significance of 
the correlation.)

The fact that the abundances of Kr seem to ``pay attention to'' the 
abundances of other elements with large values of $A_X$ may signify that Kr 
is truly depleting and not exhibiting chance deviations caused either by real 
abundance variations or errors in measurement (either for Kr or H).  This 
phenomenon was not evident in a conventional comparison of $({\rm Kr}_{\rm 
gas}/{\rm H})$ vs. $\langle n({\rm H})\rangle$ in the most recent compilation 
of Kr abundances (Cartledge et al. 2008), and it indicates that this 
noble gas might either be attached to grains via physisorption or could 
possibly be trapped in a water clathrate (recall from the discussion in 
\S\ref{oxygen} that there might possibly be enough H$_2$O ice on large grains 
in the ISM to explain the depletion of O).

\subsubsection{Trends with Condensation Temperatures}\label{cond_temp}

In trying to understand why different elements show different depletion 
strengths, a conventional approach is to compare them with their respective 
condensation temperatures.  The condensation temperatures indicate the points 
at which the elements should show an appreciable deficit in the gaseous form 
as the result of forming compounds in a chemical equilibrium.  However, apart 
from formation processes in stellar atmospheres and circumstellar envelopes, 
the formation and destruction of compounds within dust grains is not an 
equilibrium process.  Even so, the condensation temperature may still be used 
as an approximate surrogate for the relative affinity of an element to form a 
solid compound and be resistant to destruction by shocks.  As proposed by 
Dwek \& Scalo (1980), the relative ease for the 
destruction of different compounds in grains is related to their respective 
sublimation energies, which are reflected by their formation temperatures 
through the Clausius-Clapeyron relation.

\begin{figure}[ht!]
\plottwo{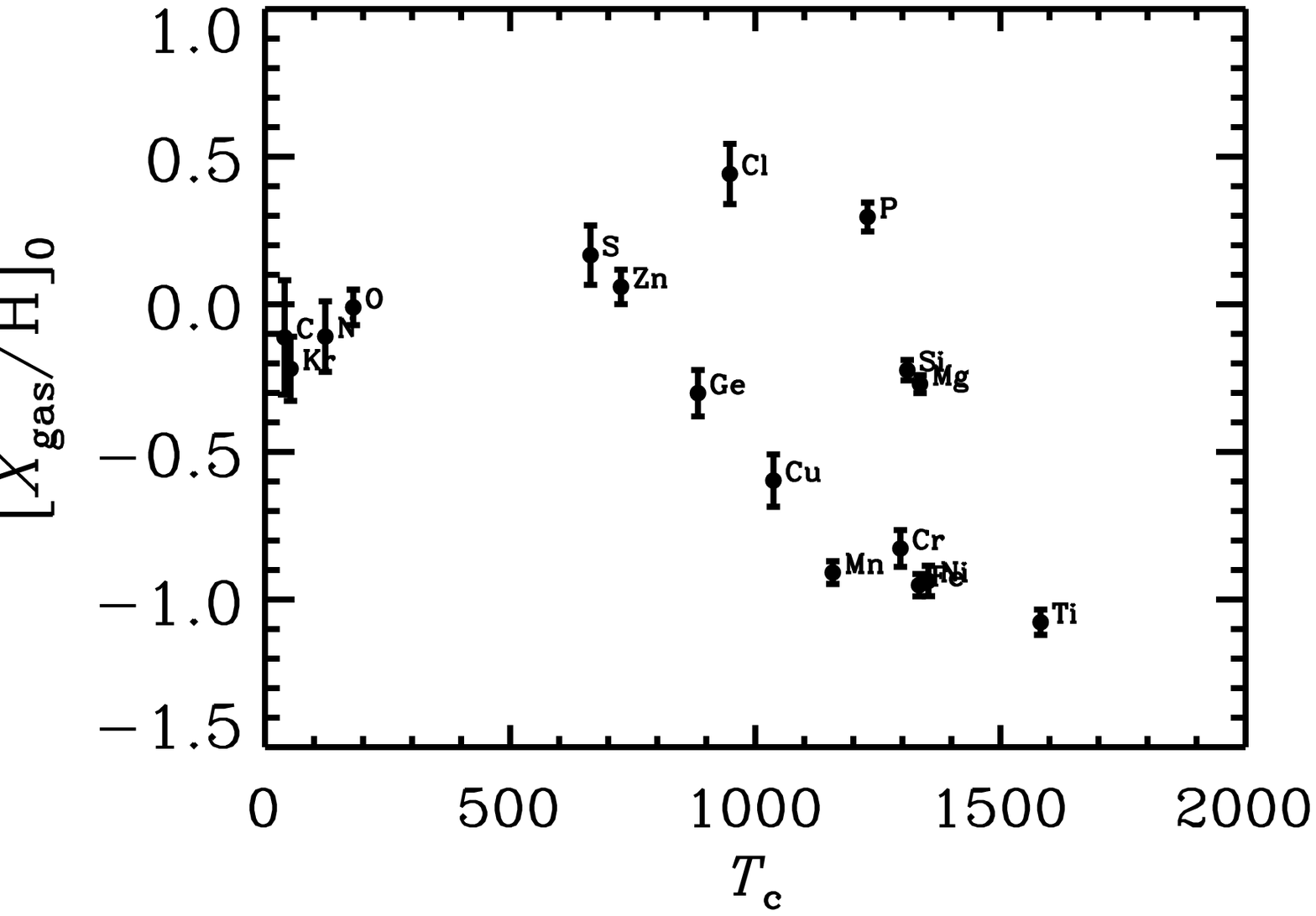}{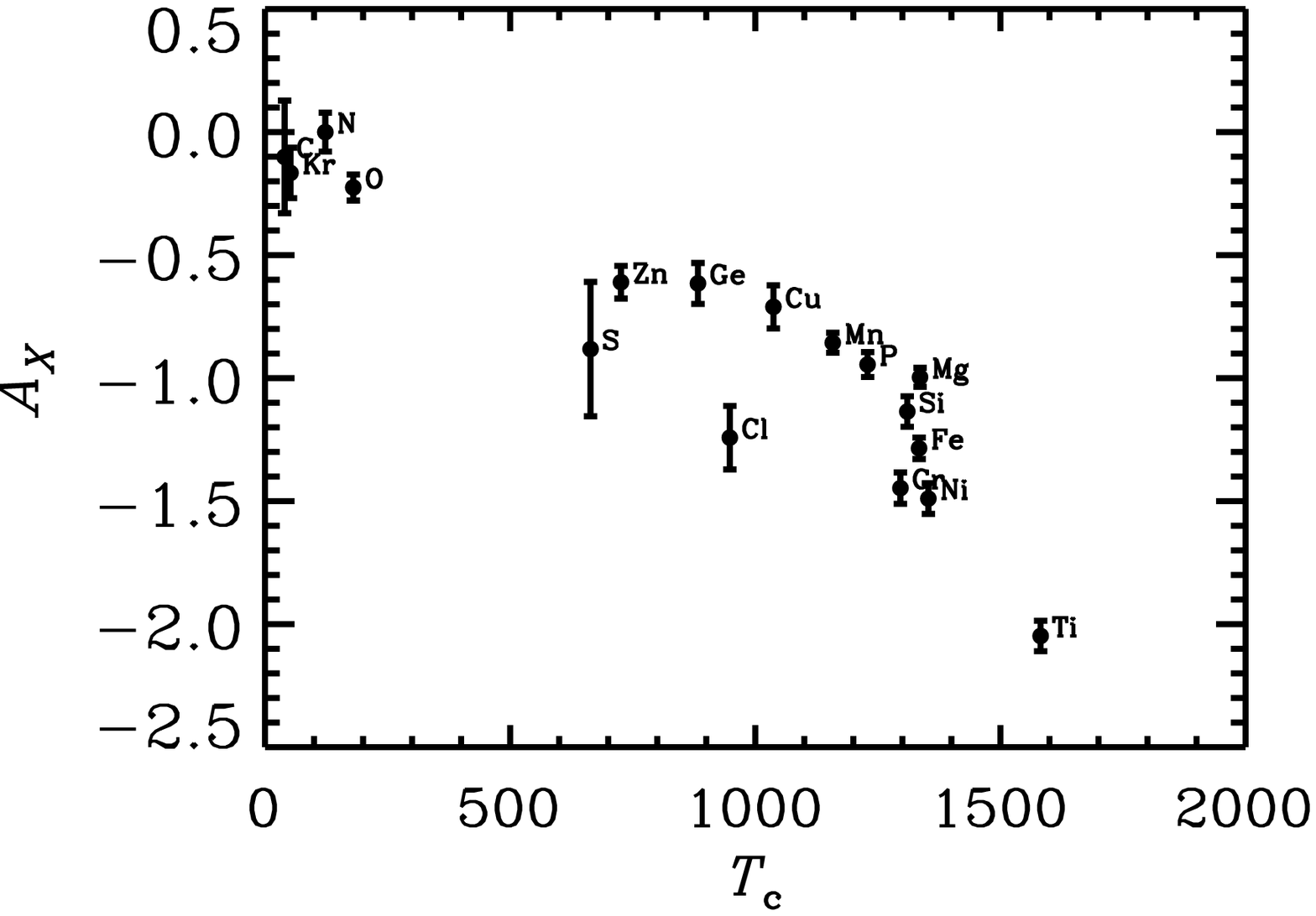}
\caption{{\it Left-hand panel:\/} The trend of initial depletions as a 
function of the respective element condensation temperatures $T_c$ listed by 
Lodders (2003).  {\it Right-hand panel:\/} Values of $A_X$ 
vs. $T_c$.\label{grain_growth}}
\end{figure}

The left-hand panel of Fig.~\ref{grain_growth} shows the magnitudes of the 
initial depletions $[X_{\rm gas}/{\rm H}]_0$ as a function of the 
condensation temperatures $T_c$ computed by Lodders (2003).  
Her values of $T_c$ apply to a 50\% decrease in the gas phase abundance at a 
pressure $10^{-4}\,$bar with a pre-solar distribution of abundances.  Values 
of $[X_{\rm gas}/{\rm H}]_0$ probably give the closest representation of the 
composition of grains that emerge from the atmospheres of late-type stars, 
supernovae, and circumstellar shells or disks. The relationship between the 
elements seen in this figure is similar to that for the highly depleted 
component toward $\zeta$~Oph: most elements show a trend of increasing 
depletion toward higher $T_c$, with the exception of P, Cl, Mg and Si, which 
seem to lie above the trend established by the other elements (Savage \&
Sembach 1996a).

\placefigure{grain_growth}

If we now examine the differential depletions, the picture is a bit 
different.  Recall that from Eq.~\ref{differential_grain_comp} that the 
differential depletion scales in proportion to $A_X(X_{\rm gas}/{\rm 
H})_{F_*}$.  If we normalize this rate to the concentration of atoms that are 
present, i.e., $(X_{\rm gas}/{\rm H})_{F_*}$, we get simply $A_X$.  In 
effect, $A_X$ represents a rate coefficient that applies to the 
quasiequilibrium state between the punctuated creation and destruction events 
of dust compounds (this is a loose concept because the condensation and 
destruction processes are physically different).  The right-hand panel of 
Fig.~\ref{grain_growth} shows the values of $A_X$ as a function of $T_c$.  
The placement of the points in this diagram seems more regular than what was 
seen for $[X_{\rm gas}/{\rm H}]_0$ (except for Cl, which seems to have 
flipped its position relative to the other elements -- but recall the remarks 
about Cl made in \S\ref{chlorine}).

\subsection{The Relationship of \boldmath $F_*$ to Other 
Variables}\label{relationship_other}

\begin{figure}[ht!]
\plottwo{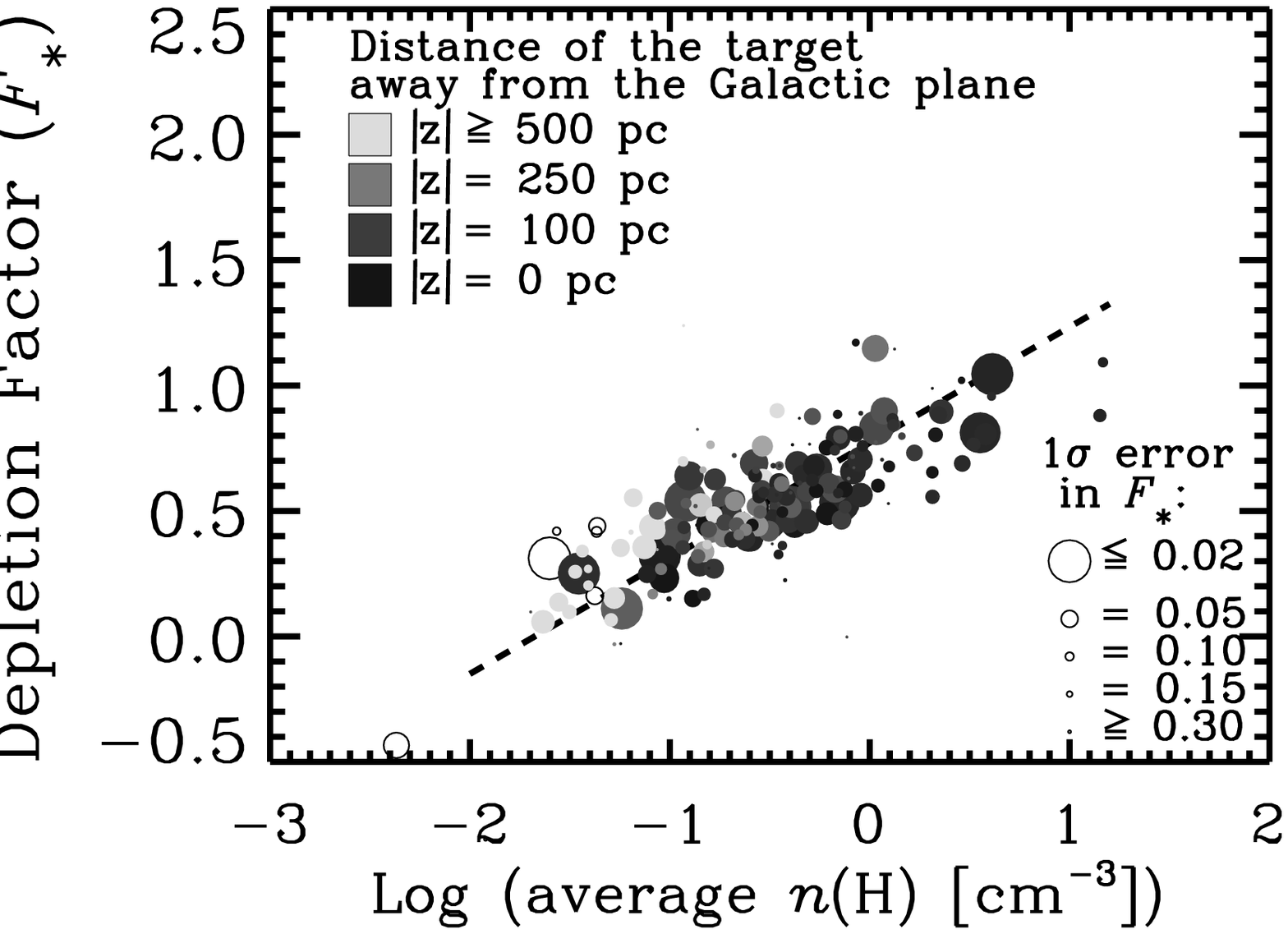}{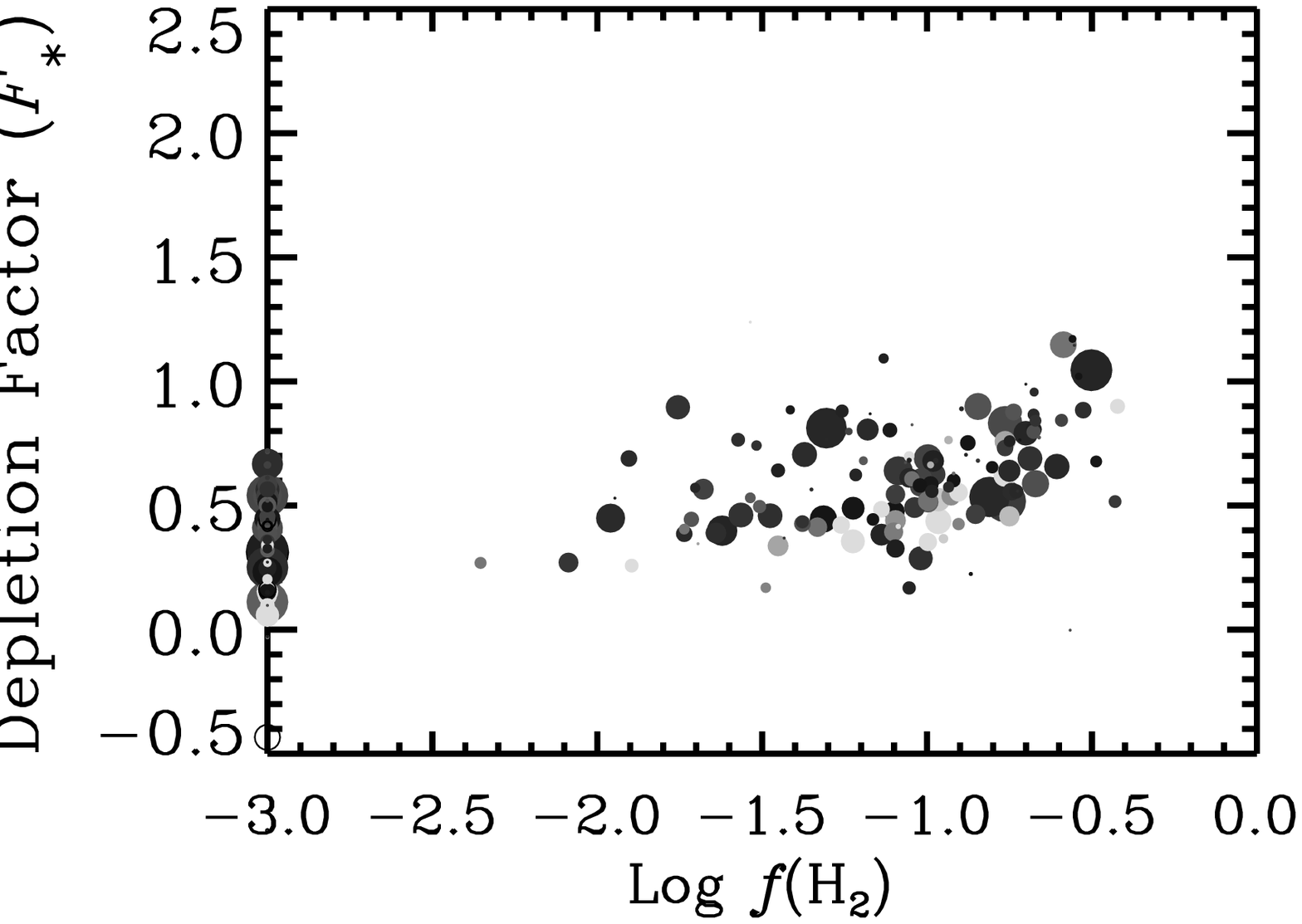}
\caption{{\it Left-hand panel:\/} The trend of $F_*$ as a function of the 
logarithm of the average density along the respective sight lines $\langle 
n({\rm H})\rangle$.  As indicated by the legends, the sizes of the circles 
indicate the errors in $F_*$, and their shades of gray indicate the distances 
of the stars from the Galactic plane. Open circles indicate cases where $\log 
N({\rm H})<19.5$. The least-squares best fit to the trend is indicated by the 
dashed line. {\it Right-hand panel:\/} The trend of $F_*$ against the 
logarithm of the fraction of hydrogen in molecular form $f({\rm H}_2)\equiv 
2N({\rm H}_2)/[N({\rm H~I})+2N({\rm H}_2)]$.  The sizes and gray levels of 
the points are the same as in the left-hand panel.  Cases where $\log f({\rm 
H}_2)< -3.0$ are all bunched together on the $y$ axis of the 
plot.\label{abundf1}}
\end{figure}

We are in a position to repeat some of the comparisons mentioned in 
\S\ref{general_findings} using our generalized depletion index $F_*$ for each 
sightline instead of just concentrating on the depletions of a single element 
(or several elements, but in an individual fashion), as has been done in the 
past. Figure~\ref{abundf1} shows the trends of $F_*$ against two popular 
extrinsic variables, the average density along the line of sight $\langle 
n({\rm H})\rangle$ and the fraction of hydrogen atoms in molecular form 
$2N({\rm H}_2)/[N({\rm H~I})+2N({\rm H}_2)]$.  This figure shows that the 
relationship of $F_*$ with the former seems more well defined than with the 
latter.  Snow, Rachford \& Figoski (2002) and Cartledge 
et al. (2006) arrived at a similar conclusion on 
the basis of their studies of the abundances of interstellar Fe, Ge, and Mg.  
As one would expect, target stars at some distance from the Galactic plane 
have lower than usual values of $\langle n(H)\rangle$, but the color coding 
of the symbols indicates that their depletion indices do not seem to show any 
distinct differences from other sight lines with the same average density.  
In contrast, Sembach \& Savage (1996) found that 
gas identified with material in the lower halo of the Galaxy had smaller 
depletions than gas at comparable density in the disk [see also Savage \& 
Sembach (1996a)].  The appearance of such a 
difference probably results from the fact that they treated separately the 
abundances in velocity components whose kinematics were consistent with being 
at large distances (i.e., far from the plane) instead of grouping all of the 
gas together, as in this study.  An illustration of the importance of this 
distinction was shown earlier in Fig.~\ref{nh_fstar_examples} for the star 
HD~116852.  Gas at progressively more negative velocities exhibited less 
depletion, and this material is farther from the Galactic plane 
 (Sembach \& Savage 1996).

\placefigure{abundf1}

A linear least-squares best fit to the trend shown in the left-hand panel of 
Fig.~\ref{abundf1} follows the formula $F_*=0.772+0.461\log\langle n({\rm 
H})\rangle$ indicated by the dashed line.  This fit was evaluated in a way 
that minimized errors both in $F_*$ and $\log\langle n({\rm H})\rangle$.  For 
the former, we take the actual estimates for the uncertainties in $F_*$ [see 
column (6) of Table~\ref{nh_fstar}], and for the latter, we assume that the 
errors are dominated by the errors in the distances, which was assumed to be 
a uniform value of 0.2~dex (mostly from a 1~mag error in $M_V$; see 
Appendix~B3 of Bowen et al. (2008) for details on the 
probable errors in distances).  In evaluating the fit, the protocol of 
accepting only sight lines where $N({\rm H})>10^{19.5}{\rm cm}^{-2}$ was 
followed (see rule nr.~3 in \S\ref{rejection}).  Such cases are indicated by 
the filled circles in the figure.  The vertical dispersion of points on 
either side of the best fit line is 0.18.  The $\chi^2$ value for the fit is 
285 for 175 degrees of freedom, indicating that the natural dispersion is 
somewhat larger than that created by our errors in $F_*$ and $\log\langle 
n({\rm H})\rangle$.

For the fit of $F_*$ vs. $\log\langle n({\rm H})\rangle$, there seems to be 
no departure from a simple linear trend, which seems contrary to the 
assertion by Cartledge et al. (2004, 2006) that a more complex association
between the two variables exists in the form of two plateaus with a
transition between them at an intermediate value of $\log\langle n({\rm
H})\rangle$ (they used a Boltzmann function to express this behavior).  In
their study of several different elements, the value of $\log\langle n({\rm
H})\rangle$ for the transition region seemed to change somewhat from one
element to the next, so the structure of this functional relationship may be
lost when the generalized depletions based on many different elements are
evaluated.  Another alternative is that the extra parameters needed to define
the Boltzmann function are not fully justified by the data, given the
uncertainties present.  The same remarks apply to the nonlinear forms shown
by Jenkins, Savage \& Spitzer (1986), who showed functions that 
fitted within the theoretical interpretation published earlier by Spitzer 
 (1985).

\begin{figure}
\epsscale{1.}
\plotone{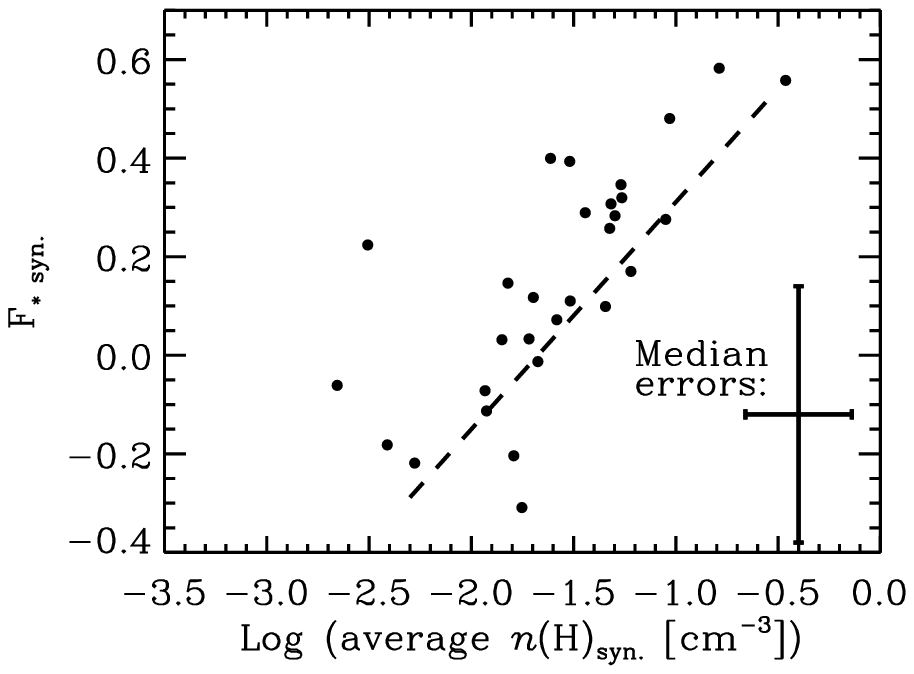}
\caption{The relationship between $F_{*\,{\rm syn.}}$ and the average sight 
line density (determined from $N({\rm H})_{\rm syn.}$) for the white dwarf 
stars in the Local Bubble, whose locations and distances are indicated in 
Fig.~\protect\ref{lism}.  Most of the estimated errors for the points in this 
figure have a size about equal to the bars shown in the lower right-hand 
portion of the plot.  The diagonal dashed line indicates the location of the 
fit to the points shown in left-hand panel of 
Fig.~\protect\ref{abundf1}.\label{lism_correl}}
\end{figure}
\begin{figure}
\plotone{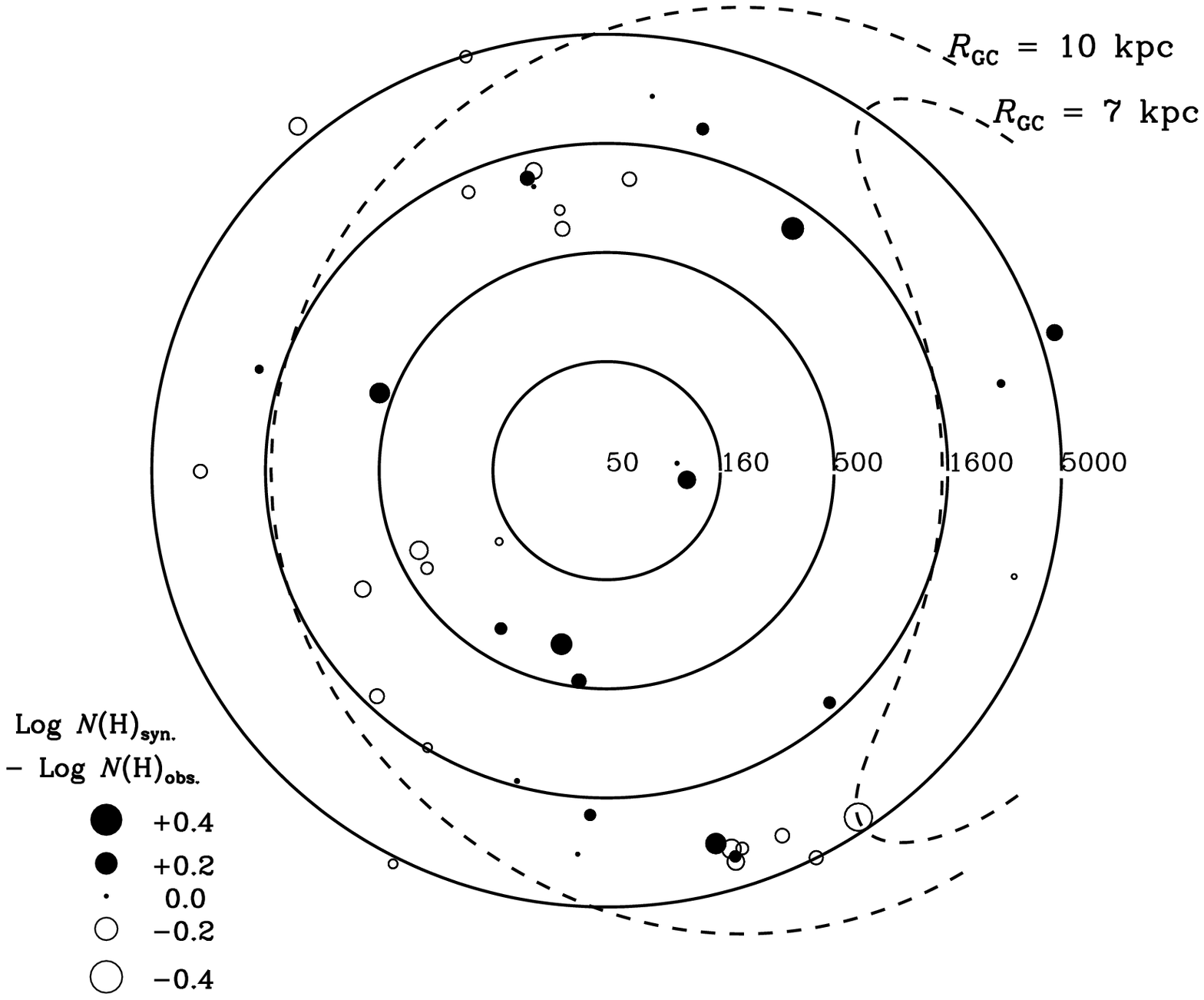}
\caption{Differences in overall metallicity (gas + dust) with respect to 
locations relative to the Sun, as indicated by the quantity $\log N({\rm 
H})_{\rm syn.} - \log N({\rm H})_{\rm obs.}$.  The only points that are shown 
are those with $N({\rm H})_{\rm obs.}>10^{19.5}{\rm cm}^{-2}$ and where the 
ratio of the two values of $N$(H) is known to an accuracy of $<\,0.1\,$dex.  
Solid circles indicate gas that is metal rich, and open ones indicate the 
opposite condition.  The Galactic center is to the right, and this polar 
representation has the distance from the Sun represented in a logarithmic 
fashion.  The dashed lines indicate locations that are 7 and 10$\,$kpc from 
the Galactic center.\label{metallicity}}
\end{figure}

We can repeat the comparison of $F_*$ against the average of $n$(H) along 
sight lines for the white dwarf star sight lines that were analyzed in 
\S\ref{WD}.  Figure~\ref{lism_correl} shows this comparison and how it 
relates to the trend line shown in the left-hand panel of Fig.~\ref{abundf1} 
for the early-type stars at greater distances.  The correlation between 
$F_{*\,{\rm syn.}}$ and $n({\rm H})_{\rm syn.}$ is clear, and most of the 
points lie above the trend that was found for the more distant stars.  Again, 
we express the caution that this difference could arise from the effects of 
photoionization.

\placefigure{lism_correl}

Finally, a definition of our $F_*$ parameter in the context of the summary of 
abundances given by Savage \& Sembach (1996a) is 
presented in Appendix~\ref{ss96_comparison}.  This comparison is presented to 
allow one to place later works that made use of these generalized abundance 
patterns into the framework of the present study.

\subsection{Regional Differences in Total Abundances}\label{regional}

In \S\ref{rejection}, we expressed reservations about using data from stars 
located at Galactocentric distances much different from that of the Sun.  It 
is of interest to see if this concern was warranted, now that we have 
determined the coefficients $F_*$ for all sight lines, but used only stars 
inside the range $7<R_{\rm GC}<10\,$kpc for determining the element 
parameters $A_X$ and $B_X$.  In essence, we wish to see if the interstellar 
line data show any hint of the gradient of overall abundances that have been 
detected for stars, planetary nebulae and H~II regions [see the references 
cited within rule nr.~7 in \S\ref{rejection} and also Table~1 of Rolleston et 
al. (2000)].

It also would be interesting to see if there are metallicity changes that 
mimic in any way the more specific abundance deviations reported in the 
literature for certain elements, such as an apparent enhancement of ${\rm 
N}_{\rm gas}/{\rm O}_{\rm gas}$ within 500~pc of the Sun reported by Knauth 
et al. (2006) or the increase in ${\rm Kr}_{\rm 
gas}/{\rm H}$ within an annulus $600 < r < 2500$~pc discussed by Cartledge et 
al. (2008).

A method for determining the overall metallicity of the gas that should be 
independent of the amount of depletion in a line of sight is to compare the 
synthetic determinations of hydrogen $N({\rm H})_{\rm syn.}$, calculated on 
the basis of only the values of $N(X)$ in conjunction with their respective 
$A_X$ and $B_X$ in \S\ref{calc}, with the observed counterparts $N({\rm 
H})_{\rm obs.}$.  The ratio of the two should indicate the relative excess or 
deficiency of the total heavy element abundances (gas + dust), compared to 
their pre-solar values.  That is, if $N({\rm H})_{\rm syn.} < N({\rm H})_{\rm 
obs.}$, we would conclude that the gas is metal poor.

Figure~\ref{metallicity} shows a depiction of excesses and deficiencies of 
$\log N({\rm H})_{\rm syn.} - \log N({\rm H})_{\rm obs.}$ in a polar 
coordinate system that is logarithmic in radius and centered on the Sun.  The 
points shown in this diagram are restricted to cases where both $N({\rm 
H})_{\rm syn.}$ and $ N({\rm H})_{\rm obs.}$ are known, the error of their 
difference is less than 0.1~dex, and $N({\rm H})_{\rm obs.}>10^{19.5}{\rm 
cm}^{-2}$.  The locations of the solid and open circles appear to be random, 
indicating that no coherent changes in metallicity appear to be detected in 
different, readily identifiable locations.  Of course, real differences may 
be masked by the averaging effect over the sight line that extends from the 
location of the Sun to that of the star.  Unfortunately, there are too few 
sight lines that extend outside the zone $7<R_{\rm GC}<10\,$kpc and that 
satisfy the restrictions given above to offer a good test of the metallicity 
gradient with galactocentric distance.

\placefigure{metallicity}

\subsection{Applications for Abundances in Quasar and GRB Absorption Line 
Systems}\label{QSOALS}

\subsubsection{Introductory Remarks}\label{intro_rem}

There have been many contemporary studies of element abundances in the Damped 
Lyman Alpha (DLA) and sub-DLA gas systems in front of quasars (Lu et al.
1996; Prochaska \& Wolfe 2002; Pettini 2003; Prochaska, Howk, \& Wolfe 2003;
Dessauges-Zavadsky et al. 2004; Wolfe, Gawiser, \& Prochaska 2005), as well
as gases within or in front of the host galaxies of gamma-ray burst (GRB)
sources that had bright afterglows in the visible part of the spectrum
(Prochaska et al. 2007; Calura et al. 2009).  A persistent problem with
attempts to derive the intrinsic abundances of the elements has been the need
to correct for the effect of depletion on the column density results.  Early
work on the abundances of these systems concentrated on the elements zinc and
chromium (Meyer \& Roth 1990; Pettini, Boksenberg, \& Hunstead 1990), whose
lines were easily accessible in the visible for low-$z$ systems (Pettini
2003).  Since it was known that the depletion of zinc is usually small and
chromium large, approximate depletion corrections could be made by comparing
the abundances of these elements to their solar abundance ratios.

Advances with larger, more sensitive telescopes and spectrographs led to 
studies of a wider range of elements, since transitions with short 
wavelengths in the rest frame could be viewed in systems at large redshifts 
in front of faint quasars.  While opening up more elements for study has led 
to significant progress in chemical evolution studies of distant systems, 
investigators have still been hampered by a near degeneracy between the 
effects of dust depletion and those arising from shifts in the ratio of 
elements arising from $\alpha$-capture processes compared to those associated 
with the iron peak (chiefly coming from Type~1a supernovae), since many of 
the iron peak elements are strongly susceptible to depletion while the 
$\alpha$-process elements are usually much less so (Dessauges-Zavadsky,
Prochaska, \& D'Odorico 2002; Prochaska \& Wolfe 2002).  One can bypass the
issue of dust by focusing on very mildly depleted elements such as C, N, O,
Ar and Zn to measure the overall metallicity of the gas, but except possibly
for Zn, there is little leverage in learning about the $\alpha$/Fe ratio,
which is important for our understanding of the stellar initial mass
functions and formation histories of the DLAs.  Likewise, one may study
systems that are known to contain very little dust, but this selection can
introduce a sample bias that could lead to false conclusions on the chemical
evolution of systems in general.  Finally, one can take a broader approach by
correcting for the effects of dust depletion, using information based on the
empirical evidence on how element abundances change with the formation of
dust in our Galaxy.  We explain how to do this here.

\subsubsection{The Current Proposal to Correct for Dust 
Depletions}\label{current_prop}

As a way to interpret the abundances observed in an absorption system with an 
unknown overall metallicity and level of dust depletion, we can start with an 
initial guess that the intrinsic abundance pattern is not much different from 
that of our Galaxy, aside from an overall elevation or depression of all 
elemental abundances with respect to hydrogen.  We then employ the method 
outlined in \S\ref{calc} to determine the severity of depletion, as 
represented by $F_*$ (the slope of the trend of $y$ vs. $x$).  If $N$(H~I) 
can be determined by observing the L$\alpha$ absorption (either from a space 
observatory for low-$z$ absorption systems or from the ground for high-$z$ 
systems), the difference between the synthetic $N$(H) and the real one 
indicates the metallicity of the system relative to that of the Galaxy.  
Next, we can examine the validity of the assumption that the pattern is not 
much different from that of our Galaxy by examining how well the element 
abundances conform to a straight line in $y$ vs. $x$, as exemplified by 
Fig.~\ref{nh_fstar_examples} for two sight lines in our Galaxy (but recall 
that these two panels show special demonstrations of poor fits).  Certain 
patterns of deviations from a straight line could serve as a warning that 
alternate intrinsic abundance patterns must be considered.  (While this may 
be true, one must be watchful that one is not being deceived by effects of 
seeing mixtures of regions with different values of $F_*$, as exemplified by 
the lower panel of Fig.~\ref{nh_fstar_examples}.  It is noteworthy that on 
the one hand Prochaska (2003) found that abundance 
variations from all possible causes for different velocity components in 13 
different DLA systems were less than 0.2~dex. On the other hand, 
Dessauges-Zavadsky et al. (2006) found 
that for some DLAs there were pronounced variations in some abundance ratios, 
sometimes $>$0.3~dex, which they interpreted to arise from changes in the 
depletion levels from one cloud to the next.)

As we move on to gas systems that we have good reason to believe to have 
intrinsic abundance patterns that are different from that of our Galaxy, we 
must rely on a more general approach, but one that still makes use of 
information provided by the current study.  It is clear from the differences 
of the slopes of the trends shown in lower panels of Figs.~\ref{panel_set_1} 
to \ref{panel_set_4} that the progressively increasing scarcity of certain 
elements starts to modify the composition of added material on the grains as 
they become larger or more numerous (i.e., as $F_*$ increases).  In such 
circumstances, it seems sensible to imagine that different atoms have their 
own proclivities to attach themselves to dust grains and form compounds, 
which forces the grain compositions to be regulated by the effective reaction 
rate constants of the elements (i.e., in a regime of only grain growth, we 
may think of such rate constants in terms of the atomic sticking efficiencies 
times their mean velocities) multiplied by their respective concentrations at 
any given time.  As we pointed out near the end of \S\ref{cond_temp}, the 
$A_X$ terms derived here represent just such a collection of rate constants.

\subsubsection{Possible Complications}\label{complications}

A few cautionary remarks are in order for gas systems whose initial 
compositions differ appreciably from those in our Galaxy.  In the following 
paragraphs, these cautions will be expressed, and it is important to 
emphasize that they apply to {\it any\/} dust correction scheme for 
abundances -- not only the one proposed here.

One could imagine that for any arbitrary mixtures of gas-phase elements, one 
could simply integrate the equations for the condensations as a function of 
time (or grain growth), using the $A_X$ values as rate constants.  However, 
such an approach invokes the assumption that the retention of atoms after an 
initial sticking is in no way driven by the composition of the existing grain 
material.  This may not be correct.  Most elements probably depend on the 
presence of others to form chemically stable compounds that are durable 
enough to remain in solid form for long periods of time.  For instance, 
Lodders (2003) presented examples where the elements Ni and 
Ge depend on the presence of a host element Fe to create an alloy.  Likewise, 
Mg, Si and O are needed to form the host minerals forsterite and enstatite 
that are pathways to synthesize the most refractory compounds of Zn 
(Zn$_2$SiO$_4$ and ZnSiO$_3$) and Mn, (Mn$_2$SiO$_4$ and MnSiO$_3$).  Thus, 
while we may note from Table~\ref{elem_parameters} that $A_{\rm Zn}$ is about 
the same as $A_{\rm Ge}$, in an environment where the $\alpha$/Fe ratio 
differs appreciably from that of our Galaxy, one or the other of these two 
elements could be more starved for its respective host element(s) and could 
thus would probably behave differently from what we have observed here.

Another complication is the possibility that there could be large differences 
in dust-to-gas ratios and grain compositions in the ejecta of various kinds 
of sources that enrich a galaxy throughout its lifetime, such as evolved 
stars or different varieties of supernovae (Dwek 1998; Kozasa et al. 2009). 
It is well known that contributions of differing proportions from these
sources throughout the history of a system's chemical evolution will change
the intrinsic mix of elements with time (Calura, Pipino, \& Matteucci 2008),
but they might also induce changes in the populations of either the primitive
grains or the resilient cores of mature grains that do not normally grow or
decline in the ISM.  These primitive grains (or grain cores) are probably
major contributors to the depletions seen at $F_*=0$.  Thus, while values of
$A_X$ might accurately describe how elements accrete onto grains as they grow
in the ISM, the overall offsets represented by $B_X$ could be influenced by
the amounts and compositions of grains ejected by the sources.

As a simple illustration of how differing properties of grains in the ejecta 
of metal sources might create misleading conclusions, we can consider some 
comparisons of Si and Fe seen in the absorption spectra of DLAs.  From 
entries in Table~\ref{elem_parameters}, we note that $A_{\rm Si}\approx 
A_{\rm Fe}$, so that differences in the depletions of these two elements do 
not change much with differing values of $F_*$.  The minimum separation 
between $[{\rm Si}_{\rm gas}/{\rm H}]$ and $[{\rm Fe}_{\rm gas}/{\rm H}]$ 
occurs at $F_*=0$ and has a value of 0.73~dex.  (A small decrease in this 
number could be realized by extrapolating the depletion trends to negative 
values of $F_*$, but not beyond that which makes $[{\rm Si}_{\rm gas}/{\rm 
H}]>0$.)  Wolfe, Gawiser \& Prochaska (Wolfe, Gawiser, \& Prochaska 2005)
show measurements of [Si/Fe] in the gas phase (i.e., without any dust
correction) as a function of [Si/H] for a collection of high quality
observations of DLAs (see their Fig.~8).  Their plot shows that $[{\rm Si/
Fe}]\approx 0.3$ for $[{\rm Si/H}]\lesssim -1.0$, and it increases somewhat
for systems that have Si abundances that approach that of our Galaxy.  On the
one hand, if we simplistically apply our minimum correction for differences
in dust depletion, we arrive at an intrinsic $[{\rm Si}/{\rm Fe}]\approx -
0.4$.  Correction factors of about the same magnitude were shown by Calura, 
Matteucci \& Vladilo (2003), but their corrected 
[Si/Fe] are not as low because their observed values of [Si/Fe] started out 
at values that were generally higher than +0.3~dex.  Of course, in this 
context it is possible that long, quiet periods between bursts of star 
formation could produce $[\alpha/{\rm Fe}]<0$ (Gilmore \& Wyse 1991), and
indeed the sequence of $[\alpha/{\rm Fe}]$ as a function of [Fe/H] for dwarf
galaxies and the LMC is $\sim 0.1-0.3\,$dex below that seen for stars in our
Galaxy (Venn et al. 2004).  On the other hand, it is quite possible that we
could be misled in our interpretation if all of the following conditions
apply: (1) Type~Ia supernovae eject most of their Fe-peak elements eventually
in the form of dust -- a prospect that seems to have no observational support
(Draine 2009), (2) this material is not significantly reprocessed through
subsequent generations of stars and (3) core collapse supernovae do not form
nearly as much dust in their ejecta as the Type~Ia supernovae.  In such
circumstances, the depletion corrections could be distorted by the differing
proportions of primitive grain production sites, compared to those in the
present-day Galaxy (where the effects Type~Ia supernovae are more influential
than in the more primitive proto-disk galaxy systems), leading to corrected
abundances that do not properly reflect the intrinsic gas + dust compositions
of these other systems.

A strategy for checking on the complications discussed above is to examine 
many different elements simultaneously.  For example, one might supplement 
the Si and Fe determinations discussed in the above paragraph with 
measurements of Ti (Dessauges-Zavadsky, Prochaska, \& D'Odorico 2002; Ledoux,
Bergeron, \& Petitjean 2002), which is an $\alpha$-process element that has
large depletion parameters, even larger than those of Fe. If the depletion of
Ti seems to be in the correct proportion to that of Fe, after correcting for
a different intrinsic $[\alpha/{\rm Fe}]$, then our misgivings about the
effects of vastly different grain productions in different sources may be
unwarranted.

Again, we suggest that to obtain the best general understanding of the 
complex processes that may influence the observed abundances in DLAs, the 
most productive insights may arise from examinations of the plots of $y$ vs. 
$x$ for individual absorbing systems.  Such plots may be far more instructive 
than a battery of correlation plots that compare for many systems the various 
combinations of element abundance ratios.

\subsubsection{Other Dust Correction Methods}\label{other_methods}

In the recent past, investigations of DLA system abundances have relied on 
different schemes for correcting for dust depletion.  For instance, Vladilo 
 (2002b) devised a method to account for dust depletion in 
DLAs, using information on dust grain compositions derived from the summary 
of interstellar abundances under different conditions in our Galaxy given by 
Savage \& Sembach (Savage \& Sembach 1996a) (see 
Appendix~\ref{ss96_comparison}).  Vladilo's abundance corrections worked with 
parameters linked to the amounts of elements within the dust grains, with 
some recognition that these processes might change with different overall 
metallicities. Since this approach has been used to correct the observed 
abundances in a number of recent studies
(Calura, Matteucci, \& Vladilo 2003; Centuri\'on et al. 2003; 
Dessauges-Zavadsky et al. 2004, 2007; Vladilo 2004; Vladilo et al. 2006; Quast,
Reimers, \& Baade 2008), it may be helpful re-express his abundance
compensation parameters in the light of the parameters in the current study,
so that we have a clearer understanding of what changes were made in the
previous investigations. This is done in Appendix~\ref{vladilo_comparison}.

In Appendix~\ref{PW02_comparison} there is a similar cross calibration for 
two depletion parameters $\kappa^{\rm Zn}$ and $\kappa^{\rm Si}$ defined by 
Prochaska \& Wolfe (2002) in their study of 
elements seen in their collection of DLAs.  It is important to realize that 
there is no fundamental reason why distant DLAs could not have values of 
$F_*$ less than zero, since their intrinsic metallicities [M/H] range from 
about $-2.5$ to $-0.5$~dex (Wolfe, Gawiser, \& Prochaska 2005).  Our
condition $F_*=0$ applies to the arbitrary condition (but a practical one in
our case) that the logarithm of the metallicity [M/H]~=~0 and $N({\rm
H})\approx 10^{19.5}{\rm cm}^{-2}$.  While $F_*<0$ might seem to be a
reasonable outcome for such systems with very low concentrations of metals,
especially those at the lower limit of $N({\rm H~I})=10^{20.3}{\rm cm}^{-2}$
for the standard definition of a DLA, in practice a comparison of our
Fig.~\ref{PW02} with Fig.~22 of Prochaska \& Wolfe (2002) indicates 
that $F_*$ is always greater than zero.

\section{Summary}\label{summary}

The principal aim of this study has been to arrive at a simple, generalized 
description of the depletions of atoms in the interstellar medium of our 
Galaxy.  The objective is not only to help us understand the elemental 
composition of dust grains, and how it changes as depletions increase, but 
also to provide the necessary guidance on how to correct for deviations in 
elemental abundances caused by dust grain depletions in distant gas systems 
that can be studied via their absorption lines in the spectra of quasars or 
GRB afterglows.

It has been known for some time that the strengths of depletions vary from 
one region to the next.  However, to a remarkable degree of uniformity, we 
find that as these general depletion strengths vary, the logarithms of the 
depletion factors of different elements $[X_{\rm gas}/{\rm H}]$ are related 
to each other in a linear fashion that can be described by an equation 
\begin{equation}
[X_{\rm gas}/{\rm H}]=B_X+A_X(F_*-z_X)
\end{equation}
where $B_X$ and $A_X$ are empirically determined constants that apply to each 
element $X$, and $F_*$ is a generalized depletion strength parameter that 
applies to the line of sight through the ISM that is under study.  The 
zero-point offset for any element $z_X$ that applies to $F_*$ is added to the 
equation simply to make the measurement errors in $B_X$ and $A_X$ independent 
of each other (and its value is governed only by the distribution of $F_*$ 
values and their errors in the sample).

A large accumulation of interstellar column densities gathered from the 
literature has been used to establish the validity of this simple model for 
depletions, as well as to provide the most likely values for the $F_*$ 
line-of-sight parameters and the two constants $B_X$ and $A_X$ for 17 
different elements. The data were screened to eliminate determinations that 
may have been compromised by uncertain corrections for line saturation, and 
in many cases the column densities were corrected in a manner to make them 
conform to a modern compilation of transition $f$-values. For all of the 
elements except sulfur, which was handled separately, the two constants are 
listed in Table~\ref{elem_parameters}.  In establishing these constants, we 
considered only those sight lines that had $N({\rm H})>10^{19.5}{\rm 
cm}^{-2}$, in order to decrease the chances that we could have been misled by 
unseen ionization stages, either for the element in question or hydrogen.  
Values of $F_*$ and/or $F_{*\,{\rm syn}}$ for 239 separate regions are listed 
in Table~\ref{nh_fstar} (these include, for a few cases, some separate 
velocity components exhibited for some single sight lines and also sight 
lines that had $N$(H) below our established column density threshold).  In a 
separate exercise, indirectly determined (synthetic) values of $N$(H) and 
$F_*$ were evaluated for 29 white dwarf stars in the Local Bubble and listed 
in Table~\ref{lism_tbl}.

In conventional investigations of the elemental composition of dust grains, 
one compares the observed values of $N$(X) to those expected from some 
adopted standard for the total abundance, most often taken from either a 
solar or meteoritic abundance (or the abundances of nearby B-type stars).  
The difference between a total abundance $(X/{\rm H})_\odot$ and $(X/{\rm 
H})_{\rm ISM}$ indicates the quantity of the element that is locked up in 
solid form.  One can apply this method to the measurements reported here, but 
the accuracy of the outcome is strongly driven by how well the adopted total 
abundance $(X/{\rm H})_\odot$ conforms to reality.  While this approach is 
needed to obtain the total makeup of the grains, another useful tactic is to 
study differences in specific elemental depletions as the overall levels of 
depletion increase.  The outcome here is entirely independent of whatever one 
adopts for the total abundances.  For any given value of $F_*$, the 
differential dust composition scales in proportion to $A_X(X_{\rm gas}/{\rm 
H})_{F_*}$.  

For any sight line where $N$(H) has not been observed, one does not have 
explicit measurements of any depletions.  Nevertheless, by making use of the 
information on how elements deplete in a collective manner, we can derive 
reasonably accurate (synthetic) values of $F_*$ and $N$(H) if we have column 
density measurements $N(X)$ for several elements that have large differences 
in $A_X$ and perform a least-squares fit of the quantities $\log N(X)-\log 
(X/{\rm H})_\odot-B_X+A_Xz_X$ against their respective values of $A_X$.  
While this is a useful tool for overcoming our inability to measure directly 
$N$(H) for any of several possible reasons, its greatest utility should be an 
application to the study of intrinsic element abundances in absorption-line 
systems seen in the spectra of distant quasars or the optical afterglows of 
GRBs.  Specifically, one can compare the measured value of $N$(H), obtained 
through an observation of the L$\alpha$ absorption, to the synthetic value of 
$N$(H) derived from the pattern of element column densities.  The ratio of 
$N({\rm H})_{\rm syn.}$ to $N({\rm H})_{\rm obs.}$ yields the metallicity of 
the system relative to that of our Galaxy.  The value of $F_{*\,{\rm syn.}}$ 
indicates the dust content of the system.  However, caution is advised for 
systems that are suspected to have a pattern of intrinsic abundances that 
differs appreciably from that of our Galaxy.  For systems outside our Galaxy 
that are not too distant (e.g. the Magellanic Clouds), one should be able to 
validate the concept of using $N({\rm H})_{\rm syn.}/N({\rm H})_{\rm obs.}$ 
to obtain a metallicity by comparing the outcome to the average metallicity 
of the embedded stars.

The above paragraphs outline the basic themes contained in this paper.  Some 
additional, more specific insights that came from this investigation are as 
follows:
\begin{enumerate}
\item Except for the elements C, N, O, P, Cl, S, and Zn, all elements show 
some measurable depletion at $F_*=0$.  For the elements Ti, Cr, Mn, Fe, and 
Ni, these base depletions $[X_{\rm gas}/{\rm H}]_0$ are of order $-1\,$dex.  
They correlate moderately well with their respective condensation 
temperatures $T_c$, but the correlation of the depletion slopes $A_X$ with 
$T_c$ is even better.
\item Nitrogen appears to be the only element that does not show 
progressively stronger depletions as $F_*$ approaches 1.  There are too few 
measurements of carbon to establish whether or not the apparent strengthening 
of its depletion with $F_*$ is real.  For both of these elements, the errors 
are large enough to permit the progressive incorporation of these atoms (by 
number) to still exceed the accumulations of P, Cl, Cr, Mn, Fe, Ni, Zn, Ti, 
Cu, Ge or Kr when $F_*\approx 0$ and also Mg or Si when $F_*=1$. 
\item Until now, the observed small variations of $[{\rm Kr}_{\rm gas}/{\rm 
H}]$ seemed random (and possibly driven either by changes in intrinsic 
abundances of this element from one place to the next or by observational 
uncertainties).  The apparent correlation of this quantity with $F_*$ in the 
current investigation suggests that the relative gas-phase abundance of this 
chemically inert element is coupled to those of other elements, but at a very 
low level.  A possible means for depleting Kr is either physisorption on the 
surfaces of dust grains or locking within water ice clathrates.
\item For chlorine, there might be a mild misrepresentation of the 
relationship for its gas-phase abundance with respect to $F_*$.  In part, the 
observational errors for $N$(Cl~II) are larger than for other elements, but a 
more important effect is reversion of some of the atoms to a neutral form 
through a series of reactions with H$_2$.  This could be especially important 
at large values of $F_*$, where the fractional abundances of H$_2$ are large.  
There are some known cases where $N({\rm Cl~I})>N({\rm Cl~II})$, according to 
JSS86.
\item The differential depletion of oxygen at low levels of $F_*$ is just 
barely consistent with the consumption of O in the form of oxides and 
silicates.  For $F_*\approx 1$ this is no longer true: the loss of O atoms 
from the gas phase far outstrips the production of silicates and oxides, 
suggesting that the formation of compounds involving abundant partner 
elements such as H or C may play an important role.  While N is abundant, it 
does not have differential depletions that are large enough to help explain 
the consumption of O.  The large loss of oxygen atoms from the gas phase 
found in the present study is very difficult to reconcile with current models 
of interstellar grains. 
\item Even though the average density $\langle n_{\rm H}\rangle=N({\rm H})/d$ 
is a crude representation for the true local densities experienced by most of 
the atoms, we find that this quantity still exhibits a tight correlation with 
$F_*$. This same correlation is seen for both the distant stars and for sight 
lines within a regime of generally low space densities within the Local 
Bubble. The surprisingly good relationship between average density and 
depletion strength supports the notion that either the lines of sight 
exhibiting high $\langle n_{\rm H}\rangle$ have gas that is contained within, 
or has recently evolved from, very dense regions where rapid grain growth can 
occur or that these regions of space are better shielded from destructive, 
high velocity shocks, or both.  The fraction of hydrogen in molecular form 
does not show a correlation that is probably any better than a secondary one 
that should arise from the correlation between $f({\rm H}_2)$ and $\langle 
n_{\rm H}\rangle$.
\item By comparing synthetic values of $N$(H) with the observed ones, we see 
no evidence for changes in the intrinsic abundances of heavy elements in 
different regions around the Sun.  The pattern of observed deviations seems 
random.
\item More needs to be done: additional data of good quality are needed to 
define better the differential depletion relationships of C, S and Kr, to see 
if their values of $A_X$ are truly nonzero.  Also, with the increase in 
sensitivity provided by the {\it Cosmic Origins Spectrograph\/} that will be 
installed on the forthcoming {\it HST\/} servicing mission, we should have an 
opportunity to observe lines of sight with greater extinctions (and hence 
probably much higher $F_*$), so that we can obtain extend our reach to denser 
clouds and obtain a better understanding of chemically active environments 
that are better protected from uv dissociation.
\end{enumerate}

\acknowledgments
This research was supported by Program number HST-AR-10279.01-A which was 
provided by NASA through a grant from the Space Telescope Science Institute, 
which is operated by the Association of Universities for Research in 
Astronomy, Incorporated, under NASA contract NAS5-26555.  The author 
acknowledges the great utility of the SIMBAD and VizieR catalog databases, 
operated at CDS, Strasbourg, France, which provided critically important 
information on the target stars included in this survey.  The author thanks 
B.~T.~Draine, J.~X.~Prochaska, B.~D.~Savage, U.~J.~Sofia and G.~Wallerstein 
for providing helpful comments after reading a draft version of this article.

{\it Facilities:\/} \facility{HST(STIS)}, \facility{FUSE}, 
\facility{Copernicus}, \facility{IUE}
\appendix
\section{Errors in Quotients}\label{quotient_errs}
For several equations in \S\ref{parameters}, we must evaluate the 
uncertainties of quotients of two terms, each of which have their own errors.  
In order to do so, we make use of Geary's (1930) 
approximation for the frequency distribution of the quotient of two 
quantities that each have normally distributed errors.  According to Geary, 
for a quotient
\begin{equation}\label{z}
z={b+y\over a+x}
\end{equation}
involving a denominator $a$ and numerator $b$ that have respective normally 
distributed errors $x$ with a standard deviation $\alpha$ and $y$ with a 
standard deviation $\beta$, the quantity
\begin{equation}\label{t}
t(z)={az-b\over \sqrt{\alpha^2z^2-2r\alpha\beta z+\beta^2}}
\end{equation}
has a normal distribution with a mean of zero and standard deviation of 
unity,
provided that $a+x$ is unlikely to be negative (i.e., $a/\alpha \gtrsim 3$).  
(The quantity $r$ in Eq.~\ref{t} is the correlation coefficient of $x$ and 
$y$, which in our applications within \S\ref{parameters} is always assumed to 
be zero.)  After squaring both sides of Eq.~\ref{t} and collecting terms in 
$z$ and $z^2$, we have a quadratic equation,
\begin{equation}\label{quadratic}
(t^2\alpha^2-a^2)z^2+(2ab-2t^2r\alpha\beta)z+(t^2\beta^2-\beta^2)=0
\end{equation}
whose roots give the extreme values of $z$ that bound the possible 
combinations of $b/a$ at the ``$t\sigma$ level of significance.''  For the 
errors that appear in Eqs.~\ref{W_X} and \ref{W_*}, we evaluate half of the 
difference between the two roots for $t=1$.

\section{A Compilation of the Basic Data and Sight-Line Depletion 
Factors}\label{basic_data}

Tables~\ref{C} through \ref{Kr}\footnote{
These tables are placed at the end of this article, immediately
after the references.  When this article is published in
the Astrophysical Journal, all of the tables
will appear as a single, long table only in the electronic edition 
in the form of a machine-readable table.}
show the basic measurements of column densities (columns~(3)$-$(5)) for each 
of the elements considered in this study, with the codes that signify the 
sources in the literature [column~(6)] that are linked to references shown in 
column (2) in Table~\ref{fvals}.  All of the logarithms of the column 
densities have been corrected for changes in $f$-values (\S\ref{fvalues}) by 
adding the factors expressed in dex for the appropriate elements given in 
Table~\ref{fvals}.  Column~(7) of the element tables shows the value of the 
depletion index $F_*$ and its error for the star in question, determined from 
Eqs.~\ref{F_*} and \ref{sigma(F_*)}, respectively.  This column is followed 
one that shows the expected depletion of the element $[X_{\rm gas}/{\rm 
H}]_{\rm fit}$, calculated from Eq.~\ref{better_equation} with the 
coefficients shown in Table~\ref{elem_parameters}. The error in this term,
\begin{equation}\label{sigma_fit}
\sigma([X_{\rm gas}/{\rm H}]_{\rm fit})=\sqrt{
\sigma(B_X)^2+[(F_*-z_X)\sigma(A_X)]^2+[A_X\sigma(F_*)]^2}
\end{equation}
combines in quadrature the various sources of errors arising from either the 
best-fit coefficients $A_X$ and $B_X$ of element $X$ or the determination of 
$F_*$.  (Recall from the discussion in \S\ref{determination_of_ABz} that 
$\sigma(B_X)$ here does not include the systematic error $\sigma(X/{\rm 
H})_\odot$.)  Column~(9) shows the amount by which the observed depletion 
$[X_{\rm gas}/{\rm H}]_{\rm obs}$ differs from $[X_{\rm gas}/{\rm H}]_{\rm 
fit}$, and column~(10) shows this number divided by the combined uncertainty 
of $[X_{\rm gas}/{\rm H}]_{\rm obs}$ and $[X_{\rm gas}/{\rm H}]_{\rm fit}$ 
(with the two added together in quadrature), so that one can easily recognize 
deviations that seem to be unacceptably large. Values of $[X_{\rm gas}/{\rm 
H}]_{\rm obs}$ can be obtained by adding together the numbers in columns~(8) 
and (9), but asymmetrical error bars in the original data are not evident.  
To reconstruct the errors in $[X_{\rm gas}/{\rm H}]_{\rm obs}$ one must take 
into account the uncertainties of both $N(X)$ and $N$(H); for the latter, see 
Table~\ref{nh_fstar}.  For each element, one can sense how well the 
calculated values $[X_{\rm gas}/{\rm H}]_{\rm fit}$ agree with the observed 
ones by examining either individual deviations shown in the last two columns 
of the element tables or by inspecting the collective statistical information 
presented in Table~\ref{elem_parameters} that was explained in 
\S\ref{solutions}.  Missing entries in columns~(7)$-$(10) are caused by a 
lack of information needed to calculate $N$(H) (see 
Table~\ref{stellar_data}).

\placetable{C}

\section{Retrospectives on Earlier Characterizations}\label{retro}

Many studies of gas abundances and dust compositions have made use of earlier 
descriptions of the depletion process.  In order to recast these previous 
investigations in the light of the present work, we present in the 
subsections below the outcomes of the earlier depletion studies in terms of 
the parameters defined in \S\ref{parameters}.  This should be useful in 
obtaining a better understanding of the earlier abundance works in the 
present context.

\subsection{A Comparison with the Results of Savage \& Sembach
(1996a)}\label{ss96_comparison}

\begin{figure}
\epsscale{0.7}
\plotone{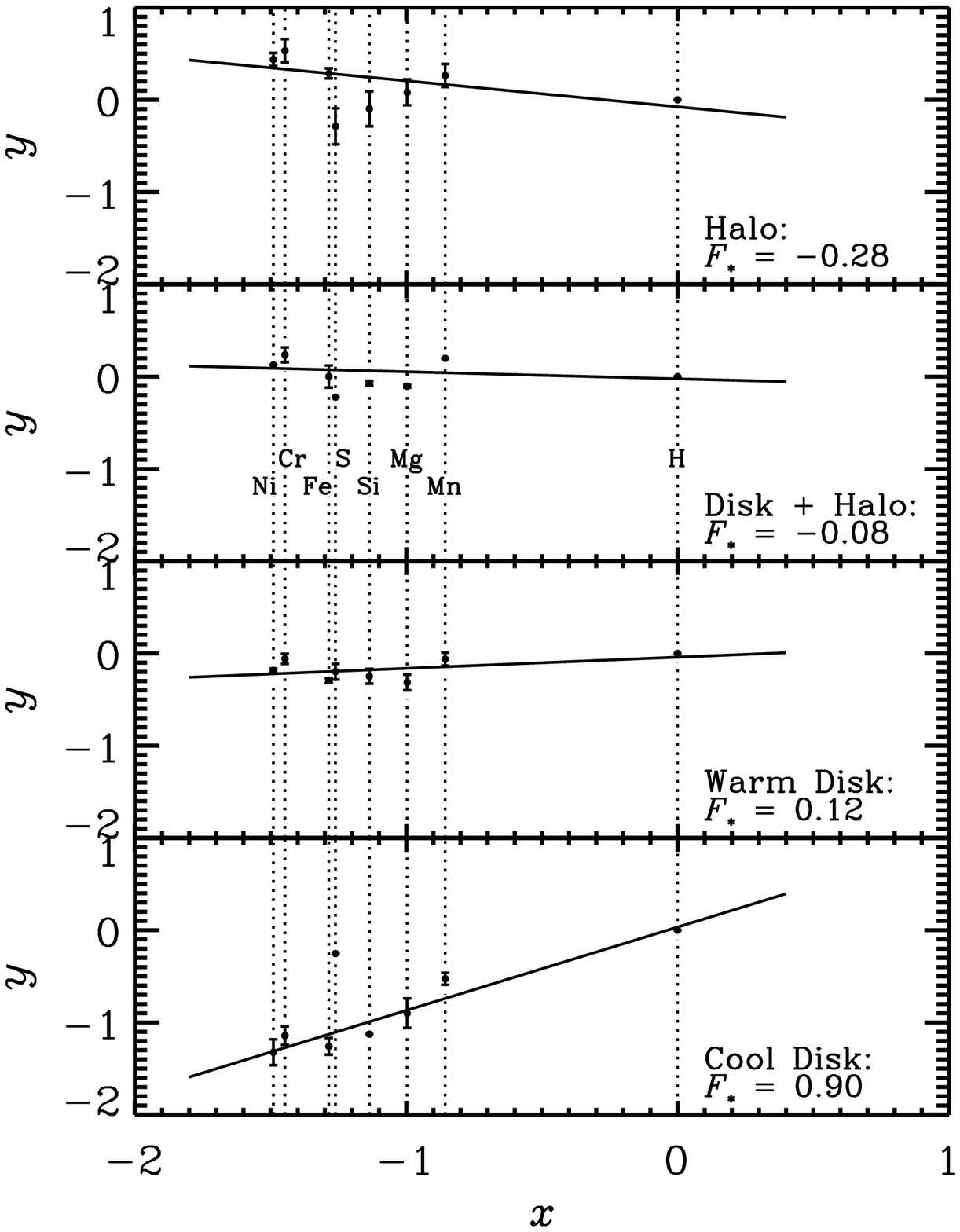}
\caption{An application of Eq.~\protect\ref{xy} to the results shown in 
Table~6 of Savage \& Sembach (1996a), so that 
their results for different environments can be calibrated to the $F_*$ index 
of the present work.  The values of $x$ for each element are identified in 
the panel second from the top, with vertical dotted lines linking them across 
the remaining panels.  In each case an unweighted least-squares linear fit to 
the points is shown by a slanted line whose slope yields the value of $F_*$ 
listed in the respective panel.  Error bars attached to some of the points do 
not represent uncertainties; instead they show the range of values that 
Savage \& Sembach found for the respective elements in each 
environment.\label{ss96}}
\end{figure}

After the publication of the review paper by Savage \& Sembach (1996a) on
interstellar abundances from absorption line measurements using the {\it
Hubble Space Telescope}, many investigations compared their own results with
the generic depletions discussed in that review.  We now offer a cross
calibration between our $F_*$ and the environments discussed by Savage \&
Sembach (1996a).  We can do this through the application of Eq.~\ref{xy} to
their results, with mean values of the numbers given in their Table~6
replacing the expression $\log N(X)-\log (X/{\rm H})_\odot$ in our equation
and by pinning $N$(H) to a value of zero.  Figure~\ref{ss96} shows this
comparison for the four different representative environments specified by
Savage \& Sembach.

\placefigure{ss96}

In order to obtain the values of $F_*$ in each case, we evaluated a 
least-squares linear fit to the points, but without any weight factors to 
account for differing errors.  For reasons given \S\ref{sulfur}, sulfur was 
not included in the fit (but it is shown at the appropriate locations in the 
panels of the figure).  Since depletions are known with respect to hydrogen, 
it is appropriate to add this element to the fit at the location $(x,y)=0$.  
In order to make the comparison accurate, we made adjustments to the numbers 
specified by Savage \& Sembach, so that we can account for (1) small 
differences between their adopted reference abundances and the ones given by 
Lodders (2003), as listed in column (2) of our 
Table~\ref{elem_parameters}, and (2) the differences between the $f$-values 
that seemed appropriate in 1996 to the more recent ones published by Morton 
 (2003) and Jenkins \& Tripp (2006); see \S\ref{fvalues} for some
considerations that applied to the $f$-value adjustments.  The net changes
are given by the following terms (expressed in dex) that were added to the
original depletion values listed by Savage \& Sembach: Mg~=~0.23, Si~=~$-
0.04$, S~=~$-0.01$, Mn~=~$-0.05$, Cr~=~0.21, Fe~=~$-0.03$, and Ni~=~0.34. 
The values of $F_*$ that apply to each environment are given in the
respective panels of Figure~\ref{ss96}.  As one would expect, there is a
regular progression in $F_*$ from the least dense environments to the most
dense ones.

\subsection{Dust Corrections of Vladilo (2002a, b)}\label{vladilo_comparison}

\begin{figure}
\epsscale{0.7}
\plotone{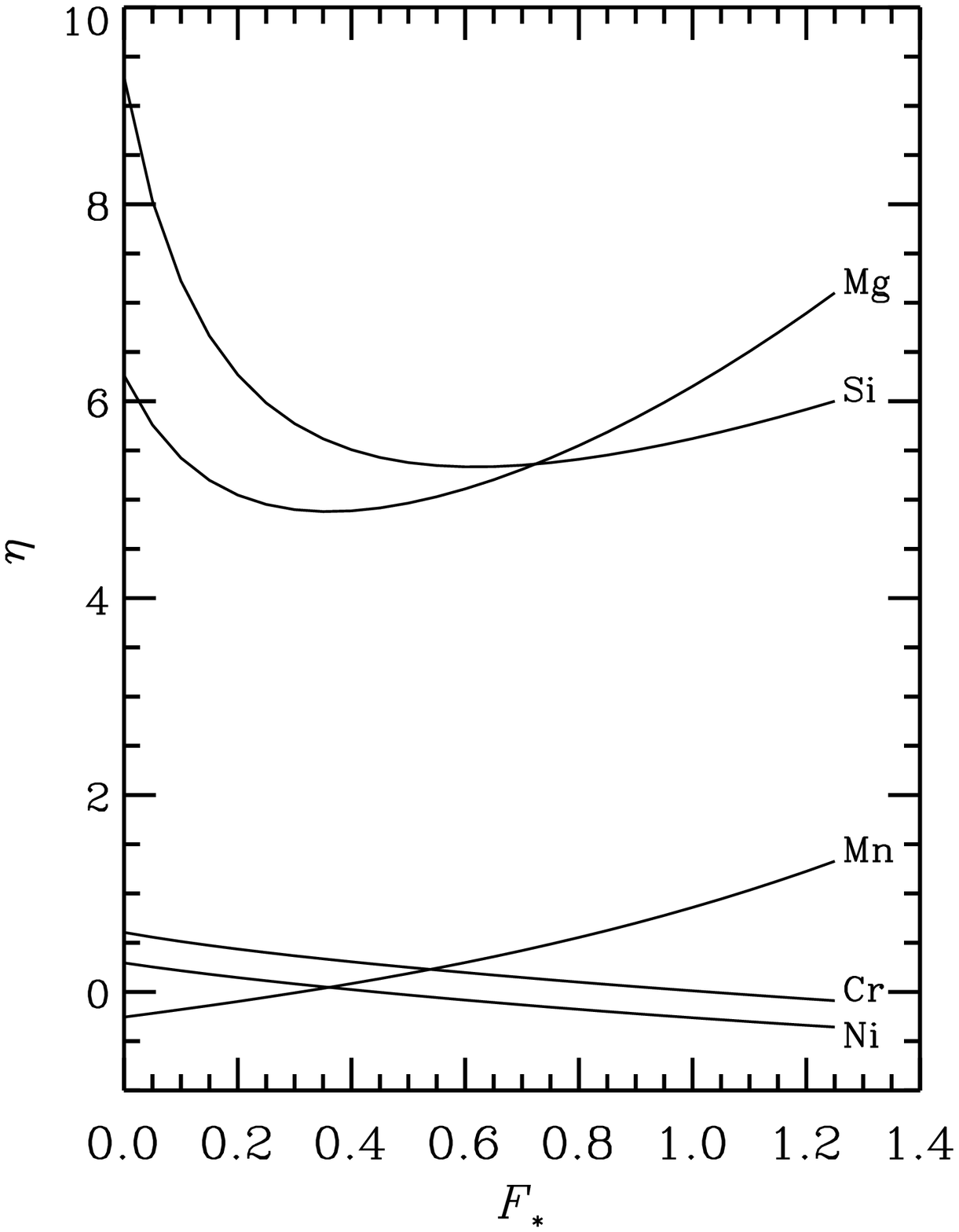}
\caption{Values of the parameter $\eta_X$ for various elements in the dust 
correction scheme of Vladilo (2002b), expressed in terms of 
our parameter $F_*$.  $\eta_{\rm Zn}$ is off scale in this plot at values of 
around 20.\label{Vladilo_eta}}
\end{figure}

Vladilo (2002b) used the summaries of element abundances in 
different environments given by Savage \& Sembach (1996a) to characterize
changes in $(X_{\rm dust}/{\rm H})$ (which he calls $p_X$) in terms of $({\rm
Fe}_{\rm dust}/{\rm H})$ (which he calls $r$) by evaluating the derivatives
of their logarithms and defining a parameter $\eta_X\equiv d\log p/d\log r$.
 We can evaluate the differentials of these two quantities with respect to
$F_*$ to obtain \begin{eqnarray}\label{eta}
\eta_X&=&{r\over p_X}~{dp_X/dF_*\over dr/dF_*}\nonumber\\
&=&(X/{\rm H})_\odot\,{({\rm Fe}_{\rm dust}/{\rm H})10^{[X/{\rm H}]}A_X({\rm 
Fe}/{\rm H})_\odot^{-1}-(X_{\rm dust}/{\rm H})10^{[{\rm Fe}/{\rm H}]}A_{\rm 
Fe}(X/{\rm H})_\odot^{-1}\over (X_{\rm dust}/{\rm H})10^{[{\rm Fe}/{\rm 
H}]}A_{\rm Fe}}~.
\end{eqnarray}
Figure~\ref{Vladilo_eta} shows how $\eta_X$ varies as a function of $F_*$ for 
the elements that Vladilo (2002a)
chose to work with in his investigation of DLA abundances.  It is clear from 
the figure that these quantities change with $F_*$, which makes it difficult 
to arrive at generalized dust correction factors for the DLA abundances using 
his method.

\placefigure{Vladilo_eta}

\subsection{Dust Corrections of Prochaska \& Wolfe (Prochaska \& Wolfe
2002)}\label{PW02_comparison}

\begin{figure}
\epsscale{0.7}
\plotone{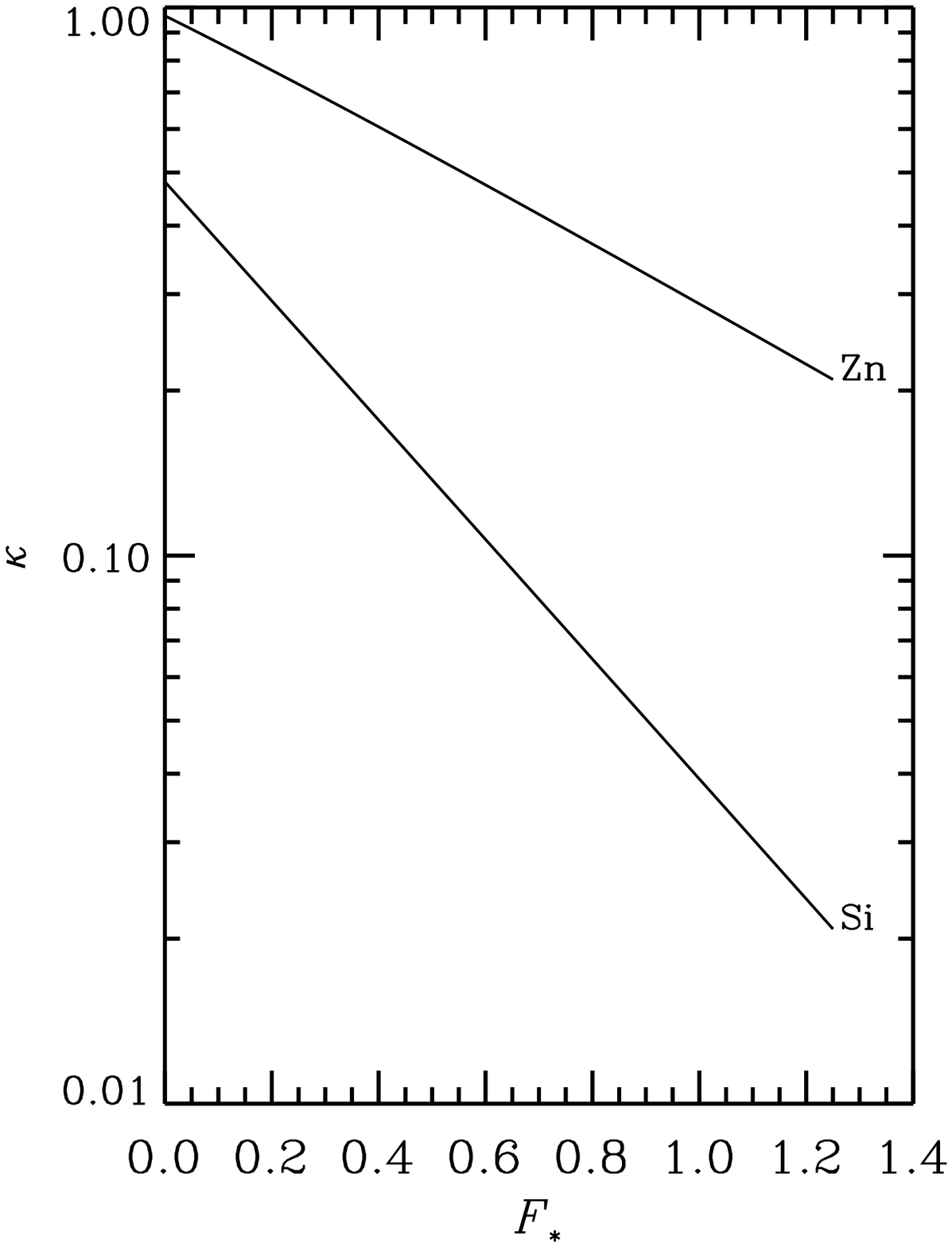}
\caption{Trends with respect to $F_*$ for the parameters $\kappa^{\rm Zn}$ 
(upper line) and $\kappa^{\rm Si}$ (lower line) used by Prochaska \& Wolfe 
 (2002) to exemplify the dust-to-gas ratios of 
different DLAs included in their investigation.\label{PW02}}
\end{figure}

In their investigation of the relative amounts of dust as a function of 
$N$(H) in the DLA systems that they studied, Prochaska \& Wolfe 
 (2002) defined two dust parameters, 
$\kappa^X\equiv (1-10^{[X/{\rm Fe}]})10^{[X/{\rm H}]}$ for $X={\rm Zn}$ and 
$X={\rm Si}$, to characterize the dust-to-gas ratios in relation to those in 
our Galaxy. Figure~\ref{PW02} shows how these two parameters behave as a 
function of $F_*$, and these trends allow one to evaluate their findings in 
the context of the present study.

\placefigure{PW02}
\clearpage
\clearpage
\begin{deluxetable}{
r  
c  
c  
c  
c  
c  
c  
c  
c  
c  
}
\tabletypesize{\footnotesize}
\tablecolumns{10}
\tablewidth{0pt}
\tablecaption{Observations and Fits for Carbon\label{C}}
\tablehead{
\colhead{} & \colhead{} & \multicolumn{4}{c}{Observed $\log N$(C II)} & \colhead{} & \colhead{} &
 \colhead{} & \colhead{Deviation\tablenotemark{e}}\\
\cline{3-6}
\colhead{HD Number\tablenotemark{a}} &
\colhead{Name} &
\colhead{l.l.} &
\colhead{best} &
\colhead{u.l.} &
\colhead{Source\tablenotemark{b}} &
\colhead{$F_*$} &
\colhead{[C$_{\rm gas}$/H]$_{\rm fit}$\tablenotemark{c}} &
\colhead{Residual\tablenotemark{d}} &
\colhead{(in $\sigma$)}\\
\colhead{(1)} &
\colhead{(2)} &
\colhead{(3)} &
\colhead{(4)} &
\colhead{(5)} &
\colhead{(6)} &
\colhead{(7)} &
\colhead{(8)} &
\colhead{(9)} &
\colhead{(10)}
}
\startdata
         24398&    $\zeta$ Per&   17.26&   17.34&   17.41&  CMJS96&   $\phm{-}  0.88\pm 0.05$&       $ -0.201 \pm   0.049$&       $ -0.111$&       $  -0.79$\\
         24534&          X Per&   17.35&   17.51&   17.62&   SFM98&   $\phm{-}  0.90\pm 0.06$&       $ -0.203 \pm   0.050$&       $ -0.085$&       $  -0.58$\\
         24912&      $\xi$ Per&   17.64&   17.78&   17.89&   C++91&   $\phm{-}  0.83\pm 0.02$&       $ -0.196 \pm   0.046$&$\phm{-}  0.225$&$\phm{-}   1.47$\\
         27778&         62 Tau& \nodata& \nodata&   17.34&  SLMC04&                   \nodata&              \nodata       &         \nodata&         \nodata\\
         34029&   $\alpha$ Aur&   14.65&   14.80&   14.95&  WRLS02&   $\phm{-}  0.44\pm 0.05$&       $ -0.157 \pm   0.095$&$\phm{-}  0.257$&$\phm{-}   1.27$\\
         35149&         23 Ori&   16.90&   16.98&   17.04&   W++99&   $\phm{-}  0.54\pm 0.04$&       $ -0.167 \pm   0.076$&$\phm{-}  0.055$&$\phm{-}   0.32$\\
         36861&$\lambda$ Ori A&   16.79&   17.00&   17.14&  CMJS96&   $\phm{-}  0.57\pm 0.04$&       $ -0.169 \pm   0.071$&       $ -0.085$&       $  -0.39$\\
         37021&$\theta^1$ Ori &   17.64&   17.82&   17.94&  SLMC04&                   \nodata&              \nodata       &        \nodata &        \nodata \\
\enddata
\tablenotetext{a}{Terms in parentheses indicate separate velocity components, if they are explicitly identified
and not grouped together; see \S\protect\ref{other_considerations}}
\tablenotetext{b}{Codes in this column are linked to references
listed in Table~\protect\ref{fvals}}
\tablenotetext{c}{The expected depletion $[{\rm C}_{\rm gas}/{\rm H}]$
computed using Eq.~\protect\ref{better_equation}.
The listed errors do not include an overall systematic uncertainty of
0.04 in the solar abundance $\sigma({\rm C}/{\rm H})_\odot$
in order to show just the formal error that arises from the
uncertainties in location of line of best fit and the value of $F_*$.}
\tablenotetext{d}{The observed $[{\rm C}_{\rm gas}/{\rm H}]$ minus 
that computed using Eq.~\protect\ref{better_equation}.}
\tablenotetext{e}{The difference shown in the previous column divided by the expected
difference due to the uncertainties in both the measured column density and the
coefficients that appear in Eq.~\protect\ref{better_equation}.}
\tablecomments{Tables~\protect\ref{N} through \protect\ref{Kr} are
published in the electronic edition of the Astrophysical Journal.
This table illustrates the content and format of these tables.}
\end{deluxetable}
\begin{deluxetable}{
r  
c  
c  
c  
c  
c  
c  
c  
c  
c  
}
\tabletypesize{\footnotesize}
\tablecolumns{10}
\tablewidth{0pt}
\tablecaption{Observations and Fits for Nitrogen\label{N}}
\tablehead{
\colhead{} & \colhead{} & \multicolumn{4}{c}{Observed $\log N$(N I)} & \colhead{} & \colhead{} &
 \colhead{} & \colhead{Deviation\tablenotemark{e}}\\
\cline{3-6}
\colhead{HD Number\tablenotemark{a}} &
\colhead{Name} &
\colhead{l.l.} &
\colhead{best} &
\colhead{u.l.} &
\colhead{Source\tablenotemark{b}} &
\colhead{$F_*$} &
\colhead{[N$_{\rm gas}$/H]$_{\rm fit}$\tablenotemark{c}} &
\colhead{Residual\tablenotemark{d}} &
\colhead{(in $\sigma$)}\\
\colhead{(1)} &
\colhead{(2)} &
\colhead{(3)} &
\colhead{(4)} &
\colhead{(5)} &
\colhead{(6)} &
\colhead{(7)} &
\colhead{(8)} &
\colhead{(9)} &
\colhead{(10)}
}
\startdata
          5394&   $\gamma$ Cas&   15.96&   16.00&   16.04&   MCS97&   $\phm{-}  0.52\pm 0.04$&       $ -0.109 \pm   0.017$&$\phm{-}  0.046$&$\phm{-}   0.46$\\
         22928&   $\delta$ Per&   15.18&   15.30&   15.39&    MY82&                   \nodata&              \nodata       &        \nodata &        \nodata \\
         23180&         o  Per&   16.87&   17.02&   17.17&   Y++83&   $\phm{-}  0.84\pm 0.06$&       $ -0.109 \pm   0.029$&$\phm{-}  0.039$&$\phm{-}   0.21$\\
         24534&          X Per&   17.14&   17.22&   17.30&   JRS07&   $\phm{-}  0.90\pm 0.06$&       $ -0.109 \pm   0.032$&$\phm{-}  0.087$&$\phm{-}   0.94$\\
         24760& $\epsilon$ Per&   16.22&   16.28&   16.35&    MY82&   $\phm{-}  0.68\pm 0.04$&       $ -0.109 \pm   0.020$&       $ -0.007$&       $  -0.06$\\
         27778&         62 Tau&   16.95&   17.11&   17.27&   JRS07&                   \nodata&              \nodata       &        \nodata &        \nodata \\
         34029&   $\alpha$ Aur&   13.78&   13.87&   13.96&  WRLS02&   $\phm{-}  0.44\pm 0.05$&       $ -0.109 \pm   0.019$&       $ -0.164$&       $  -1.24$\\
         35149&         23 Ori&   16.47&   16.57&   16.65&   W++99&   $\phm{-}  0.54\pm 0.04$&       $ -0.109 \pm   0.017$&$\phm{-}  0.150$&$\phm{-}   0.92$\\
\enddata
\tablenotetext{a}{Terms in parentheses indicate separate velocity components, if they are explicitly identified
and not grouped together; see \S\protect\ref{other_considerations}}
\tablenotetext{b}{Codes in this column are linked to references
listed in Table~\protect\ref{fvals}}
\tablenotetext{c}{The expected depletion $[{\rm N}_{\rm gas}/{\rm H}]$
computed using Eq.~\protect\ref{better_equation}.
The listed errors do not include an overall systematic uncertainty of
0.11 in the solar abundance $\sigma({\rm N}/{\rm H})_\odot$
in order to show just the formal error that arises from the
uncertainties in location of line of best fit and the value of $F_*$.}
\tablenotetext{d}{The observed $[{\rm N}_{\rm gas}/{\rm H}]$ minus 
that computed using Eq.~\protect\ref{better_equation}.}
\tablenotetext{e}{The difference shown in the previous column divided by the expected
difference due to the uncertainties in both the measured column density and the
coefficients that appear in Eq.~\protect\ref{better_equation}.}
\end{deluxetable}
\begin{deluxetable}{
r  
c  
c  
c  
c  
c  
c  
c  
c  
c  
}
\tabletypesize{\footnotesize}
\tablecolumns{10}
\tablewidth{0pt}
\tablecaption{Observations and Fits for Oxygen\label{O}}
\tablehead{
\colhead{} & \colhead{} & \multicolumn{4}{c}{Observed $\log N$(O I)} & \colhead{} & \colhead{} &
 \colhead{} & \colhead{Deviation\tablenotemark{e}}\\
\cline{3-6}
\colhead{HD Number\tablenotemark{a}} &
\colhead{Name} &
\colhead{l.l.} &
\colhead{best} &
\colhead{u.l.} &
\colhead{Source\tablenotemark{b}} &
\colhead{$F_*$} &
\colhead{[O$_{\rm gas}$/H]$_{\rm fit}$\tablenotemark{c}} &
\colhead{Residual\tablenotemark{d}} &
\colhead{(in $\sigma$)}\\
\colhead{(1)} &
\colhead{(2)} &
\colhead{(3)} &
\colhead{(4)} &
\colhead{(5)} &
\colhead{(6)} &
\colhead{(7)} &
\colhead{(8)} &
\colhead{(9)} &
\colhead{(10)}
}
\startdata
          1383&        HD 1383&   18.09&   18.15&   18.21&  CLMS04&   $\phm{-}  0.61\pm 0.04$&       $ -0.148 \pm   0.014$&$\phm{-}  0.034$&$\phm{-}   0.35$\\
          5394&   $\gamma$ Cas&   16.72&   16.76&   16.80&   MJC98&   $\phm{-}  0.52\pm 0.04$&       $ -0.127 \pm   0.015$&       $ -0.032$&       $  -0.32$\\
         12323&       HD 12323&   17.96&   18.02&   18.08&  CLMS04&   $\phm{-}  0.52\pm 0.04$&       $ -0.128 \pm   0.014$&$\phm{-}  0.103$&$\phm{-}   1.08$\\
         13268&       HD 13268&   18.07&   18.13&   18.19&  CLMS04&   $\phm{-}  0.51\pm 0.04$&       $ -0.126 \pm   0.015$&$\phm{-}  0.079$&$\phm{-}   0.82$\\
         14434&       HD 14434&   18.10&   18.18&   18.26&  CLMS04&   $\phm{-}  0.52\pm 0.04$&       $ -0.127 \pm   0.015$&$\phm{-}  0.080$&$\phm{-}   0.73$\\
         22586&       HD 22586&   15.45&$\infty$&$\infty$&    JW96&   $\phm{-}  0.34\pm 0.07$&       $ -0.087 \pm   0.023$&         \nodata&         \nodata\\
         23180&         o  Per&   17.82&   17.93&   18.02&   MJC98&   $\phm{-}  0.84\pm 0.06$&       $ -0.200 \pm   0.022$&$\phm{-}  0.180$&$\phm{-}   1.24$\\
         23478&       HD 23478&   17.70&   17.79&   17.88&   C++08&          $ -0.00\pm 0.50$&       $ -0.010 \pm   0.116$&       $ -0.006$&       $  -0.03$\\
\enddata
\tablenotetext{a}{Terms in parentheses indicate separate velocity components, if they are explicitly identified
and not grouped together; see \S\protect\ref{other_considerations}}
\tablenotetext{b}{Codes in this column are linked to references
listed in Table~\protect\ref{fvals}}
\tablenotetext{c}{The expected depletion $[{\rm O}_{\rm gas}/{\rm H}]$
computed using Eq.~\protect\ref{better_equation}.
The listed errors do not include an overall systematic uncertainty of
0.05 in the solar abundance $\sigma({\rm O}/{\rm H})_\odot$
in order to show just the formal error that arises from the
uncertainties in location of line of best fit and the value of $F_*$.}
\tablenotetext{d}{The observed $[{\rm O}_{\rm gas}/{\rm H}]$ minus 
that computed using Eq.~\protect\ref{better_equation}.}
\tablenotetext{e}{The difference shown in the previous column divided by the expected
difference due to the uncertainties in both the measured column density and the
coefficients that appear in Eq.~\protect\ref{better_equation}.}
\end{deluxetable}
\begin{deluxetable}{
r  
c  
c  
c  
c  
c  
c  
c  
c  
c  
}
\tabletypesize{\footnotesize}
\tablecolumns{10}
\tablewidth{0pt}
\tablecaption{Observations and Fits for Magnesium\label{Mg}}
\tablehead{
\colhead{} & \colhead{} & \multicolumn{4}{c}{Observed $\log N$(Mg II)} & \colhead{} & \colhead{} &
 \colhead{} & \colhead{Deviation\tablenotemark{e}}\\
\cline{3-6}
\colhead{HD Number\tablenotemark{a}} &
\colhead{Name} &
\colhead{l.l.} &
\colhead{best} &
\colhead{u.l.} &
\colhead{Source\tablenotemark{b}} &
\colhead{$F_*$} &
\colhead{[Mg$_{\rm gas}$/H]$_{\rm fit}$\tablenotemark{c}} &
\colhead{Residual\tablenotemark{d}} &
\colhead{(in $\sigma$)}\\
\colhead{(1)} &
\colhead{(2)} &
\colhead{(3)} &
\colhead{(4)} &
\colhead{(5)} &
\colhead{(6)} &
\colhead{(7)} &
\colhead{(8)} &
\colhead{(9)} &
\colhead{(10)}
}
\startdata
          1383&        HD 1383&   16.27&   16.36&   16.45&  CLMS06&   $\phm{-}  0.61\pm 0.04$&       $ -0.880 \pm   0.044$&$\phm{-}  0.115$&$\phm{-}   0.92$\\
          2905&   $\kappa$ Cas&   15.93&   16.00&   16.42&   JSS86&   $\phm{-}  0.58\pm 0.06$&       $ -0.849 \pm   0.059$&       $ -0.060$&       $  -0.21$\\
          5394&   $\gamma$ Cas&   14.86&   14.91&   14.96&   JSS86&   $\phm{-}  0.52\pm 0.04$&       $ -0.784 \pm   0.043$&       $ -0.085$&       $  -0.76$\\
         12323&       HD 12323&   15.99&   16.04&   16.09&  CLMS06&   $\phm{-}  0.52\pm 0.04$&       $ -0.792 \pm   0.037$&       $ -0.074$&       $  -0.77$\\
         13268&       HD 13268&   16.19&   16.24&   16.29&  CLMS06&   $\phm{-}  0.51\pm 0.04$&       $ -0.783 \pm   0.041$&       $ -0.015$&       $  -0.15$\\
         13745&       HD 13745&   16.11&   16.18&   16.24&   JS07b&   $\phm{-}  0.43\pm 0.07$&       $ -0.694 \pm   0.069$&       $ -0.120$&       $  -0.99$\\
         14434&       HD 14434&   16.22&   16.27&   16.32&  CLMS06&   $\phm{-}  0.52\pm 0.04$&       $ -0.787 \pm   0.043$&       $ -0.031$&       $  -0.31$\\
         15137&       HD 15137&   16.13&   16.15&   16.17&   JS07b&   $\phm{-}  0.37\pm 0.09$&       $ -0.636 \pm   0.089$&       $ -0.055$&       $  -0.36$\\
\enddata
\tablenotetext{a}{Terms in parentheses indicate separate velocity components, if they are explicitly identified
and not grouped together; see \S\protect\ref{other_considerations}}
\tablenotetext{b}{Codes in this column are linked to references
listed in Table~\protect\ref{fvals}}
\tablenotetext{c}{The expected depletion $[{\rm Mg}_{\rm gas}/{\rm H}]$
computed using Eq.~\protect\ref{better_equation}.
The listed errors do not include an overall systematic uncertainty of
0.02 in the solar abundance $\sigma({\rm Mg}/{\rm H})_\odot$
in order to show just the formal error that arises from the
uncertainties in location of line of best fit and the value of $F_*$.}
\tablenotetext{d}{The observed $[{\rm Mg}_{\rm gas}/{\rm H}]$ minus 
that computed using Eq.~\protect\ref{better_equation}.}
\tablenotetext{e}{The difference shown in the previous column divided by the expected
difference due to the uncertainties in both the measured column density and the
coefficients that appear in Eq.~\protect\ref{better_equation}.}
\end{deluxetable}
\begin{deluxetable}{
r  
c  
c  
c  
c  
c  
c  
c  
c  
c  
}
\tabletypesize{\footnotesize}
\tablecolumns{10}
\tablewidth{0pt}
\tablecaption{Observations and Fits for Silicon\label{Si}}
\tablehead{
\colhead{} & \colhead{} & \multicolumn{4}{c}{Observed $\log N$(Si II)} & \colhead{} & \colhead{} &
 \colhead{} & \colhead{Deviation\tablenotemark{e}}\\
\cline{3-6}
\colhead{HD Number\tablenotemark{a}} &
\colhead{Name} &
\colhead{l.l.} &
\colhead{best} &
\colhead{u.l.} &
\colhead{Source\tablenotemark{b}} &
\colhead{$F_*$} &
\colhead{[Si$_{\rm gas}$/H]$_{\rm fit}$\tablenotemark{c}} &
\colhead{Residual\tablenotemark{d}} &
\colhead{(in $\sigma$)}\\
\colhead{(1)} &
\colhead{(2)} &
\colhead{(3)} &
\colhead{(4)} &
\colhead{(5)} &
\colhead{(6)} &
\colhead{(7)} &
\colhead{(8)} &
\colhead{(9)} &
\colhead{(10)}
}
\startdata
         18100&       HD 18100&   15.31&   15.34&   15.37&   SS96b&   $\phm{-}  0.14\pm 0.04$&       $ -0.378 \pm   0.056$&       $ -0.046$&       $  -0.31$\\
         22586&       HD 22586&   15.26&   15.31&   15.51&    JW96&   $\phm{-}  0.34\pm 0.07$&       $ -0.610 \pm   0.078$&       $ -0.053$&       $  -0.24$\\
         27778&         62 Tau& \nodata& \nodata&   15.94&   M++07&                   \nodata&              \nodata       &         \nodata&         \nodata\\
         34029&   $\alpha$ Aur&   13.16&   13.20&   13.24&  WRLS02&   $\phm{-}  0.44\pm 0.05$&       $ -0.722 \pm   0.062$&$\phm{-}  0.076$&$\phm{-}   0.64$\\
         35149&         23 Ori&   15.36&   15.44&   15.51&   W++99&   $\phm{-}  0.54\pm 0.04$&       $ -0.835 \pm   0.050$&$\phm{-}  0.034$&$\phm{-}   0.21$\\
         37021&$\theta^1$ Ori &   16.04&   16.16&   16.24&   M++07&                   \nodata&              \nodata       &        \nodata &        \nodata \\
         37061&      $\nu$ Ori&   16.10&   16.16&   16.21&   M++07&                   \nodata&              \nodata       &        \nodata &        \nodata \\
         38666&      $\mu$ Col&   15.14&   15.16&   15.18&   HSF99&   $\phm{-}  0.11\pm 0.01$&       $ -0.349 \pm   0.026$&$\phm{-}  0.040$&$\phm{-}   1.05$\\
\enddata
\tablenotetext{a}{Terms in parentheses indicate separate velocity components, if they are explicitly identified
and not grouped together; see \S\protect\ref{other_considerations}}
\tablenotetext{b}{Codes in this column are linked to references
listed in Table~\protect\ref{fvals}}
\tablenotetext{c}{The expected depletion $[{\rm Si}_{\rm gas}/{\rm H}]$
computed using Eq.~\protect\ref{better_equation}.
The listed errors do not include an overall systematic uncertainty of
0.02 in the solar abundance $\sigma({\rm Si}/{\rm H})_\odot$
in order to show just the formal error that arises from the
uncertainties in location of line of best fit and the value of $F_*$.}
\tablenotetext{d}{The observed $[{\rm Si}_{\rm gas}/{\rm H}]$ minus 
that computed using Eq.~\protect\ref{better_equation}.}
\tablenotetext{e}{The difference shown in the previous column divided by the expected
difference due to the uncertainties in both the measured column density and the
coefficients that appear in Eq.~\protect\ref{better_equation}.}
\end{deluxetable}
\begin{deluxetable}{
r  
c  
c  
c  
c  
c  
c  
c  
c  
c  
}
\tabletypesize{\footnotesize}
\tablecolumns{10}
\tablewidth{0pt}
\tablecaption{Observations and Fits for Phosphorus\label{P}}
\tablehead{
\colhead{} & \colhead{} & \multicolumn{4}{c}{Observed $\log N$(P II)} & \colhead{} & \colhead{} &
 \colhead{} & \colhead{Deviation\tablenotemark{e}}\\
\cline{3-6}
\colhead{HD Number\tablenotemark{a}} &
\colhead{Name} &
\colhead{l.l.} &
\colhead{best} &
\colhead{u.l.} &
\colhead{Source\tablenotemark{b}} &
\colhead{$F_*$} &
\colhead{[P$_{\rm gas}$/H]$_{\rm fit}$\tablenotemark{c}} &
\colhead{Residual\tablenotemark{d}} &
\colhead{(in $\sigma$)}\\
\colhead{(1)} &
\colhead{(2)} &
\colhead{(3)} &
\colhead{(4)} &
\colhead{(5)} &
\colhead{(6)} &
\colhead{(7)} &
\colhead{(8)} &
\colhead{(9)} &
\colhead{(10)}
}
\startdata
          2905&   $\kappa$ Cas&   14.43&   14.78&$\infty$&   JSS86&   $\phm{-}  0.58\pm 0.06$&       $ -0.253 \pm   0.057$&         \nodata&         \nodata\\
          5394&   $\gamma$ Cas&   13.33&   13.41&   13.51&   JSS86&   $\phm{-}  0.52\pm 0.04$&       $ -0.192 \pm   0.042$&       $ -0.100$&       $  -0.75$\\
         18100&       HD 18100&   13.65&   13.72&   13.79&   SS96b&   $\phm{-}  0.14\pm 0.04$&$\phm{-}  0.167 \pm   0.047$&       $ -0.138$&       $  -0.88$\\
         21278&       HD 21278&   13.29&   13.54&$\infty$&   JSS86&                   \nodata&              \nodata       &         \nodata&         \nodata\\
         21856&       HD 21856&   14.08&   14.38&$\infty$&   JSS86&   $\phm{-}  0.67\pm 0.39$&       $ -0.340 \pm   0.368$&         \nodata&         \nodata\\
         22928&   $\delta$ Per&   13.31&   13.61&   13.78&   JSS86&                   \nodata&              \nodata       &        \nodata &        \nodata \\
         22951&         40 Per&   13.85&   14.03&   14.15&   JSS86&   $\phm{-}  0.73\pm 0.05$&       $ -0.396 \pm   0.051$&       $ -0.341$&       $  -1.78$\\
         23180&         o  Per&   14.22&   14.43&   14.64&     S76&   $\phm{-}  0.84\pm 0.06$&       $ -0.501 \pm   0.065$&$\phm{-}  0.199$&$\phm{-}   0.82$\\
\enddata
\tablenotetext{a}{Terms in parentheses indicate separate velocity components, if they are explicitly identified
and not grouped together; see \S\protect\ref{other_considerations}}
\tablenotetext{b}{Codes in this column are linked to references
listed in Table~\protect\ref{fvals}}
\tablenotetext{c}{The expected depletion $[{\rm P}_{\rm gas}/{\rm H}]$
computed using Eq.~\protect\ref{better_equation}.
The listed errors do not include an overall systematic uncertainty of
0.04 in the solar abundance $\sigma({\rm P}/{\rm H})_\odot$
in order to show just the formal error that arises from the
uncertainties in location of line of best fit and the value of $F_*$.}
\tablenotetext{d}{The observed $[{\rm P}_{\rm gas}/{\rm H}]$ minus 
that computed using Eq.~\protect\ref{better_equation}.}
\tablenotetext{e}{The difference shown in the previous column divided by the expected
difference due to the uncertainties in both the measured column density and the
coefficients that appear in Eq.~\protect\ref{better_equation}.}
\end{deluxetable}
\begin{deluxetable}{
r  
c  
c  
c  
c  
c  
c  
c  
c  
c  
}
\tabletypesize{\footnotesize}
\tablecolumns{10}
\tablewidth{0pt}
\tablecaption{Observations and Fits for Sulfur\label{S}}
\tablehead{
\colhead{} & \colhead{} & \multicolumn{4}{c}{Observed $\log N$(S II)} & \colhead{} & \colhead{} &
 \colhead{} & \colhead{Deviation\tablenotemark{e}}\\
\cline{3-6}
\colhead{HD Number\tablenotemark{a}} &
\colhead{Name} &
\colhead{l.l.} &
\colhead{best} &
\colhead{u.l.} &
\colhead{Source\tablenotemark{b}} &
\colhead{$F_*$} &
\colhead{[S$_{\rm gas}$/H]$_{\rm fit}$\tablenotemark{c}} &
\colhead{Residual\tablenotemark{d}} &
\colhead{(in $\sigma$)}\\
\colhead{(1)} &
\colhead{(2)} &
\colhead{(3)} &
\colhead{(4)} &
\colhead{(5)} &
\colhead{(6)} &
\colhead{(7)} &
\colhead{(8)} &
\colhead{(9)} &
\colhead{(10)}
}
\startdata
         22586&       HD 22586&   15.05&   15.10&   15.30&    JW96&   $\phm{-}  0.34\pm 0.07$&       $ -0.186 \pm   0.091$&       $ -0.333$&       $  -1.47$\\
         22928&   $\delta$ Per&   14.75&   14.82&   14.88&    MY82&                   \nodata&              \nodata       &        \nodata &        \nodata \\
         34029&   $\alpha$ Aur& \nodata& \nodata&   13.59&  WRLS02&   $\phm{-}  0.44\pm 0.05$&       $ -0.312 \pm   0.081$&         \nodata&         \nodata\\
         38666&      $\mu$ Col&   15.19&   15.21&   15.23&   HSF99&   $\phm{-}  0.11\pm 0.01$&$\phm{-}  0.102 \pm   0.028$&       $ -0.010$&       $  -0.26$\\
         44743&    $\beta$ CMa&   14.50&   14.54&   14.58&    DG98&          $ -0.43\pm 0.03$&$\phm{-}  0.790 \pm   0.111$&$\phm{-}  0.186$&$\phm{-}   1.41$\\
         49798&       HD 49798&   15.25&   15.30&   15.50&    JW96&                   \nodata&              \nodata       &        \nodata &        \nodata \\
         66811&    $\zeta$ Pup&   15.17&   15.21&   15.25&     M78&   $\phm{-}  0.32\pm 0.02$&       $ -0.164 \pm   0.042$&$\phm{-}  0.158$&$\phm{-}   2.24$\\
         68273& $\gamma^2$ Vel&   14.95&   15.02&   15.11&    FS94&   $\phm{-}  0.25\pm 0.02$&       $ -0.074 \pm   0.037$&$\phm{-}  0.120$&$\phm{-}   1.24$\\
\enddata
\tablenotetext{a}{Terms in parentheses indicate separate velocity components, if they are explicitly identified
and not grouped together; see \S\protect\ref{other_considerations}}
\tablenotetext{b}{Codes in this column are linked to references
listed in Table~\protect\ref{fvals}}
\tablenotetext{c}{The expected depletion $[{\rm S}_{\rm gas}/{\rm H}]$
computed using Eq.~\protect\ref{better_equation}.
The listed errors do not include an overall systematic uncertainty of
0.04 in the solar abundance $\sigma({\rm S}/{\rm H})_\odot$
in order to show just the formal error that arises from the
uncertainties in location of line of best fit and the value of $F_*$.}
\tablenotetext{d}{The observed $[{\rm S}_{\rm gas}/{\rm H}]$ minus 
that computed using Eq.~\protect\ref{better_equation}.}
\tablenotetext{e}{The difference shown in the previous column divided by the expected
difference due to the uncertainties in both the measured column density and the
coefficients that appear in Eq.~\protect\ref{better_equation}.}
\end{deluxetable}
\begin{deluxetable}{
r  
c  
c  
c  
c  
c  
c  
c  
c  
c  
}
\tabletypesize{\footnotesize}
\tablecolumns{10}
\tablewidth{0pt}
\tablecaption{Observations and Fits for Chlorine\label{Cl}}
\tablehead{
\colhead{} & \colhead{} & \multicolumn{4}{c}{Observed $\log N$(Cl II)} & \colhead{} & \colhead{} &
 \colhead{} & \colhead{Deviation\tablenotemark{e}}\\
\cline{3-6}
\colhead{HD Number\tablenotemark{a}} &
\colhead{Name} &
\colhead{l.l.} &
\colhead{best} &
\colhead{u.l.} &
\colhead{Source\tablenotemark{b}} &
\colhead{$F_*$} &
\colhead{[Cl$_{\rm gas}$/H]$_{\rm fit}$\tablenotemark{c}} &
\colhead{Residual\tablenotemark{d}} &
\colhead{(in $\sigma$)}\\
\colhead{(1)} &
\colhead{(2)} &
\colhead{(3)} &
\colhead{(4)} &
\colhead{(5)} &
\colhead{(6)} &
\colhead{(7)} &
\colhead{(8)} &
\colhead{(9)} &
\colhead{(10)}
}
\startdata
          2905&   $\kappa$ Cas&   13.85&   14.21&   15.11&   JSS86&   $\phm{-}  0.58\pm 0.06$&       $ -0.280 \pm   0.076$&       $ -0.134$&       $  -0.21$\\
          5394&   $\gamma$ Cas&   12.62&   13.05&   13.25&   JSS86&   $\phm{-}  0.52\pm 0.04$&       $ -0.199 \pm   0.059$&       $ -0.243$&       $  -0.73$\\
         21278&       HD 21278&   14.25&   14.55&$\infty$&   JSS86&                   \nodata&              \nodata       &         \nodata&         \nodata\\
         21856&       HD 21856& \nodata&   13.48&   13.95&   JSS86&   $\phm{-}  0.67\pm 0.39$&       $ -0.394 \pm   0.484$&         \nodata&         \nodata\\
         22928&   $\delta$ Per& \nodata& \nodata&   12.73&   JSS86&                   \nodata&              \nodata       &         \nodata&         \nodata\\
         22951&         40 Per&   14.24&   14.33&   14.54&   JSS86&   $\phm{-}  0.73\pm 0.05$&       $ -0.467 \pm   0.070$&$\phm{-}  0.241$&$\phm{-}   1.23$\\
         23180&         o  Per& \nodata&   13.41&   13.73&   JSS86&   $\phm{-}  0.84\pm 0.06$&       $ -0.606 \pm   0.090$&         \nodata&         \nodata\\
         23408&         20 Tau&   12.76&   14.06&   15.14&   JSS86&                   \nodata&              \nodata       &        \nodata &        \nodata \\
\enddata
\tablenotetext{a}{Terms in parentheses indicate separate velocity components, if they are explicitly identified
and not grouped together; see \S\protect\ref{other_considerations}}
\tablenotetext{b}{Codes in this column are linked to references
listed in Table~\protect\ref{fvals}}
\tablenotetext{c}{The expected depletion $[{\rm Cl}_{\rm gas}/{\rm H}]$
computed using Eq.~\protect\ref{better_equation}.
The listed errors do not include an overall systematic uncertainty of
0.06 in the solar abundance $\sigma({\rm Cl}/{\rm H})_\odot$
in order to show just the formal error that arises from the
uncertainties in location of line of best fit and the value of $F_*$.}
\tablenotetext{d}{The observed $[{\rm Cl}_{\rm gas}/{\rm H}]$ minus 
that computed using Eq.~\protect\ref{better_equation}.}
\tablenotetext{e}{The difference shown in the previous column divided by the expected
difference due to the uncertainties in both the measured column density and the
coefficients that appear in Eq.~\protect\ref{better_equation}.}
\end{deluxetable}
\begin{deluxetable}{
r  
c  
c  
c  
c  
c  
c  
c  
c  
c  
}
\tabletypesize{\footnotesize}
\tablecolumns{10}
\tablewidth{0pt}
\tablecaption{Observations and Fits for Titanium\label{Ti}}
\tablehead{
\colhead{} & \colhead{} & \multicolumn{4}{c}{Observed $\log N$(Ti II)} & \colhead{} & \colhead{} &
 \colhead{} & \colhead{Deviation\tablenotemark{e}}\\
\cline{3-6}
\colhead{HD Number\tablenotemark{a}} &
\colhead{Name} &
\colhead{l.l.} &
\colhead{best} &
\colhead{u.l.} &
\colhead{Source\tablenotemark{b}} &
\colhead{$F_*$} &
\colhead{[Ti$_{\rm gas}$/H]$_{\rm fit}$\tablenotemark{c}} &
\colhead{Residual\tablenotemark{d}} &
\colhead{(in $\sigma$)}\\
\colhead{(1)} &
\colhead{(2)} &
\colhead{(3)} &
\colhead{(4)} &
\colhead{(5)} &
\colhead{(6)} &
\colhead{(7)} &
\colhead{(8)} &
\colhead{(9)} &
\colhead{(10)}
}
\startdata
          2905&   $\kappa$ Cas&   12.03&   12.09&   12.15&     S78&   $\phm{-}  0.58\pm 0.06$&       $ -2.266 \pm   0.120$&$\phm{-}  0.069$&$\phm{-}   0.39$\\
          5394&   $\gamma$ Cas&   11.01&   11.09&   11.17&     S78&   $\phm{-}  0.52\pm 0.04$&       $ -2.133 \pm   0.087$&$\phm{-}  0.067$&$\phm{-}   0.45$\\
         18100&       HD 18100& \nodata& \nodata&   10.93&   A++93&   $\phm{-}  0.14\pm 0.04$&       $ -1.356 \pm   0.094$&         \nodata&         \nodata\\
         22586&       HD 22586&   11.38&   11.58&   11.78&   JW96*&   $\phm{-}  0.34\pm 0.07$&       $ -1.773 \pm   0.136$&       $ -0.006$&       $  -0.02$\\
         22928&   $\delta$ Per&   10.45&   10.62&   10.89&     S78&                   \nodata&              \nodata       &        \nodata &        \nodata \\
         22951&         40 Per&   11.70&   11.77&   11.84&     S78&   $\phm{-}  0.73\pm 0.05$&       $ -2.575 \pm   0.107$&$\phm{-}  0.126$&$\phm{-}   0.75$\\
         23180&         o  Per&   11.30&   11.43&   11.56&     S78&   $\phm{-}  0.84\pm 0.06$&       $ -2.804 \pm   0.136$&$\phm{-}  0.045$&$\phm{-}   0.21$\\
         23630&     $\eta$ Tau&   10.69&   10.84&   10.99&     S78&                   \nodata&              \nodata       &        \nodata &        \nodata \\
\enddata
\tablenotetext{a}{Terms in parentheses indicate separate velocity components, if they are explicitly identified
and not grouped together; see \S\protect\ref{other_considerations}}
\tablenotetext{b}{Codes in this column are linked to references
listed in Table~\protect\ref{fvals}}
\tablenotetext{c}{The expected depletion $[{\rm Ti}_{\rm gas}/{\rm H}]$
computed using Eq.~\protect\ref{better_equation}.
The listed errors do not include an overall systematic uncertainty of
0.03 in the solar abundance $\sigma({\rm Ti}/{\rm H})_\odot$
in order to show just the formal error that arises from the
uncertainties in location of line of best fit and the value of $F_*$.}
\tablenotetext{d}{The observed $[{\rm Ti}_{\rm gas}/{\rm H}]$ minus 
that computed using Eq.~\protect\ref{better_equation}.}
\tablenotetext{e}{The difference shown in the previous column divided by the expected
difference due to the uncertainties in both the measured column density and the
coefficients that appear in Eq.~\protect\ref{better_equation}.}
\end{deluxetable}
\begin{deluxetable}{
r  
c  
c  
c  
c  
c  
c  
c  
c  
c  
}
\tabletypesize{\footnotesize}
\tablecolumns{10}
\tablewidth{0pt}
\tablecaption{Observations and Fits for Chromium\label{Cr}}
\tablehead{
\colhead{} & \colhead{} & \multicolumn{4}{c}{Observed $\log N$(Cr II)} & \colhead{} & \colhead{} &
 \colhead{} & \colhead{Deviation\tablenotemark{e}}\\
\cline{3-6}
\colhead{HD Number\tablenotemark{a}} &
\colhead{Name} &
\colhead{l.l.} &
\colhead{best} &
\colhead{u.l.} &
\colhead{Source\tablenotemark{b}} &
\colhead{$F_*$} &
\colhead{[Cr$_{\rm gas}$/H]$_{\rm fit}$\tablenotemark{c}} &
\colhead{Residual\tablenotemark{d}} &
\colhead{(in $\sigma$)}\\
\colhead{(1)} &
\colhead{(2)} &
\colhead{(3)} &
\colhead{(4)} &
\colhead{(5)} &
\colhead{(6)} &
\colhead{(7)} &
\colhead{(8)} &
\colhead{(9)} &
\colhead{(10)}
}
\startdata
         18100&       HD 18100&   12.87&   12.91&   12.95&   SS96b&   $\phm{-}  0.14\pm 0.04$&       $ -1.024 \pm   0.071$&$\phm{-}  0.061$&$\phm{-}   0.39$\\
         23630&     $\eta$ Tau&   11.81&   11.86&   11.91&    RB95&                   \nodata&              \nodata       &        \nodata &        \nodata \\
         24912&      $\xi$ Per&   12.88&   12.90&   12.91&   C++91&   $\phm{-}  0.83\pm 0.02$&       $ -2.030 \pm   0.047$&       $ -0.080$&       $  -0.88$\\
         35149&         23 Ori&   12.73&   12.74&   12.75&   W++99&   $\phm{-}  0.54\pm 0.04$&       $ -1.607 \pm   0.059$&       $ -0.004$&       $  -0.02$\\
         36486& $\delta$ Ori A&   12.04&   12.09&   12.14&    RB95&   $\phm{-}  0.54\pm 0.02$&       $ -1.609 \pm   0.036$&       $ -0.213$&       $  -2.89$\\
         37128& $\epsilon$ Ori&   12.40&   12.45&   12.50&    RB95&   $\phm{-}  0.54\pm 0.03$&       $ -1.609 \pm   0.048$&       $ -0.113$&       $  -1.00$\\
         38666&      $\mu$ Col&   12.60&   12.63&   12.66&   HSF99&   $\phm{-}  0.11\pm 0.01$&       $ -0.988 \pm   0.034$&$\phm{-}  0.036$&$\phm{-}   0.73$\\
         47839&         15 Mon&   12.81&   12.86&   12.91&    RB95&   $\phm{-}  0.25\pm 0.05$&       $ -1.187 \pm   0.072$&$\phm{-}  0.015$&$\phm{-}   0.11$\\
\enddata
\tablenotetext{a}{Terms in parentheses indicate separate velocity components, if they are explicitly identified
and not grouped together; see \S\protect\ref{other_considerations}}
\tablenotetext{b}{Codes in this column are linked to references
listed in Table~\protect\ref{fvals}}
\tablenotetext{c}{The expected depletion $[{\rm Cr}_{\rm gas}/{\rm H}]$
computed using Eq.~\protect\ref{better_equation}.
The listed errors do not include an overall systematic uncertainty of
0.05 in the solar abundance $\sigma({\rm Cr}/{\rm H})_\odot$
in order to show just the formal error that arises from the
uncertainties in location of line of best fit and the value of $F_*$.}
\tablenotetext{d}{The observed $[{\rm Cr}_{\rm gas}/{\rm H}]$ minus 
that computed using Eq.~\protect\ref{better_equation}.}
\tablenotetext{e}{The difference shown in the previous column divided by the expected
difference due to the uncertainties in both the measured column density and the
coefficients that appear in Eq.~\protect\ref{better_equation}.}
\end{deluxetable}
\begin{deluxetable}{
r  
c  
c  
c  
c  
c  
c  
c  
c  
c  
}
\tabletypesize{\footnotesize}
\tablecolumns{10}
\tablewidth{0pt}
\tablecaption{Observations and Fits for Manganese\label{Mn}}
\tablehead{
\colhead{} & \colhead{} & \multicolumn{4}{c}{Observed $\log N$(Mn II)} & \colhead{} & \colhead{} &
 \colhead{} & \colhead{Deviation\tablenotemark{e}}\\
\cline{3-6}
\colhead{HD Number\tablenotemark{a}} &
\colhead{Name} &
\colhead{l.l.} &
\colhead{best} &
\colhead{u.l.} &
\colhead{Source\tablenotemark{b}} &
\colhead{$F_*$} &
\colhead{[Mn$_{\rm gas}$/H]$_{\rm fit}$\tablenotemark{c}} &
\colhead{Residual\tablenotemark{d}} &
\colhead{(in $\sigma$)}\\
\colhead{(1)} &
\colhead{(2)} &
\colhead{(3)} &
\colhead{(4)} &
\colhead{(5)} &
\colhead{(6)} &
\colhead{(7)} &
\colhead{(8)} &
\colhead{(9)} &
\colhead{(10)}
}
\startdata
          1383&        HD 1383&   13.54&   13.60&   13.66&  CLMS06&   $\phm{-}  0.61\pm 0.04$&       $ -1.432 \pm   0.039$&       $ -0.054$&       $  -0.52$\\
          2905&   $\kappa$ Cas&   13.36&   13.41&   13.85&   JSS86&   $\phm{-}  0.58\pm 0.06$&       $ -1.406 \pm   0.051$&       $ -0.058$&       $  -0.21$\\
          5394&   $\gamma$ Cas&   12.44&   12.50&   12.57&   JSS86&   $\phm{-}  0.52\pm 0.04$&       $ -1.350 \pm   0.037$&$\phm{-}  0.108$&$\phm{-}   0.92$\\
         12323&       HD 12323&   13.45&   13.53&   13.61&  CLMS06&   $\phm{-}  0.52\pm 0.04$&       $ -1.357 \pm   0.033$&$\phm{-}  0.019$&$\phm{-}   0.17$\\
         13268&       HD 13268&   13.55&   13.59&   13.63&  CLMS06&   $\phm{-}  0.51\pm 0.04$&       $ -1.349 \pm   0.036$&       $ -0.060$&       $  -0.66$\\
         14434&       HD 14434&   13.53&   13.60&   13.67&  CLMS06&   $\phm{-}  0.52\pm 0.04$&       $ -1.352 \pm   0.038$&       $ -0.097$&       $  -0.90$\\
         18100&       HD 18100&   12.71&   12.74&   12.77&   SS96b&   $\phm{-}  0.14\pm 0.04$&       $ -1.025 \pm   0.043$&$\phm{-}  0.036$&$\phm{-}   0.25$\\
         21278&       HD 21278& \nodata&   11.90&   14.15&   JSS86&                   \nodata&              \nodata       &         \nodata&         \nodata\\
\enddata
\tablenotetext{a}{Terms in parentheses indicate separate velocity components, if they are explicitly identified
and not grouped together; see \S\protect\ref{other_considerations}}
\tablenotetext{b}{Codes in this column are linked to references
listed in Table~\protect\ref{fvals}}
\tablenotetext{c}{The expected depletion $[{\rm Mn}_{\rm gas}/{\rm H}]$
computed using Eq.~\protect\ref{better_equation}.
The listed errors do not include an overall systematic uncertainty of
0.03 in the solar abundance $\sigma({\rm Mn}/{\rm H})_\odot$
in order to show just the formal error that arises from the
uncertainties in location of line of best fit and the value of $F_*$.}
\tablenotetext{d}{The observed $[{\rm Mn}_{\rm gas}/{\rm H}]$ minus 
that computed using Eq.~\protect\ref{better_equation}.}
\tablenotetext{e}{The difference shown in the previous column divided by the expected
difference due to the uncertainties in both the measured column density and the
coefficients that appear in Eq.~\protect\ref{better_equation}.}
\end{deluxetable}
\begin{deluxetable}{
r  
c  
c  
c  
c  
c  
c  
c  
c  
c  
}
\tabletypesize{\footnotesize}
\tablecolumns{10}
\tablewidth{0pt}
\tablecaption{Observations and Fits for Iron\label{Fe}}
\tablehead{
\colhead{} & \colhead{} & \multicolumn{4}{c}{Observed $\log N$(Fe II)} & \colhead{} & \colhead{} &
 \colhead{} & \colhead{Deviation\tablenotemark{e}}\\
\cline{3-6}
\colhead{HD Number\tablenotemark{a}} &
\colhead{Name} &
\colhead{l.l.} &
\colhead{best} &
\colhead{u.l.} &
\colhead{Source\tablenotemark{b}} &
\colhead{$F_*$} &
\colhead{[Fe$_{\rm gas}$/H]$_{\rm fit}$\tablenotemark{c}} &
\colhead{Residual\tablenotemark{d}} &
\colhead{(in $\sigma$)}\\
\colhead{(1)} &
\colhead{(2)} &
\colhead{(3)} &
\colhead{(4)} &
\colhead{(5)} &
\colhead{(6)} &
\colhead{(7)} &
\colhead{(8)} &
\colhead{(9)} &
\colhead{(10)}
}
\startdata
          2905&   $\kappa$ Cas&   14.93&   15.01&   15.46&   JSS86&   $\phm{-}  0.58\pm 0.06$&       $ -1.698 \pm   0.076$&       $ -0.126$&       $  -0.43$\\
          5394&   $\gamma$ Cas&   14.01&   14.11&   14.21&   JSS86&   $\phm{-}  0.52\pm 0.04$&       $ -1.614 \pm   0.055$&$\phm{-}  0.022$&$\phm{-}   0.15$\\
         12323&       HD 12323&   15.09&   15.16&   15.23&   JS07a&   $\phm{-}  0.52\pm 0.04$&       $ -1.624 \pm   0.048$&       $ -0.039$&       $  -0.35$\\
         13745&       HD 13745&   15.45&   15.53&   15.62&   JS07a&   $\phm{-}  0.43\pm 0.07$&       $ -1.498 \pm   0.089$&$\phm{-}  0.115$&$\phm{-}   0.80$\\
         15137&       HD 15137&   15.31&   15.39&   15.49&   JS07a&   $\phm{-}  0.37\pm 0.09$&       $ -1.422 \pm   0.115$&$\phm{-}  0.054$&$\phm{-}   0.28$\\
         18100&       HD 18100&   14.63&   14.67&   14.71&   SS96b&   $\phm{-}  0.14\pm 0.04$&       $ -1.126 \pm   0.060$&$\phm{-}  0.102$&$\phm{-}   0.67$\\
         21278&       HD 21278&   14.36&   15.01&$\infty$&   JSS86&                   \nodata&              \nodata       &         \nodata&         \nodata\\
         21856&       HD 21856&   14.88&   15.06&$\infty$&   JSS86&   $\phm{-}  0.67\pm 0.39$&       $ -1.816 \pm   0.501$&         \nodata&         \nodata\\
\enddata
\tablenotetext{a}{Terms in parentheses indicate separate velocity components, if they are explicitly identified
and not grouped together; see \S\protect\ref{other_considerations}}
\tablenotetext{b}{Codes in this column are linked to references
listed in Table~\protect\ref{fvals}}
\tablenotetext{c}{The expected depletion $[{\rm Fe}_{\rm gas}/{\rm H}]$
computed using Eq.~\protect\ref{better_equation}.
The listed errors do not include an overall systematic uncertainty of
0.03 in the solar abundance $\sigma({\rm Fe}/{\rm H})_\odot$
in order to show just the formal error that arises from the
uncertainties in location of line of best fit and the value of $F_*$.}
\tablenotetext{d}{The observed $[{\rm Fe}_{\rm gas}/{\rm H}]$ minus 
that computed using Eq.~\protect\ref{better_equation}.}
\tablenotetext{e}{The difference shown in the previous column divided by the expected
difference due to the uncertainties in both the measured column density and the
coefficients that appear in Eq.~\protect\ref{better_equation}.}
\end{deluxetable}%

\clearpage
\begin{references}

\reference{6787} Abt, H. A. 1981, \apjs,  45, 437

\reference{5228} Adams, W. S. 1949, \apj,  109, 354

\reference{6805} Afflerbach, A., Churchwell, E., \& Werner, M. W. 1997, 
\apj,  478, 190

\reference{2527} Albert, C. E., Blades, J. C., Morton, D. C., Lockman, F. 
J., Proulx, M., \& Ferrarese, L. 1993, \apjs,  88, 81

\reference{2346} Allen, M. M., Jenkins, E. B., \& Snow, T. P. 1992, 
\apjs,  83, 261

\reference{6815} Allende Prieto, C. 2008, in 14th Cambridge Workshop on 
Cool Stars, Stellar Systems, and the Sun, ed. G. T. van Belle (San Francisco: 
Astr. Soc. Pacific), p. 39

\reference{5009} Allende Prieto, C., Lambert, D. L., \& Asplund, M. 2002, 
\apjl,  573, L137

\reference{5264} Aloisi, A., Savaglio, S., Heckman, T. M., Hoopes, C. G., 
Leitherer, C., \& Sembach, K. R. 2003, \apj,  595, 760

\reference{6783} Andersen, J., Clausen, J. V., Nordstrom, B., \& Popper, 
D. M. 1985, \aap,  151, 329

\reference{5144} Andr\'e, M., Oliveira, C., Howk, J. C., Ferlet, R., 
D\'esert, J. M., H\'ebrard, G., Lacour, S., Lecavelier des \'Etangs, A., 
Vidal-Madjar, A., \& Moos, H. W. 2003, \apj,  591, 1000

\reference{} Antia, H. M., \& Basu, S. 2005, \apj, 620, L129

\reference{} ---. 2006, \apj, 644, 1292

\reference{5780} Asplund, M., Grevesse, N., Sauval, A. J., Allende Prieto, 
C., \& Blomme, R. 2005, \aap, 431, 693

\reference{5499} Asplund, M., Grevesse, N., Sauval, A. J., Allende Prieto, 
C., \& Kiselman, D. 2004b, \aap,  417, 751

\reference{5789} Badnell, N. R., Bautista, M. A., Butler, K., Delahaye, F.,
 Mendoza, C., Palmeri, P., Zeippen, C. J., \& Seaton, M. J. 2005, \mnras,  
360, 458

\reference{} Bahcall, J. N., Basu, S., Pinsonneault, M., \&
Serenelli, A. M. 2005a, \apj, 618, 1049

\reference{} Bahcall, J. N., Serenelli, A. M., \& Basu, S. 2005b,
\apj, 621, L85

\reference{2257} Barker, E. S., Lugger, P. M., Weiler, E. J., \& York, D. 
G. 1984, \apj,  280, 600

\reference{4928} Bergh\"ofer, T. W., \& Breitschwerdt, D. 2002, \aap,  
390, 299

\reference{6770} Bertiau, F. C. 1958, \apj,  128, 533

\reference{6785} Bidelman, W. P. 1954, \pasp,  66, 249

\reference{6846} Boesgaard, A. M. 1985, \pasp,  97, 37

\reference{1026} Bohlin, R. C., Savage, B. D., \& Drake, J. F. 1978, 
\apj,  224, 132

\reference{1048} Bohlin, R. C., Hill, J. K., Jenkins, E. B., Savage, B. D.,
 Snow, T. P., Spitzer, L., \& York, D. G. 1983, \apjs,  51, 277

\reference{6029} Bowen, D. V., Jenkins, E. B., Pettini, M., \& Tripp, T. 
M. 2005, \apj,  635, 880

\reference{6557} Bowen, D. V., Jenkins, E. B., Tripp, T. M., Sembach, K. 
R., Savage, B. D., Moos, H. W., Oegerle, W. R., Friedman, S. D., Gry, C., 
Kruk, J. W., Murphy, E., Sankrit, R., Shull, J. M., Sonneborn, G., \& York, 
D. G. 2008, \apjs,  176, 59

\reference{6236} Breitschwerdt, D., \& de Avillez, M. A. 2006, \aap,  
452, L1

\reference{3357} Breitschwerdt, D., Egger, R., Freyberg, M. J., Frisch, P. 
C., \& Vallerga, J. V. 1996, \ssr,  78, 183

\reference{6446} Burgh, E. B., France, K., \& McCandliss, S. R. 2007, 
\apj,  658, 446

\reference{6963} Caffau, E., Maiorca, E., Bonifacio, P., Faraggiana, R., 
Steffen, M., Ludwig, H.-G., Kamp, I., \& Busso, M. 2009,, 0903.3406,

\reference{5096} Calura, F., Matteucci, F., \& Vladilo, G. 2003, \mnras, 
 340, 59

\reference{6934} Calura, F., Pipino, A., \& Matteucci, F. 2008, \aap,  
479, 669

\reference{6932} Calura, F., Dessauges-Zavadsky, M., Prochaska, J. X., \& 
Matteucci, F. 2009, \apj, 693, 1236

\reference{6972} Cannon, J. M., Skillman, E. D., Sembach, K. R., \& 
Bomans, D. J. 2005, \apj,  618, 247

\reference{2682} Cardelli, J. A. 1994, Sci,  265, 209

\reference{4254} Cardelli, J. A., \& Meyer, D. M. 1997, \apjl,  477, L57

\reference{2886} Cardelli, J. A., Sembach, K. R., \& Savage, B. D. 1995, 
\apj,  440, 241

\reference{1966} Cardelli, J. A., Savage, B. D., Bruhweiler, F. C., Smith, 
A. M., Ebbets, D. C., Sembach, K. R., \& Sofia, U. J. 1991, \apjl,  377, L57

\reference{2371} Cardelli, J. A., Mathis, J. S., Ebbets, D. C., \& Savage,
 B. D. 1993, \apjl,  402, L17

\reference{2645} Cardelli, J. A., Sofia, U. J., Savage, B. D., Keenan, F. 
P., \& Dufton, P. L. 1994, \apjl,  420, L29

\reference{3346} Cardelli, J. A., Meyer, D. M., Jura, M., \& Savage, B. D.
 1996, \apj,  467, 334

\reference{5288} Cartledge, S. I. B., Meyer, D. M., \& Lauroesch, J. T. 
2003, \apj,  597, 408

\reference{4775} Cartledge, S. I. B., Meyer, D. M., Lauroesch, J. T., \& 
Sofia, U. J. 2001, \apj,  562, 394

\reference{5599} Cartledge, S. I. B., Lauroesch, J. T., Meyer, D. M., \& 
Sofia, U. J. 2004, \apj,  613, 1037

\reference{6140} --- 2006, \apj,  641, 327

\reference{6759} Cartledge, S. I. B., Lauroesch, J. T., Meyer, D. M., 
Sofia, U. J., \& Clayton, G. C. 2008, \apj,  687, 1043

\reference{5158} Cassinelli, J. P., Cohen, D. H., MacFarlane, J. J., Drew, 
J. E., Lynas-Gray, A. E., Hoare, M. G., Vallerga, J. V., Welsh, B. Y., Vedder,
 P. W., Hubeny, I., \& Lanz, T. 1995, \apj,  438, 932

\reference{5157} Cassinelli, J. P., Cohen, D. H., MacFarlane, J. J., Drew, 
J. E., Lynas-Gray, A. E., Hubeny, I., Vallerga, J. V., Welsh, B. Y., \& 
Hoare, M. G. 1996, \apj,  460, 949

\reference{6655} Centeno, R., \& Socas-Navarro, H. 2008, \apj, 682, L61

\reference{5165} Centuri\'on, M., Molaro, P., Vladilo, G., P\'eroux, C., 
Levshakov, S. A., \& D'Odorico, V. 2003, \aap,  403, 55

\reference{5546} Chambaud, G., Launay, J. M., Levy, B., Millie, P., Roueff,
 E., \& Tran Minh, F. 1980, J. Phys. B,  13, 4205

\reference{5065} Collins, J. A., Shull, J. M., \& Giroux, M. L. 2003, 
\apj,  585, 336

\reference{6042} Costantini, E., Freyberg, M. J., \& Predehl, P. 2005, 
\aap,  444, 187

\reference{1894} Cox, D. P., \& Reynolds, R. J. 1987, \araa,  25, 303

\reference{2746} Crinklaw, G., Federman, S. R., \& Joseph, C. L. 1994, 
\apj,  424, 748

\reference{6857} Cunningham, N. J., McCray, R., \& Snow, T. P. 2004, 
\apj,  611, 353

\reference{5185} Daflon, S., Cunha, K., Smith, V. V., \& Butler, K. 2003, 
\aap,  399, 525

\reference{1409} de Boer, K. S., Lenhart, H., Van Der Hucht, K. A., 
Kamperman, T. M., Kondo, Y., \& Bruhweiler, F. C. 1986, \aap,  157, 119

\reference{4542} Deharveng, L., Pe\~na, M., Caplan, J., \& Costero, R. 
2000, \mnras,  311, 329

\reference{6943} Dessauges-Zavadsky, M., Prochaska, J. X., \& D'Odorico, 
S. 2002, \aap,  391, 801

\reference{5596} Dessauges-Zavadsky, M., Calura, F., Prochaska, J. X., 
D'Odorico, S., \& Matteucci, F. 2004, \aap,  416, 79

\reference{6099} Dessauges-Zavadsky, M., Prochaska, J. X., D'Odorico, S., 
Calura, F., \& Matteucci, F. 2006, \aap,  445, 93

\reference{6847} Dessauges-Zavadsky, M., Calura, F., Prochaska, J. X., 
D'Odorico, S., \& Matteucci, F. 2007, \aap,  470, 431

\reference{2712} Diplas, A., \& Savage, B. D. 1994, \apjs,  93, 211

\reference{5231} Draine, B. T. 2003a, \araa,  41, 241

\reference{5234} --- 2003b, \apj,  598, 1017

\reference{6808} --- 2003c, \apj,  598, 1026

\reference{5246} --- 2004, in Origin and Evolution of the Elements, ed. A. 
Mc William \& M. Rauch (Cambridge: Cambridge Univ. Press), p. 317

\reference{6955} --- 2009,  in Cosmic Dust, Near and Far,
ed. T. Henning, E. Gr\"un \& J. Steinacker (San Francisco: Ast.
 Soc. Pacific), in press (also arXiv 0903.1658)

\reference{6866} Draine, B. T., \& Tan, J. C. 2003, \apj,  594, 347

\reference{6792} Dufton, P. L., Smartt, S. J., Lee, J. K., Ryans, R. S. I.,
 Hunter, L., Evans, C. J., Herrero, A., Trundle, C., Lennon, D. J., Irwin, M. 
J., \& Daufer, A. 2006, \aap,  457, 265

\reference{6811} Dunham, T. 1939, proc am phil soc,  81, 277

\reference{4337} Dupin, O., \& Gry, C. 1998, \aap,  335, 661

\reference{4256} Dwek, E. 1997, \apj,  484, 779

\reference{4294} --- 1998, \apj,  501, 643

\reference{6816} Dwek, E., \& Scalo, J. M. 1980, \apj,  239, 193

\reference{6865} Dwek, E., Zubko, V., Arendt, R. G., \& Smith, R. K. 2004,
 in Astrophysics of Dust, ed. A. N. Witt, G. C. Clayton \& B. T. Draine (San 
Francisco: Astr. Soc. Pacific), p. 499

\reference{3427} Eiroa, C., \& Hodapp, K.-W. 1989, \aap,  210, 345

\reference{6462} Ellison, S. L., Prochaska, J. X., \& Lopez, S. 2007, 
\mnras,  380, 1245

\reference{6772} Feast, M. W. 1958, \mnras,  118, 618

\reference{6775} Feast, M. W., Thackeray, A. D., \& Wesselink, A. J. 1955,
 \memras,  67, 51

\reference{6451} Federman, S. R., Brown, M., Torok, S., Cheng, S., Irving, 
R. E., Schectman, R. M., \& Curtis, L. J. 2007, \apj,  660, 919

\reference{3804} Ferlet, R. 1999, \aapr,  9, 153

\reference{2878} Field, G. B. 1974, \apj,  187, 453

\reference{3406} Field, G. B., \& Steigman, G. 1971, \apj,  166, 59

\reference{386} Fitzpatrick, E. L. 1996, \apjl,  473, L55

\reference{5159} Fitzpatrick, E. L., \& Massa, D. 1990, \apjs,  72, 163

\reference{2701} Fitzpatrick, E. L., \& Spitzer, L. 1994, \apj,  427, 232

\reference{4207} --- 1997, \apj,  475, 623

\reference{4962} Friedman, S. D., Howk, J. C., Chayer, P., Tripp, T. M., 
H\'ebrard, G., Andr\'e, M., Oliveira, C., Jenkins, E. B., Moos, H. W., 
Oegerle, W. R., Sonneborn, G., Lamontagne, R., Sembach, K. R., \& Vidal-
Madjar, A. 2002, \apjs,  140, 37

\reference{6497} Frisch, P. C. 2007, \ssr, 130, 355

\reference{4537} Frisch, P. C., Dorschner, J. M., Geiss, J., Greenberg, J. 
M., Gr\"un, E., Landgraf, M., Hoppe, P., Jones, A. P., Kr\"atschmer, W., 
Linde, T. J., Morfill, G. E., Reach, W., Slavin, J. D., Svestka, J., Witt, A. 
N., \& Zank, G. P. 1999, \apj,  525, 492

\reference{6364} Fuchs, B., Breitschwerdt, D., de Avillez, M. A., Dettbarn,
 C., \& Flynn, C. 2006, \mnras,  373, 993

\reference{6829} Gail, H.-P., \& Sedlmayr, E. 1986, \aap,  166, 225

\reference{6784} Garrison, R. F., \& Gray, R. O. 1994, \aj,  107, 1556

\reference{6788} Garrison, R. F., \& Kormendy, J. 1976, \pasp,  88, 865

\reference{6537} Geary, R. C. 1930, J. Royal Stat. Soc.,  93, 442

\reference{2943} Gibson, B. K., Giroux, M. L., Penton, S. V., Stocke, J. 
T., Shull, J. M., \& Tumlinson, J. 2001, \aj,  122, 3280

\reference{3428} Gillett, F. C., Jones, T. W., Merrill, K. M., \& Stein, 
W. A. 1975, \aap,  45, 77

\reference{6944} Gilmore, G., \& Wyse, R. F. G. 1991, \apjl,  367, L55

\reference{4877} Giveon, U., Sternberg, A., Lutz, D., Feuchtgruber, H., \&
 Pauldrach, A. W. A. 2002, \apj,  566, 880

\reference{6928} Golay, M., Mandwewala, N., \& Bartholdi, P. 1977, \aap, 
 60, 181

\reference{6870} Greenberg, J. M. 1982, in Comets, ed. L. L. Wilkening 
(Tucson: Univ. Arizona Press), p. 131

\reference{6807} --- 1989, in Interstellar Dust, ed. L. J. Allamandola \& 
A. G. G. M. Tielens (Dordrecht: Kluwer), p. 345

\reference{5525} Grevesse, N., \& Sauval, A. J. 1998, \ssr,  85, 161

\reference{6804} Gummershach, C. A., Kaufer, A., Sch\"afer, D. R., 
Szeifert, T., \& Wolf, B. 1998, \aap,  338, 881

\reference{6809} Habing, H. 1969, \bain,  20, 177

\reference{1117} Harris, A. W., Gry, C., \& Bromage, G. E. 1984, \apj,  
284, 157

\reference{6929} Hauck, B., \& Mermilliod, M. 1998, \aaps,  129, 431

\reference{4281} H\'ebrard, G., Lemoine, M., Ferlet, R., \& Vidal-Madjar, 
A. 1997, \aap,  324, 1145

\reference{4523} H\'ebrard, G., Mallouris, C., Ferlet, R., Koester, D., 
Lemoine, M., Vidal-Madjar, A., \& York, D. 1999, \aap,  350, 643

\reference{2772} Herbig, G. H. 1968, \zap,  68, 243

\reference{3873} Hill, P. W. 1970, \mnras,  150, 23

\reference{6794} Hiltner, W. A. 1956, \apjs,  2, 389

\reference{6791} Hiltner, W. A., Garrison, R. F., \& Schild, R. E. 1969, 
\apj,  157, 313

\reference{1100} Hobbs, L. M., York, D. G., \& Oegerle, W. 1982, \apj,  
252, L21

\reference{2482} Hobbs, L. M., Welty, D. E., Morton, D. C., Spitzer, L., 
\& York, D. G. 1993, \apj,  411, 750

\reference{6771} Hoffleit, D. 1956, \apj,  124, 61

\reference{4778} Holweger, H. 2001, in Solar and Galactic Composition, A 
Joint SOHO/ACE Workshop, ed. R. F. Wimmer-Schweingruber (New York: AIP), p. 23

\reference{5094} Hoopes, C. G., Sembach, K. R., H\'ebrard, G., Moos, H. W.,
 \& Knauth, D. C. 2003, \apj,  586, 1094

\reference{4506} Howk, J. C., \& Savage, B. D. 1999, \apj,  517, 746

\reference{3692} Howk, J. C., Savage, B. D., \& Fabian, D. 1999, \apj,  
525, 253

\reference{3721} Howk, J. C., \& Sembach, K. R. 1999, \apjl,  523, L141

\reference{5056} Howk, J. C., Sembach, K. R., \& Savage, B. D. 2003, 
\apj,  586, 249

\reference{6051} --- 2006, \apj,  637, 333

\reference{6750} Ioppolo, S., Cuppen, H. M., Romanzin, C., van Dishoeck, E.
 F., \& Linnartz, H. 2008, \apj,  686, 1474

\reference{6782} Jaschek, C., \& Jaschek, M. 1992, \aaps,  95, 535

\reference{1369} Jenkins, E. B. 1970, in Ultraviolet Stellar Spectra and 
Related Ground-Based Observations, ed. L. Houziaux \& H. Butler (Dordrecht: 
Reidel), p. 281

\reference{1312} --- 1971, \apj,  169, 25

\reference{1355} --- 1986, \apj,  304, 739

\reference{1388} --- 1987, in Interstellar Processes, ed. D. J. Hollenbach 
\& H. A. Thronson Jr. (Dordrecht: Reidel), p. 533

\reference{3184} --- 1996, \apj,  471, 292

\reference{5247} --- 2004, in Origin and Evolution of the Elements, ed. A. 
Mc William \& W. Rauch (Cambridge: Cambridge Univ. Press), p. 336

\reference{6939} --- 2009, Physica Scripta,   in press.

\reference{3724} Jenkins, E. B., Gry, C., \& Dupin, O. 2000, \aap,  354, 
253

\reference{1359} Jenkins, E. B., \& Savage, B. D. 1974, \apj,  187, 243

\reference{1063} Jenkins, E. B., Savage, B. D., \& Spitzer, L. 1986, 
\apj,  301, 355

\reference{1330} Jenkins, E. B., Silk, J., \& Wallerstein, G. 1976, 
\apjs,  32, 681

\reference{6035} Jenkins, E. B., \& Tripp, T. M. 2006, \apj,  637, 548

\reference{3197} Jenkins, E. B., \& Wallerstein, G. 1996, \apj,  462, 758

\reference{1003} Jenkins, E. B., Drake, J. F., Morton, D. C., Rogerson, J. 
B., Spitzer, L., \& York, D. G. 1973, \apjl,  181, L122

\reference{4135} Jenkins, E. B., Tripp, T. M., Fitzpatrick, E. L., Lindler,
 D., Danks, A. C., Beck, T. L., Bowers, C. W., Joseph, C. L., Kaiser, M. E., 
Kimble, R. A., Kraemer, S. B., Robinson, R. D., Timothy, J. G., Valenti, J. 
A., \& Woodgate, B. E. 1998, \apjl,  492, L147

\reference{3639} Jenkins, E. B., Tripp, T. M., Wozniak, P. R., Sofia, U. 
J., \& Sonneborn, G. 1999, \apj,  520, 182

\reference{5655} Jensen, A. G., Rachford, B. L., \& Snow, T. P. 2005, 
\apj,  619, 891

\reference{6395} --- 2007, \apj,  654, 955

\reference{6558} Jensen, A. G., \& Snow, T. P. 2007a, \apj,  669, 378

\reference{6559} --- 2007b, \apj,  669, 401

\reference{6786} Jensen, K. S. 1981, \aaps,  45, 455

\reference{3855} Johnson, H. L., \& Morgan, W. W. 1953, \apj,  117, 313

\reference{6875} Johnson, H. L., Mitchell, R. I., Iriarte, B., \& 
Wisniewski, W. Z. 1966, Comm. Lunar and Planetary Lab.,  4, 99

\reference{6806} Jones, A. P., Duley, W. W., \& Williams, D. A. 1990, 
\qjras,  31, 567

\reference{1948} Joseph, C. L. 1988, \apj,  335, 157

\reference{1728} Joseph, C. L., \& Jenkins, E. B. 1991, \apj,  368, 201

\reference{5468} Juett, A. M., Schulz, N. S., \& Chakrabarty, D. 2004, 
\apj,  612, 308

\reference{2774} Jura, M. 1974, \apjl,  190, L33

\reference{1023} Jura, M., \& York, D. G. 1978, \apj,  219, 861

\reference{6918} Kaltcheva, N., \& Knude, J. 2002, \aap,  385, 1107

\reference{5677} Khare, P., Kulkarni, V. P., Lauroesch, J. T., York, D. G.,
 Crotts, A. P. S., \& Nakamura, O. 2004, \apj,  616, 86

\reference{6777} Kilkenny, D., \& Muller, S. 1989, South African. Astr. 
Obs. Circ,  13, 69

\reference{3282} Kim, S. H., \& Martin, P. G. 1996, \apj,  462, 296

\reference{5169} Kimura, H., Mann, I., \& Jessberger, E. K. 2003, \apj,  
582, 846

\reference{6330} Knauth, D. C., Meyer, D. M., \& Lauroesch, J. T. 2006, 
\apjl,  647, L115

\reference{5266} Knauth, D., Andersson, B.-G., McCandliss, S., \& Moos, H.
 W. 2003, \apj,  596, L51

\reference{5514} Knauth, D. C., Andersson, B.-G., McCandliss, S. R., \& 
Moos, H. W. 2004, \nat,  429, 636

\reference{6938} --- 2006, in Astrophysics in the Far Ultraviolet, Five 
Years of Discovery with FUSE, ed. G. Sonneborn, H. W. Moos \& B.-G. 
Andersson (San Francisco: Ast. Soc. Pacific), p. 421

\reference{6927} Kohoutek, L., \& Wehmeyer, R. 1999, \aaps,  134, 255

\reference{6953} Kozasa, T., Nozawa, T., Tominaga, N., Umeda, H., Maeda, 
K., \& Nomoto, K. 2009, arXiv 0903.0217

\reference{6867} Kr\"uger, H. et al. 2006, \planss,  54, 932

\reference{4963} Kruk, J. W., Howk, J. C., Andr\'e, M., Moos, H. W., 
Oegerle, W. R., Oliveira, C., Sembach, K. R., Chayer, P., Linsky, J. L., Wood,
 B. E., Ferlet, R., H\'ebrard, G., Lemoine, M., Vidal-Madjar, A., \& 
Sonneborn, G. 2002, \apjs,  140, 19

\reference{5717} Kulkarni, V. P., Fall, S. M., Lauroesch, J. T., York, D. 
G., Welty, D. E., Khare, P., \& Truran, J. W. 2005, \apj,  618, 68

\reference{6606} Lallement, R., H\'ebrard, G., \& Welsh, B. Y. 2008, 
\aap, 481, 381

\reference{5533} Lallement, R., Welsh, B. Y., Vergely, J. L., Crifo, F., 
\& Sfeir, D. 2003, \aap,  411, 447

\reference{6869} Landgraf, M., Baggaley, W. J., Gr\"un, E., Kr\"uger, H., 
\& Linkert, G. 2000, \jgr,  105, 10343

\reference{1784} Laurent, C., Vidal-Madjar, A., \& York, D. G. 1979, 
\apj,  229, 923

\reference{6020} Lebouteiller, V., Kuassivi, \& Ferlet, R. 2005, \aap,  
443, 509

\reference{4901} Ledoux, C., Bergeron, J., \& Petitjean, P. 2002, \aap,  
385, 802

\reference{5145} Lehner, N., Jenkins, E. B., Gry, C., Moos, H. W., Chayer, 
P., \& Lacour, S. 2003, \apj,  595, 858

\reference{3885} Lesh, J. R. 1968, \apjs,  17, 371

\reference{6778} Levato, H. 1972, \pasp,  84, 584

\reference{6779} Levato, H., \& Abt, H. A. 1976, \pasp,  88, 712

\reference{5924} Li, A. 2005, \apj,  622, 965

\reference{6863} Ling, Z., Zhang, S. N., Xiang, J., \& Tang, S. 2009, 
\apj, 690, 224

\reference{3085} Linsky, J. L., Diplas, A., Wood, B. E., Brown, A., Ayres, 
T. R., \& Savage, B. D. 1995, \apj,  451, 335

\reference{256} Lipman, K., \& Pettini, M. 1995, \apj,  442, 628

\reference{5604} Lodders, K. 2003, \apj,  591, 1220

\reference{3446} Lu, L., Sargent, W. L. W., Barlow, T. A., Churchill, C. 
W., \& Vogt, S. S. 1996, \apjs,  107, 475

\reference{6731} Luck, R. E., Kovtyukh, V. V., \& Andrievsky, S. M. 2006, 
\aj,  132, 902

\reference{1027} Lugger, P. M., York, D. G., Blanchard, T., \& Morton, D. 
C. 1978, \apj,  224, 1059

\reference{3329} Ma\'{\i}z-Apell\'aniz, J. 2001, \apj,  560, L83

\reference{5205} Mallouris, C. 2003, \apjs,  147, 265

\reference{1042} Martin, E. R., \& York, D. G. 1982, \apj,  257, 135

\reference{4867} Mart\'{\i}n-Hern\'andez, N. L., Peeters, E., Morisset, C.,
 Tielens, A. G. G. M., Cox, P., Roelfsema, P. R., Baluteau, J.-P., Schaerer, 
D., Mathis, J. S., Damour, F., Churchwell, E., \& Kessler, M. F. 2002, \aap,
  381, 606

\reference{4765} Martins, L. P., \& Viegas, S. M. M. 2000, \aap,  361, 
1121

\reference{3609} Massa, D., Van Steenberg, M. E., Oliversen, N., \& 
Lawton, P. 1998, in Ultraviolet Astrophysics Beyond the IUE Final Archive, ed.
 W. Wamsteker \& R. Gonzalez Riestra (Noordwijk: ESA), p. 723

\reference{3275} Mathis, J. S. 1990, \araa,  28, 37

\reference{380} --- 1996, \apj,  472, 643

\reference{6766} Mel\'endez, J., \& Asplund, M. 2008, \aap, 490, 817

\reference{5227} Merrill, P. W., Sanford, R. F., Wilson, O. C., \& 
Burwell, C. G. 1937, \apj,  86, 274

\reference{3466} Meyer, D. M., Cardelli, J. A., \& Sofia, U. J. 1997, 
\apjl,  490, L103

\reference{4196} Meyer, D. M., Jura, M., \& Cardelli, J. A. 1998, \apj,  
493, 222

\reference{1830} Meyer, D. M., \& Roth, K. C. 1990, \apj,  363, 57

\reference{2858} Meyer, D. M., Jura, M., Hawkins, I., \& Cardelli, J. A. 
1994, \apjl,  437, L59

\reference{6439} Miller, A., Lauroesch, J. T., Sofia, U. J., Cartledge, S. 
I. B., \& Meyer, D. M. 2007, \apj,  659, 441

\reference{3872} Morgan, W. W., Code, A. D., \& Whitford, A. E. 1955, 
\apjs,  2, 41

\reference{6781} Morgan, W. W., Hiltner, W. A., \& Garrison, R. F. 1971, 
\aj,  76, 242

\reference{6776} Morgan, W. W., \& Roman, N. G. 1950, \apj,  112, 362

\reference{6789} Morris, P. M. 1961, \mnras,  122, 325

\reference{1203} Morton, D. C. 1978, \apj,  222, 863

\reference{96} --- 1991, \apjs,  77, 119

\reference{5404} --- 2003, \apjs,  149, 205

\reference{1772} Morton, D. C., \& Dinerstein, H. L. 1976, \apj,  204, 1

\reference{1357} Morton, D. C., Jenkins, E. B., \& Bohlin, R. C. 1968, 
\apj,  154, 661

\reference{6970} Morton, D. C., \& Spitzer, L. 1966, \apj,  144, 1

\reference{1198} Morton, D. C., Drake, J. F., Jenkins, E. B., Rogerson, J. 
B., Spitzer, L., \& York, D. G. 1973, \apjl,  181, L103

\reference{6226} M\"unch, G. 1957, \apj,  125, 42

\reference{6225} M\"unch, G., \& Zirin, H. 1961, \apj,  133, 11

\reference{4611} Murphy, E. M., Sembach, K. R., Gibson, B. K., Shull, J. 
M., Savage, B. D., Roth, K. C., Moos, H. W., Green, J. C., York, D. G., \& 
Wakker, B. P. 2000, \apj,  538, L35

\reference{6598} Nieva, M. F., \& Przybilla, N. 2008a, \aap, 481, 199

\reference{6587} --- 2008b, \rmxaa, 33, 35

\reference{6814} Oegerle, W. R., \& Polidan, R. S. 1984, \apj,  285, 648

\reference{6332} Oliveira, C. M., \& H\'ebrard, G. 2006, \apj,  653, 345

\reference{5134} Oliveira, C. M., H\'ebrard, G., Howk, J. C., Kruk, J. W., 
Chayer, P., \& Moos, H. W. 2003, \apj,  587, 235

\reference{4789} Paerels, F., Brinkman, A. C., van der Meer, R. L. J., 
Kaastra, J. S., Kuulkers, E., den Boggende, A. J. F., Predehl, P., Drake, J. 
J., Kahn, S. M., Savin, D. W., \& McLaughlin, B. M. 2001, \apj,  546, 338

\reference{6823} Palumbo, M. E. 2006, \aap,  453, 903

\reference{6150} P\'eroux, C., Kulkarni, V. P., Meiring, J., Ferlet, R., 
Khare, P., Lauroesch, J. T., Vladilo, G., \& York, D. G. 2006a, \aap,  450, 
53

\reference{6336} P\'eroux, C., Meirling, J. D., Kulkarni, V. P., Ferlet, 
R., Khare, P., Lauroesch, J. T., Vladilo, G., \& York, D. G. 2006b, \mnras, 
 372, 369

\reference{6665} P\'eroux, C., Meiring, J. D., Kulkarni, V. P., Khare, P., 
Lauroesch, J. T., Vladilo, G., \& York, D. G. 2008, \mnras,  386, 2209

\reference{5202} Pettini, M. 2003, in Cosmochemistry: The Melting Pot of 
the Elements, ed. C. Esteban, R. J. Garci, A. H. L\'opez \& F. S\'anchez 
(Cambridge: Cambridge Univ.), p. 257

\reference{1162} Pettini, M., Boksenberg, A., \& Hunstead, R. W. 1990, 
\apj,  348, 48

\reference{2665} Pettini, M., Smith, L. J., Hunstead, R. W., \& King, D. 
L. 1994, \apj,  426, 79

\reference{4094} Pettini, M., King, D. L., Smith, L. J., \& Hunstead, R. 
W. 1997, \apj,  478, 536

\reference{3635} Pettini, M., Ellison, S. L., Steidel, C. C., \& Bowen, D.
 V. 1999, \apj,  510, 576

\reference{5540} Pettini, M., Ellison, S. L., Bergeron, J., \& Petitjean, 
P. 2002, \aap,  391, 21

\reference{6774} Popper, D. M. 1943, \apj,  97, 394

\reference{3558} Press, W. H., Teukolsky, S. A., Vetterling, W. T., \& 
Flannery, B. P. 2007, Numerical Recipes, The Art of Scientific Computing,  
3rd ed., (Cambridge: Cambridge Univ. Press)

\reference{6937} Prochaska, J. X. 2003, \apj,  582, 49

\reference{5393} --- 2004, in Origin and Evolution of the Elements, ed. A. 
McWilliam \& M. Rauch (Cambridge: Cambridge Univ. Press), p. 455

\reference{5238} Prochaska, J. X., Howk, J. C., \& Wolfe, A. M. 2003, 
\nat,  423, 57

\reference{5846} Prochaska, J. X., Tripp, T. M., \& Howk, J. C. 2005, 
\apjl,  620, L39

\reference{4796} Prochaska, J. X., \& Wolfe, A. M. 2002, \apj,  566, 68

\reference{5595} Prochaska, J. X., Howk, J. C., O'Meara, J. M., Tytler, D.,
 Wolfe, A. M., Kirkman, D., Lubin, D., \& Suzuki, N. 2002, \apj,  571, 693

\reference{6472} Prochaska, J. X., Chen, H.-W., Desauges-Zavadsky, M., \& 
Bloom, J. S. 2007, \apj,  666, 267

\reference{6872} Przybilla, N., Nieva, M.-F., \& Butler, K. 2008, \apj, 688, L103

\reference{6849} Quast, R., Reimers, D., \& Baade, R. 2008, \aap,  477, 
443

\reference{4821} Rachford, B., Snow, T. P., Tumlinson, J., Shull, J. M., 
Blair, W. P., Ferlet, R., Friedman, S. D., Gry, C., Jenkins, E. B., Morton, D.
 C., Savage, B. D., Sonnentrucker, P., Vidal-Madjar, A., Welty, D. E., \& 
York, D. G. 2002, \apj,  577, 221

\reference{4660} Rachford, B. L., Snow, T. P., Tumlinson, J., Shull, J. M.,
 Roueff, E., Andre, M., Desert, J.-M., Ferlet, R., Vidal-Madjar, A., \& York,
 D. G. 2001, \apj,  555, 839

\reference{6803} Rachford, B. L., Snow, T. P., Destree, J. D., Ross, T. L.,
 Ferlet, R., Friedman, S. D., Gry, C., Jenkins, E. B., Morton, D. C., Savage, 
B. D., Shull, J. M., Sonnentrucker, P., Tumlinson, J., Vidal-Madjar, A., 
Welty, D. E., \& York, D. G. 2009, \apjs,  180, 125

\reference{5392} Redfield, S., \& Linsky, J. L. 2004a, \apj,  602, 776

\reference{5603} --- 2004b, \apj,  613, 1004

\reference{3980} Richter, P., Sembach, K. R., Wakker, B. P., Savage, B. D.,
 Tripp, T. M., Murphy, E. M., Kalberla, P. M. W., \& Jenkins, E. B. 2001, 
\apj,  559, 318

\reference{1393} Rogerson, J. B., Spitzer, L., Drake, J. F., Dressler, K., 
Jenkins, E. B., Morton, D. C., \& York, D. G. 1973a, \apjl,  181, L97

\reference{1365} Rogerson, J. B., York, D. G., Drake, J. F., Jenkins, E. 
B., Morton, D. C., \& Spitzer, L. 1973b, \apjl,  181, L110

\reference{6831} Rolleston, W. R. J., Smartt, S. J., Dufton, P. L., \& 
Ryans, R. S. I. 2000, \aap,  363, 537

\reference{296} Roth, K. C., \& Blades, J. C. 1995, \apjl,  445, L95

\reference{4110} --- 1997, \apjl,  474, L95

\reference{2391} Routly, P. M., \& Spitzer, L. 1952, \apj,  115, 227

\reference{4543} Ryu, K. S., Dixon, W. V., Hurwitz, M., Seon, K. I., Min, 
K. W., \& Edelstein, J. 2000, \apj,  529, 251

\reference{4158} Sahu, M. S., \& Blades, J. C. 1997, \apjl,  484, L125

\reference{5173} Sarlin, S. P. 1998, Ph.D. Thesis, University of Colorado, 
Boulder.

\reference{1142} Savage, B. D., \& Bohlin, R. C. 1979, \apj,  229, 136

\reference{2369} Savage, B. D., Cardelli, J. A., \& Sofia, U. J. 1992, 
\apj,  401, 706

\reference{1000} Savage, B. D., \& Jenkins, E. B. 1972, \apj,  172, 491

\reference{5970} Savage, B. D., \& Lehner, N. 2006, \apjs,  162, 134

\reference{1169} Savage, B. D., Massa, D., \& Meade, M. 1985, \apjs,  59,
 397

\reference{3966} Savage, B. D., Meade, M. R., \& Sembach, K. R. 2001, 
\apjs,  136, 631

\reference{1138} Savage, B. D., \& Panek, R. J. 1974, \apj,  191, 659

\reference{110} Savage, B. D., \& Sembach, K. R. 1991, \apj,  379, 245

\reference{328} --- 1996a, \araa,  34, 279

\reference{3559} --- 1996b, \apj,  470, 893

\reference{1141} Savage, B. D., Bohlin, R. C., Drake, J. F., \& Budich, W.
 1977, \apj,  216, 291

\reference{6876} Schrijver, H. 1997, The Hipparcos and Tycho Catalogues, 
(ESA Publications, SP-1200),  (Noordwijk: European Space Agency)

\reference{6852} Schulz, N. S., Cui, W., Canizares, C. R., Marshall, H. L.,
 Lee, J. C., Miller, J. M., \& Lewin, W. H. G. 2002, \apj,  565, 1141

\reference{2960} Sembach, K. R., \& Savage, B. D. 1996, \apj,  457, 211

\reference{4085} Sembach, K. R., Savage, B. D., \& Tripp, T. M. 1997, 
\apj,  480, 216

\reference{1868} Shaver, P. A., McGee, R. X., Newton, L. M., Danks, A. C., 
\& Pottasch, S. R. 1983, \mnras,  204, 53

\reference{6517} Sheffer, Y., Rogers, M., Federman, S. R., Lambert, D. L., 
\& Gredel, R. 2007, \apj,  667, 1002

\reference{6751} Sheffer, Y., Rogers, M., Federman, S. R., Abel, N. P., 
Gredel, R., Lambert, D. L., \& Shaw, G. 2008, \apj,  687, 1075

\reference{1800} Shull, J. M., \& McKee, C. F. 1979, \apj,  227, 131

\reference{1058} Shull, J. M., \& Van Steenberg, M. E. 1985, \apj,  294, 
599

\reference{1770} Shull, J. M., \& York, D. G. 1977, \apj,  211, 803

\reference{6793} Simonson, S. C. 1968, \apj,  154, 923

\reference{2996} Slettebak, A. 1982, \apjs,  50, 55

\reference{3426} Smith, R. G., Sellgren, K., \& Brooke, T. Y. 1993, 
\mnras,  263, 749

\reference{6862} Smith, R. K. 2008, \apj,  681, 343

\reference{4424} Smith, R. K., \& Dwek, E. 1998, \apj,  503, 831

\reference{6861} Smith, R. K., Dame, T. M., Costantini, E., \& Predehl, P.
 2006, \apj,  648, 452

\reference{1255} Snow, T. P. 1976, \apj,  204, 759

\reference{1258} --- 1977, \apj,  216, 724

\reference{1262} Snow, T. P., Peters, G. J., \& Mathieu, R. D. 1979, 
\apjs,  39, 359

\reference{4971} Snow, T. P., Rachford, B. L., \& Figoski, L. 2002, \apj,
  573, 662

\reference{3433} Snow, T. P., \& Witt, A. N. 1996, \apjl,  468, L65

\reference{3317} Snow, T. P., Black, J. H., van Dishoeck, E. F., Burks, G.,
 Crutcher, R. M., Lutz, B. L., Hanson, M. M., \& Shuping, R. Y. 1996, \apj, 
 465, 245

\reference{4613} Snow, T. P., Rachford, B. L., Tumlinson, J., Shull, J. M.,
 Welty, D. E., Blair, W. P., Ferlet, R., Friedman, S. D., Gry, C., Jenkins, E.
 B., Lecavelier, A., Lemoine, M., Morton, D. C., Savage, B. D., Sembach, K. 
R., Vidal-Madjar, A., York, D. G., Andersson, B.-G., Feldman, P. D., \& Moos,
 H. W. 2000, \apj,  538, L65

\reference{6450} Socas-Navarro, H., \& Norton, A. A. 2007, \apjl,  660, 
L153

\reference{6895} Socrates, A., \& Draine, B. T. 2008, arXiv, 0812.3913,

\reference{2722} Sofia, U. J., Cardelli, J. A., \& Savage, B. D. 1994, 
\apj,  430, 650

\reference{3820} Sofia, U. J., Fitzpatrick, E. L., \& Meyer, D. M. 1998, 
\apjl,  504, L47

\reference{4310} Sofia, U. J., \& Jenkins, E. B. 1998, \apj,  499, 951

\reference{2850} Sofia, U. J., \& Meyer, D. M. 2001, \apj,  554, L221

\reference{6887} Sofia, U. J., \& Parvathi, V. S. 2009, in Cosmic Dust - 
Near and Far, ed. T. Henning, E. Gr\"un \& J. Steinacker (San Francisco: Ast.
 Soc. Pacific), in press.

\reference{4257} Sofia, U. J., Cardelli, J. A., Guerin, K. P., \& Meyer, 
D. M. 1997, \apjl,  482, L105

\reference{5491} Sofia, U. J., Lauroesch, J. T., Meyer, D. M., \& 
Cartledge, S. I. B. 2004, \apj,  605, 272

\reference{6971} Sofia, U. J., Gordon, K. D., Clayton, G. C., Misselt, K., 
Wolff, M. J., Cox, N. L. J., \& Ehrenfreund, P. 2006, \apj,  636, 753

\reference{3842} Sonneborn, G., Tripp, T. M., Ferlet, R., Jenkins, E. B., 
Sofia, U. J., Vidal-Madjar, A., \& Wozniak, P. R. 2000, \apj,  545, 277

\reference{5207} Sonnentrucker, P., Friedman, S. D., Welty, D. E., York, D.
 G., \& Snow, T. P. 2003, \apj,  596, 350

\reference{2773} Spitzer, L. 1985, \apjl,  290, L21

\reference{1015} Spitzer, L., Cochran, W. D., \& Hirshfeld, A. 1974, 
\apjs,  28, 373

\reference{6810} Spitzer, L., \& Field, G. B. 1955, \apj,  121, 300

\reference{2462} Spitzer, L., \& Fitzpatrick, E. L. 1993, \apj,  409, 299

\reference{250} --- 1995, \apj,  445, 196

\reference{1326} Spitzer, L., \& Jenkins, E. B. 1975, \araa,  13, 133

\reference{1002} Spitzer, L., Drake, J. F., Jenkins, E. B., Morton, D. C., 
Rogerson, J. B., \& York, D. G. 1973, \apjl,  181, L116

\reference{6638} Stancil, P. C., Schultz, D. R., Kimura, M., Gu, J.-P., 
Hirsch, G., \& Buenker, R. J. 1999, \aaps,  140, 225

\reference{1028} Stokes, G. M. 1978, \apjs,  36, 115

\reference{6780} Strassmeier, K. G., \& Fekel, F. C. 1990, \aap,  230, 
389

\reference{3101} Str\"omgren, B. 1948, \apj,  108, 242

\reference{6917} Sudzius, J., \& Bobinas, V. 1992, Bull. Vilnius Astr. 
Obs.,  86, 59

\reference{5403} Takei, Y., Fujimoto, R., Mitsuda, K., \& Onaka, T. 2002, 
\apj,  581, 307

\reference{6790} Thackeray, A. D. 1971, \mnras,  154, 103

\reference{4309} Tielens, A. G. G. M. 1998, \apj,  499, 267

\reference{6921} Tobin, W. 1985, \aap,  142, 189

\reference{6919} Tobin, W., \& Kaufmann, J. P. 1984, \mnras,  207, 369

\reference{6920} Tobin, W., Viton, M., \& Sivan, J.-P. 1994, \aaps,  107,
 385

\reference{5160} Tripp, T. M., Wakker, B. P., Jenkins, E. B., Bowers, C. 
W., Danks, A. C., Green, R. F., Heap, S. R., Joseph, C. L., Kaiser, M. E., 
Linsky, J. L., \& Woodgate, B. E. 2003, \aj,  125, 3122

\reference{6858} Ueda, Y., Mitsuda, K., Murakami, H., \& Matsushita, K. 
2005, \apj,  620, 274

\reference{3494} Vallerga, J. 1996, \ssr,  78, 277

\reference{6826} van Dishoeck, E. F. 1998, Faraday Disc.,  109, 31

\reference{6825} --- 2004, \araa,  42, 119

\reference{6798} van Leeuwen, F. 2007, \aap,  474, 653

\reference{2017} Van Steenberg, M. E., \& Shull, J. M. 1988, \apjs,  67, 
225

\reference{6945} Venn, K. A., Krwin, M., Shetrone, M. D., Tout, C. A., 
Hill, V., \& Tolstoy, E. 2004, \aj,  128, 1177

\reference{4965} Vidal-Madjar, A., \& Ferlet, R. 2002, \apj,  571, L169

\reference{3509} Vidal-Madjar, A., Ferlet, R., Laurent, C., \& York, D. G.
 1982, \apj,  260, 128

\reference{5213} Vladilo, G. 2002a, \aap,  391, 407

\reference{4905} --- 2002b, \apj,  569, 295

\reference{5519} --- 2004, \aap,  421, 479

\reference{6242} Vladilo, G., Centuri\'on, M., Levshakov, S. A., P\'eroux, 
C., Khare, P., Kulkarni, V. P., \& York, D. G. 2006, \aap,  454, 151

\reference{6601} Vladilo, J. 2008, in Pathways through an Exlectic Universe,
ed. J. H. Knapen, T. J. Mahoney, and A. Vazdakis, (San Francisco: Ast. Soc. Pacific),
p. 562

\reference{3962} Wakker, B. P. 2001, \apjs,  136, 463

\reference{3868} Wakker, B. P., \& Mathis, J. S. 2000, \apj,  544, L107

\reference{4815} Wakker, B. P., Howk, J. C., Savage, B. D., van Woerden, 
H., Tufte, S. L., Schwarz, U. J., Benjamin, R., Reynolds, R. J., Peletier, R. 
F., \& Kalberla, P. M. W. 1999, \nat,  402, 388

\reference{3874} Walborn, N. R. 1971, \apjs,  23, 257

\reference{3870} --- 1972, \aj,  77, 312

\reference{2111} Wallerstein, G., \& Gilroy, K. K. 1992, \aj,  103, 1346

\reference{2807} Wallerstein, G., \& Goldsmith, D. 1974, \apj,  187, 237

\reference{6070} Wegner, W. 1994, \mnras,  270, 229

\reference{3471} Welsh, B. Y., Sasseen, T., Craig, N., Jelinsky, S., \& 
Albert, C. E. 1997, \apjs,  112, 507

\reference{4046} Welty, D. E., Lauroesch, J. T., Blades, J. C., Hobbs, L. 
M., \& York, D. G. 1997, \apj,  489, 672

\reference{4524} Welty, D. E., Hobbs, L. M., Lauroesch, J. T., Morton, D. 
C., Spitzer, L., \& York, D. G. 1999, \apjs,  124, 465

\reference{5126} Welty, D. E., Lauroesch, J. T., Blades, J. C., Hobbs, L. 
M., \& York, D. G. 2001, \apjl,  554, L75

\reference{3425} Whittet, D. C. B., Bode, M. F., Longmore, A. J., Adamson, 
A. J., McFadzean, A. D., Aitken, D. K., \& Roche, P. F. 1988, \mnras,  233, 
321

\reference{6874} Whittet, D. C. B., Boogert, A. C. A., Gerakines, P. A., 
Schutte, W., Tielens, A. G. G. M., de Graauw, T., Prusti, T., van Dishoeck, E.
 F., Wesselius, P. R., \& Wright, C. M. 1997, \apj,  490, 729

\reference{4752} Witt, A. N., Smith, R. K., \& Dwek, E. 2001, \apj,  550,
 L201

\reference{6064} Wolfe, A. M., Gawiser, E., \& Prochaska, J. X. 2005, 
\araa,  43, 861

\reference{4500} Wolff, B., Koester, D., \& Lallement, R. 1999, \aap,  
346, 969

\reference{4258} Wood, B. E., \& Linsky, J. L. 1997, \apjl,  474, L39

\reference{5062} Wood, B. E., Redfield, S., Linsky, J. L., \& Sahu, M. S. 
2002, \apj,  581, 1169

\reference{6860} Xiang, J., Zhang, S. N., \& Yao, Y. 2005, \apj,  628, 
769

\reference{6073} Yao, Y., \& Wang, Q. D. 2006, \apj,  641, 930

\reference{6948} Yao, Y., Schulz, N. S., Gu, M. F., Nowak, M. A., \& 
Canizares, C. R. 2009, arXiv 0902.2778

\reference{1176} York, D. G. 1976, \apj,  204, 750

\reference{1184} --- 1983, \apj,  264, 172

\reference{1783} York, D. G., \& Kinahan, B. F. 1979, \apj,  228, 127

\reference{1175} York, D. G., \& Rogerson, J. B. 1976, \apj,  203, 378

\reference{1073} York, D. G., Spitzer, L., Bohlin, R. C., Hill, J., 
Jenkins, E. B., Savage, B. D., \& Snow, T. P. 1983, \apj,  266, L55

\end{references}
\end{document}